\documentclass[preprint]{aastex6}
\usepackage{graphicx}
\usepackage{epsfig}
\usepackage{multirow}
\usepackage{footnote}
\usepackage{url}

\shorttitle{New Gas Disk White Dwarfs}
\shortauthors{C. Melis {\it et al.}}
\begin{document}


\title{\large \bf Serendipitous Discovery of Nine White Dwarfs With Gaseous Debris Disks}

%% Use \author, \affil, plus the \and command to format author and affiliation 
%% information.  If done correctly the peer review system will be able to
%% automatically put the author and affiliation information from the manuscript
%% and save the corresponding author the trouble of entering it by hand.
%%
%% The \affil should be used to document primary affiliations and the
%% \altaffil should be used for secondary affiliations, titles, or email.

%% Authors with the same affiliation can be grouped in a single
\author{\large Carl Melis}
\affil{email: cmelis@ucsd.edu \\
Center for Astrophysics and Space Sciences, University of California, San Diego, CA 92093-0424, USA} 

\author{\large Beth Klein}
\affil{Department of Physics and Astronomy, University of California, Los Angeles, CA 90095-1562, USA}

\author{\large Alexandra E.\ Doyle}
\affil{Department of Earth, Planetary, and Space Sciences, University of California, Los Angeles, CA 90095, USA}

\author{\large Alycia Weinberger}
\affil{Earth and Planets Laboratory, Carnegie Institution for Science, 5241 Broad Branch Rd NW, Washington, DC 20015, USA}

\author{\large B.\ Zuckerman}
\affil{Department of Physics and Astronomy, University of California, Los Angeles, CA 90095-1562, USA}

\and

\author{\large Patrick Dufour}
\affil{D\'{e}partement de Physique, Universit\'{e} de Montr\'{e}al, Montr\'{e}al, QC H3C 3J7, Canada}

%\affil{American Astronomical Society \\
%2000 Florida Ave., NW, Suite 300 \\
%Washington, DC 20009-1231, USA}

%\author{Butler Burton\altaffilmark{3}}
%\affil{National Radio Astronomy Observatory}

%\affil{TeXnology Inc}

%\author{Julie Steffen\altaffilmark{4}}
%\affil{American Astronomical Society \\
%2000 Florida Ave., NW, Suite 300 \\
%Washington, DC 20009-1231, USA}

%% Use the \and command so offset the last author.
%\and

%\author{Jeff Lewandowski\altaffilmark{5}}
%\affil{IOP Publishing, Washington, DC 20005}


%\altaffiltext{1}{AAS Journals Data Scientist}
%\altaffiltext{2}{greg.schwarz@aas.org}
%\altaffiltext{3}{AAS Journals Associate Editor-in-Chief}
%\altaffiltext{4}{AAS Director of Publishing}
%\altaffiltext{5}{IOP Senior Publisher for the AAS Journals}

\begin{abstract}

\large{
Optical spectroscopic observations of white dwarf stars selected from catalogs based
on the {\it Gaia} DR2 database
reveal nine new gaseous debris disks that orbit single white dwarf stars, 
about a factor of two increase over the
previously known sample. For each source we present gas emission
lines identified and basic stellar parameters, including abundances for lines seen
with low-resolution spectroscopy. 
Principle discoveries include:
(1) the coolest white dwarf (T$_{\rm eff}$$\approx$12,720\,K) with a gas disk; this star, 
WD0145+234, has been reported to have undergone a recent infrared outburst;
(2) co-location in velocity space of gaseous emission from multiple elements, suggesting that
different elements are well-mixed;
(3) highly asymmetric emission structures toward SDSS\,J0006+2858, and possibly
asymmetric structures for two other systems;
(4) an overall sample composed of approximately 25\% DB and 75\% DA white dwarfs, consistent with the overall distribution of primary atmospheric types found in the field population; and
(5) never-before-seen emission lines from Na in the spectra of 
Gaia\,J0611$-$6931, semi-forbidden Mg, Ca, and Fe lines toward WD\,0842+572,
and Si in both stars.
The currently known sample of gaseous debris disk systems is significantly skewed towards 
northern hemisphere stars, suggesting a dozen or so emission line stars are waiting to be found
in the southern hemisphere.\\*[1.0mm]}

\end{abstract}

\keywords{Circumstellar matter (241) --- Exoplanet systems (484) --- Stellar abundances (1577) --- White dwarf stars (1799)}



\large{

% introduction - gaseous debris disks and theories for them, origin of material
\section{\large \bf Introduction}

The photopsheres of white dwarf stars are ``polluted'' by remnant solid material from planetary
systems that otherwise stably orbited their host star while it was on the main
sequence (e.g., \citealt{zuckerman07}; \citealt{jura08}; \citealt{farihi09}; 
\citealt{dufour10}; \citealt{klein10}; \citealt{melis10}; \citealt{gaensicke12};
\citealt{xu13}, and references therein).
Prior to being accreted, this solid material must find its way into the host star's Roche
radius and be tidally shredded into disks of dust and gas 
(e.g., \citealt{debes02}; \citealt{jura03b}; \citealt{frewen14}; 
\citealt{vanderburg15}; \citealt{veras16}; \citealt{manser19}).

While the most heavily polluted white dwarf stars known are typically host to
infrared excess emission and hence orbiting dust disks (e.g., \citealt{farihi16} and references
therein), a subset of these objects additionally are host to disks of gaseous metallic material 
%other gas disk discovery/observation papers?
(\citealt{gaensicke06,gaensicke07,gaensicke08}; \citealt{melis10,melis12}; 
\citealt{brinkworth12}; \citealt{farihi12a}; \citealt{hartmann16}; \citealt{xu16}; 
\citealt{manser16b,manser16,manser20}).
Models of disk formation and evolution suggest that dusty material is a necessary
pre-condition to producing gaseous material in white dwarf 
debris disks.  The gaseous material may form
either from repeated impacts of solids within the dust disk or from sublimation of
dusty material at the inner edge of the debris disk; the gas
subsequently viscously spreads throughout the disk
%any more theory papers?
(e.g., \citealt{jura08}; \citealt{melis10}; \citealt{rafikov11a,rafikov11b};
\citealt{bochkarev11}; \citealt{hartmann11,hartmann16}; \citealt{metzger12}; \citealt{bear13}; 
\citealt{kenyon17}).

Gaseous material may act to aerodynamically drag dust particles, removing angular 
momentum and transporting them to the inner disk and eventually the star.
Detailed study of dust and gas disk components around single white dwarf stars
can provide insight into the fate of solid material in planetary systems and hence
what may very well happen to our own solar system billions of years from now
%more Veras and Mustill papers on post-MS simulations?
(e.g., \citealt{debes02};
\citealt{farihi09};
\citealt{frewen14}; \citealt{veras16}; \citealt{manser16}; \citealt{xu16}; \citealt{cauley18};
\citealt{mustill18};
\citealt{grishin19}; \citealt{malamud20}; \citealt{maldonado20a,maldonado20b}). Additionally,
intensive monitoring of gas-disk white dwarf stars can possibly 
provide a way to elucidate the size, density, and orbit of disintegrating 
rocky bodies around white dwarf stars; for example what appears to be part of a core from a
differentiated rocky object orbiting the heavily polluted star SDSSJ1228+1040 \citep{manser19}
and an intact gas giant planet being accreted by WD\,J091405.30+191412.25 \citep{gaensicke19}.

In this paper we report on the discovery of a set of nine new white dwarf stars that host
gas disk emission lines in their optical spectra. Included in this set is the recently reported
infrared ``outburst'' object WD0145+234 \citep{wang19}. 
In addition to detection of gas disk emission lines
in this system (confirming it to be host to gaseous and dusty disk components), we also
identify atmospheric pollution indicating that this material is being accreted by the host
star. Below we describe our target selection strategy, observations conducted
for newly discovered gas disk systems, 
measurements made, target-specific discussion, and an overall sample
discussion.

\section{\large \bf Target Selection and Observations}
\label{secobs}

Stars were observed as part of a large-scale survey to identify the most
heavily polluted white dwarfs through low-resolution optical spectroscopy.
We began with the sample of \citet{gentilefusillo19} which reports the probability of
an object being a white dwarf and stellar parameters 
(e.g., T$_{\rm eff}$ and log$g$) based on fits to photometric
data. These determinations are heavily informed by $Gaia$ DR2 data, thus we
refer to it as the ``$Gaia$ DR2 sample''.
Targets with evidence for infrared excess emission or having a helium-dominated
atmosphere received higher priority for follow-up. 
Infrared excess emission data came from
2MASS, VISTA, and {\it WISE} (e.g., \citealt{xu20}) while $GALEX$ colors 
were used to identify helium-dominated atmosphere white dwarfs
(where available; helium-dominated atmosphere targets can have
more readily-identifiable atmospheric pollution at low spectral resolution than can
hydrogen-dominated atmosphere objects).
In this paper we do not assess the reality of any putative infrared excess emission detected
for our target stars nor do we report on any infrared emission characteristics; this will be the
subject of future work. 
In sum, we have observed $\approx$500 white dwarf stars as a part of this survey.
After serendipitously finding two
gas-disk hosting white dwarf stars, we began routinely using a spectral setup that 
covered the Ca~II infrared triplet (IRT) region where the strongest gas disk emission lines
are typically seen (e.g., \citealt{melis10}; \citealt{manser16}, and references therein).

Table \ref{tabobs} lists all observation dates and resulting data properties. 
Grating resolution optical spectroscopy was performed at
Lick Observatory with the Kast Double Spectrograph mounted
on the Shane 3\,m telescope and at Gemini-South with GMOS (\citealt{hook04}; hereafter GMOS-S).
In general, all Kast observations employed the blue and red arms
with light split by the d57 dichroic around 5700\,\AA.
After splitting, blue light was passed through the 600/4310 grism while
red light was in most observations passed through the 830/8460 grating (and
in a few, the 600/7500 grating which provided broader wavelength coverage). 
Slit widths of 1, 1.5, and 2$'$$'$ were
used depending on seeing and cloud extinction; integration times ranged between
30-60~minutes depending on target brightness and conditions.
GMOS-S observations used the B600 grating, a slit  size of 0.5$''$, were centered 
at a wavelength of 525\,nm, and exposed for 15-20~minutes.
GMOS-S spectra are recorded onto three detectors with a gap in wavelength coverage
between each detector; we did not use multiple setups to cover these gaps.
Final GMOS-S wavelength coverage is given in Table \ref{tabobs}.
Grating resolution data are reduced using standard
{\sf IRAF} long-slit tasks including bias subtraction, flat-fielding, wavelength calibration with
arc lamps, and instrumental response calibration via observations
of flux calibration standard stars. Arc lamp frames are not obtained close in time to science
frames and as such the zero-point of the wavelength scale is not accurate.

Higher resolution observations were obtained for new gas disk white dwarf
systems with the Keck~I telescope and HIRES at Mauna Kea Observatory
\citep{vogt94}
%not using MIKE data for J0644-0352 (Debes' data)
%with the Clay telescope and MIKE at Las Campanas Observatory,
and with the Baade telescope and MagE at Las Campanas Observatory
\citep{marshall08}.
HIRES data were taken with the C5 decker (1.148$''$ slit width),
had exposure times of 30-80~minutes, and are reduced using the
{\sf MAKEE} software package which outputs heliocentric velocity-corrected spectra
shifted to vacuum wavelengths. 
While HIRES wavelength coverage is
quoted as continuous in Table \ref{tabobs}, there are gaps in coverage between each of the 
three CCDs and sometimes between red orders.
%MIKE and 
MagE data were taken with the 0.5$''$ slit, integrated for 60~minutes, and
are reduced with the facility Carnegie Python pipeline 
\citep{kelson00,kelson03}.
After reduction and extraction, 
polynomials are fit to each order to bring overlapping 
order segments into agreement before combining all orders of every
exposure to generate a final spectrum for analysis. More details
about reducing echelle data in the presence of gas disk emission lines
can be found in \citet{melis10}.

%% In this section, we use  the \subsection command to set off
%% a subsection.  \footnote is used to insert a footnote to the text.


%% This section also includes several of the displayed math environments
%% mentioned in the Author Guide.

% modeling+results - Patrick blurb, table of target properties+gas line meas, figure of gas disk lines
\section{\large \bf Measurements}
\label{secmodres}

\subsection{\large \bf Atmospheric Parameters and Abundances}

For each target we use low-resolution data to obtain atmospheric parameters
(e.g., T$_{\rm eff}$, log$g$) 
and then measure abundances for any detected metallic absorption lines. 
In this paper we only report on elements detected in the low-resolution spectra,
results from higher-resolution spectra will appear in future publications
dedicated to atmospheric pollution.
Fitting proceeds as in 
\citet{melis11}, \citet{melis17}, and references therein.
We briefly summarize this process here.

To obtain atmospheric parameters we perform model fits to hydrogen Balmer lines 
and helium lines in the low resolution data. The method follows the ``spectroscopic technique'' 
described by \citet{bergeron92} and described at length in \citet{liebert05} 
and references therein. Gaia DR2 distances are used to help constrain the stellar radius
and hence log$g$.
Uncertainties for resulting T$_{\rm eff}$ and log$g$ values are not calculated 
individually for each star, but are typically
$\pm$1,000\,K for T$_{\rm eff}$ and $\pm$0.1\,dex for log$g$
(e.g.,  \citealt{bergeron19}; \citealt{genestbeaulieu19a,genestbeaulieu19b}
and references therein).

When metallic absorption lines are present in the low-resolution spectra, we measure
abundances by fitting synthetic model spectra to the data.
A grid of local thermodynamic equillibrium (LTE) 
model atmospheres are generated with code similar to that 
described in \citet{dufour05,dufour07} where
absorption line data are taken from the Vienna Atomic Line Database.
Synthetic spectra in the grid cover a range of abundances typically 
from log[$n$(Z)/$n$(H)]= $-$3.0 to
$-$8.0 in steps of 0.5 dex. We then determine the abundance of each
element by fitting the various observed lines using a similar method
to that described in \citet{dufour05}. This is done by
minimizing the value of $\chi$$^2$ taken as the sum over all frequencies
of the difference between the normalized observed and model fluxes
(the synthetic spectra are multiplied by a constant factor to account for the solid angle 
and the slope of the spectra locally are allowed to vary by a first order polynomial
to account for residuals from the spectral normalization procedure), 
all frequency points being given an equal weight. 
Interpolation between grid points allows us to achieve individual line abundances 
accurate to $<$0.05 dex. Uncertainties are conservatively set at 0.2 dex.

Table \ref{tabpars} reports modeled T$_{\rm eff}$, log$g$, and abundances
for Ca and/or Mg (the most readily detectable elements in low resolution optical spectra);
corresponding white dwarf masses and radii are obtained from the 
MWDD\footnote{\url{http://dev.montrealwhitedwarfdatabase.org/evolution.html}}
evolutionary models \citep{bedard20}.
For the two helium-dominated atmosphere stars we also report measured hydrogen abundances
in their atmospheres. 
For Gaia\,J0611$-$6931 we report on the additional detection of Si
in Section \ref{sec0611}.

\subsection{\large \bf Gaseous Emission Lines}
\label{secgasres}

Broad emission lines from several atomic transitions are detected in our optical spectra.
Figure \ref{figcaiiirts} shows Ca~II IRT emission from one epoch for each source
while Figures \ref{fig0006l1} through \ref{fig2100l4} show most other emission lines 
detected for each source and compare multiple epochs when such data are available.
In many cases identifications for emission lines seen
could be taken from past studies of white dwarfs with gaseous disks (e.g.,
\citealt{gaensicke06}; \citealt{manser16}, and references therein).
In some cases lines seen did not have corresponding identifications
in the literature and we sought to identify them. The procedure used for
obtaining and assessing new line identifications is given below
after discussion regarding basic measurements made for each line.

%measurements from lines
For each feature we make several measurements.
First is the emission line equivalent width which is
uncorrected for line absorption in the feature. 
From this value, one may calculate a line flux; we do not do that here.
Equivalent width uncertainties are obtained by taking the standard deviation of several 
measurements which differ by anchoring the continuum points for the
measurement at multiple locations consistent within the spectrum noise level.

Next, we measure maximum gas velocities in the blue and red wings of the detected emission 
line and/or the full velocity width of a feature at zero intensity (effectively red$-$blue
extent).
The highest velocity gas emission (reported as v$_{\rm max}$sin$i$) corresponds to the 
innermost orbit of emitting gas-phase metals around each white dwarf (e.g., \citealt{horne86}). 
This value is quoted as a product with sin$i$ where $i$ is the unknown inclination angle of
the disk ($i$ of 0$^{\circ}$ would be obtained for a face-on disk). 
Maximum gas velocities are calculated relative to the white dwarf systemic motion.
To derive each object's systemic motion, we measure radial velocities for metal and
hydrogen lines in high-resolution spectra and then correct them for
the computed gravitational redshifts at the white
dwarf photosphere. Gravitational redshifts are based on measured atmospheric 
parameters reported in Table \ref{tabpars} and have uncertainties of up
to $\sim$15\,km\,s$^{-1}$ due to uncertainties in atmospheric parameter
determinations; we do not include this additional error term in any analysis.
Systemic velocities are reported in Table \ref{tabkins} along with other astrometric
and kinematic quantities for each star.

Last, if possible, we measure the velocity separation between peaks for double-peaked 
profiles.
Peak separation gives a rough characterization of the outermost orbit of emitting gas-phase
metals and is obtained by taking the difference of each transition's peak centroid wavelengths.
These are measured by making several Gaussian fits to each peak, changing between each fit
the anchor points; the average is the adopted value and standard deviation the uncertainty.

Measurements are only made for 
lines with sufficiently good signal-to-noise such that both peaks are well-detected and/or where
one can clearly identify the line edges (where the line emission reaches zero intensity and the
continuum emission level is recovered). 
In lower-resolution data, where the absolute wavelength scale is not accurate, we do not
report maximum gas velocities relative to the white dwarf systemic motion.
Full widths can be measured in all spectra as those are derived from the
difference between blue and red extents of a line and are not affected by absolute
wavelength scale inaccuracies.
When there is the possibility of contamination by
nearby transitions (whether in emission or absorption) we do not make measurements.
For example, we typically cannot make measurements for the Fe~I $\lambda$5169 
and Mg~I triplet features around 5170\,\AA\ (e.g., see Figure \ref{fig0842l5}).

%lines detected and identification of previously unseen lines
For new and unidentified lines we follow a two-step, iterative process in establishing and
assessing line identifications; this procedure relies upon high-resolution data and thus was not
attempted with Kast or GMOS-S spectra. We begin with the selection of lines whose identifications
we believe are unambiguous (e.g., Ca~II IRT, O~I, select Fe~II, and Ca~II H+K lines).
From this selection of lines we establish a range of transition characteristics (lower and upper
energy levels, Einstein coefficients, and oscillator strengths) and measured line extents
in the red and blue wings (v$_{\rm max}$sin$i$). We then search near the wavelength of
an unidentified line for any transition with comparable transition characteristics to the known
lines that also produces v$_{\rm max}$sin$i$ values reasonably in agreement with those seen
for that target star. We used the NIST Atomic Spectra Database Lines
Form\footnote{\url{https://physics.nist.gov/PhysRefData/ASD/lines_form.html}}
and the database of 
\citet{vanhoof18}\footnote{\url{http://www.pa.uky.edu/~peter/newpage/}}
to obtain line parameters.
With ``candidate'' line identifications obtained in this manner, we then perform a self-consistency 
check of seeking out other lines from the same element and ionization state that we might 
expect to see based on energy levels, transition probabilities, and oscillator strengths.
Confirmation of such lines solidifies a line identification.

We have newly identified several Fe~I (including some semi-forbidden)
and Fe~II lines, a semi-forbidden Ca~I line, a Si~I line, 
a semi-forbidden Mg~I line, and Na D doublet emission.
Ambiguous and unidentified lines remain; many likely originate from iron
but it is not clear which specific transition is responsible.
There remain two lines for which identifications are not possible because they were covered 
only with low-resolution spectra (see Section \ref{sec0510}).

Table \ref{tablines} reports all transitions detected for each source in any given epoch.
Measured values for lines are reported in Tables \ref{tabline0006} through \ref{tabline2100}.

\section{\large \bf Individual Systems}
\label{secind}

Here we briefly discuss results for each target, including any literature work that may
exist for each system.

\subsection{SDSS\,J0006+2858}
\label{sec0006}

SDSS\,J0006+2858 was one of two stars (the other being GaiaJ2100+2122)
serendipitously found to host unambiguous emission lines early in our Kast
polluted white dwarf survey and prompted a shift in our observing strategy to
cover Ca~II IRT lines.

SDSS\,J0006+2858 hosts a rich emission spectrum with lines from oxygen,
calcium, and iron seen (Figures \ref{fig0006l1}-\ref{fig0006l5}).
Emission structures in the 6200-6500\,\AA\ region similar to those seen in the spectrum
of Gaia\,J0510+2315 are seen (Section \ref{sec0510}), it is not clear what transitions
they originate from.
Possible contributions from magnesium may also be present
near 5170\,\AA , but are inconclusive due to the presence of strong
Fe~II $\lambda$5169 and $\lambda$5197 emission. 
SDSS\,J0006+2858 hosts Ca~II IRT line strengths comparable to
some of the strongest known emitters (e.g., SDSS\,J1228+1040 and 
Gaia\,J0611$-$6931). Curiously, it also appears to host a third peak in its IRT
lines that falls on the blue shoulder of the blue-ward major peak. Such a structure
has not been previously seen for any other gas disk white dwarf star.

Of special note for this system is a significant asymmetry in the maximum blue and red
velocities of its emission lines. On average, the blue line extent is roughly double that
seen in the red ($-$800\,km\,s$^{-1}$ vs +400\,km\,s$^{-1}$; see Table \ref{tabline0006} and
Figure \ref{fig0006lv}). 
Uncertainties in stellar atmospheric parameters (and hence gravitational redshift) combined
with line measurement uncertainties are not capable of accounting for this
difference. While this asymmetry is seen in all lines, the strength of the ``extended''  portion
in the blue wing is suppressed in the Ca~II IRT relative to what is seen for other transitions
(including Ca~II H+K). This extension in the blue wing is stable over at least a week timescale
(07 July 2019 and 16 July 2019 HIRES measurements) and possibly over six month timescales
(Figures \ref{fig0006l1} and \ref{fig0006l5}).

\subsection{WD0145+234}
\label{sec0145}

WD0145+234 was first reported as a spectroscopically confirmed
white dwarf star with spectral type DA by \citet{mccook87}.
It was further studied in \citet{gianninas11} and \citet{limoges15} where they measured 
from spectroscopic observations 
a DA spectral type, an effective temperature of $\approx$13,000$\pm$200\,K, 
and log$g$ of $\approx$8.12$\pm$0.05.
\citet{gianninas06} performed high-speed optical photometric monitoring of WD0145+234 over
a couple hours in one night, finding no
obvious variability over period ranges of 20-2000~seconds with a detection limit of 0.06\%;
they also derived from spectra T$_{\rm eff}$$=$12,470\,K and log$g$ of 8.06. Similar results
in optical monitoring were found in the study performed by \citet{bognar18}.
\citet{rebassa19} identify WD0145+234 as an infrared excess candidate and associate
its excess emission to a putative debris disk.

More remarkably,
WD0145+234 was recently reported in the literature to have undergone an outburst in the 
mid-infrared \citep{wang19}. Pre-outburst archival observations demonstrate that it
was host to atmospheric pollution (but past studies of this star did not note
atmospheric pollution)
and infrared excess emission indicating that the white dwarf star is in the process of 
consuming a rocky body from its remnant planetary system (\citealt{wang19}; Melis et al., in prep). 
The infrared excess emission is now brighter by over a magnitude, likely due to a 
fresh disintegration event off of a rocky body.

The infrared outburst at WD0145+234 began in mid-2018  \citep{wang19}; the
observations of gas emission were in late 2019 (Table \ref{tabobs}).
Since no strong emission lines other than the Ca~II IRT are seen 
(Table \ref{tabline0145} and Figure \ref{fig0145l1}), 
and since no archival observations covering the Ca~II IRT exist that we are aware of,
it is not possible to comment
on whether or not gaseous emission lines were present in spectra taken before the infrared
outburst began. No obvious variability is seen in the gas emission lines (strength or structure)
between the epochs presented here (Table \ref{tabline0145} and Figure \ref{fig0145l1}).

An exhaustive study on the time variability of absorption lines and the
composition of the parent body source of the material being accreted
by WD0145+234 will appear elsewhere (Melis et al., in prep.).
%Atmospheric pollution has been seen at a consistent level (as determined through measurements of Ca~II K line equivalent widths and associated uncertainties) in data taken in December 2010 \citep{limoges15} and through 2018-2019 (results presented herein). No obvious variability is seen for the Ca~II IRT gas emission lines on inter- and intranight timescales that were probed herein (Figure \ref{fig0145} and Table \ref{tabline0145}).
 
\subsection{SDSS\,J0347+1624}
\label{sec0347}

SDSS\,J0347+1624 was brought to our attention by Dennihy et al.\ (2020, submitted) as a new
gas disk candidate. We had obtained observations of it in 2016 and obtained a further spectrum
in 2019 to aid in the variability study being conducted by 
Dennihy et al. We present the Kast spectra obtained
in both epochs here but defer to Dennihy et al. (2020, submitted) for a full discussion of this source.

Kast spectra from both epochs show clear emission from oxygen, calcium, and many iron
lines (Figures \ref{figcaiiirts} and \ref{fig0347l1}-\ref{fig0347l4}); unidentified emission
in the 6200-6500\,\AA\ region similar to that seen for Gaia\,J0510+2315 may also be present
(Section \ref{sec0510}).
The Kast data do not reveal any detectable variation between iron lines in the 2016 and
2019 epochs (2016 epoch spectra did not cover beyond 7700\,\AA\ and hence only
covered iron emission line regions in common with the 2019 epoch; Tables \ref{tabobs}
and \ref{tabline0347}).

\subsection{Gaia\,J0510+2315}
\label{sec0510}

Emission lines from oxygen, magnesium, calcium, and iron are seen
toward Gaia\,J0510+2315 (Figures \ref{fig0510l1}-\ref{fig0510l6}).
Gaia\,J0510+2315 is the only star in the sample to host clear emission from
Mg~II, in this case near 7900\,\AA\ (Figure \ref{fig0510l5}). It is also impressive
in that both O~I emission features are comparable in strength
to the Ca~II IRT emission lines (Table \ref{tabline0510} and Figures \ref{fig0510l5}
and \ref{fig0510l6}).

No obvious variability is seen between epochs.
Two emission structures that we can't identify appear in Kast data at wavelengths that are not covered
by the HIRES spectra (Table \ref{tablines},
Figures \ref{fig0510l1} and \ref{fig0510l5}).
Emission in the 6200-6500\,\AA\ region, likely from iron, is seen in Kast
and HIRES spectra (Figure \ref{fig0510l4}).
%A feature is seen surrounding the Si~II $\lambda$6347 wavelength, but only one edge
%is clearly discernable (Figure \ref{fig0510l4}).
%The maximum velocity of the blue wing of this line is consistent
%with that seen for other lines, so a tentative identification with Si~II is made but
%should be verified with higher signal-to-noise data.

Apparent asymmetry is seen in the blue and red wings
of emission lines when one considers all maximum velocity measurements together
(Figure \ref{fig0006lv} and Table \ref{tabline0510}).
Curiously, and unique to Gaia\,J0510+2315, the O~I triplet near 7772\,\AA\ 
(and possibly the O~I complex near 8446\,\AA) hosts 
a red extent that greatly exceeds measurements for any other line
(Table \ref{tabline0510} and Figures \ref{fig0006lv} and \ref{fig0510l5}).
It is not clear if some blended emission lines are responsible for this or if the distribution of
oxygen gas near Gaia\,J0510+2315 is different from other elements (in this case
extending much closer to the star).

\subsection{Gaia\,J0611$-$6931}
\label{sec0611}

GMOS-S and MagE spectra indicate that Gaia\,J0611$-$6931 hosts strong atmospheric metal
pollution and hitherto unseen emission from the Na D doublet (Figure \ref{fig0611l3}).
Also anomalous was the identification of Si~I $\lambda$3905 emission which is otherwise
only seen in WD0842+572 (Table \ref{tablines}).
Gaia\,J0611$-$6931 otherwise hosts familiar transitions from oxygen,
magnesium, calcium, and iron (Figures \ref{fig0611l1}-\ref{fig0611l5}).

No variability is seen between emission lines in the GMOS-S and MagE epochs
(Table \ref{tabline0611}).
Possible emission is seen in a variety of places in the spectra, but is hard to confirm
due to signal-to-noise constraints and systematics present in the data including
fringing.
Weak emission may be present near the core of the H$\delta$ photospheric absorption
line, a feature we cannot verify nor identify with the data in hand (Figure \ref{fig0611l1};
it is possible that similar emission may also be present for SDSS\,J0006+2858
and Gaia\,J2100+2122).
The appearance of the line in the core of H$\delta$ is reminiscent of what is seen
in the core of the He~I $\lambda$5876 line for helium-dominated atmosphere white dwarfs
(\citealt{klein20}), but without similar cores in other Balmer transitions the self-reversal is probably not due to hydrogen.
%There appears to be emission near the Fe~II $\lambda$5317 line (Figure \ref{fig0611l2}),
%but the line structure is not similar to that  seen for other iron lines.
In general there appears to be multiple emission complexes between 
%5200 and 5450\,\AA\  as well as between 
6100 and 6300\,\AA\ (Figure \ref{fig0611l3}),
but it is not clear if the emission is real and similar to what is seen in this region
for other stars (Table \ref{tablines}, Figure \ref{fig0510l4}).
%Emission near 8250\,\AA\ and beyond 8850\,\AA\ may also be present, but fall in regions
%of the spectra with low signal-to-noise and where fringing is problematic
%(Figures \ref{fig0611l4} and \ref{fig0611l5}).

Beyond the unusual volatile emission from sodium, Gaia\,J0611$-$6931 also stands out
as having some of the strongest Ca~II IRT emission lines seen in any gas disk star
(Table \ref{tabline0611})
and the second-highest magnesium abundance in the sample presented herein
(Table \ref{tabpars}). Additionally, we see even in low-resolution spectra 
photospheric absorption lines
from Si~II near 6350 and 6370\,\AA\ (Figure \ref{fig0611l3}). 
We fit these lines in addition to Ca~II and Mg~II absorption
lines finding a silicon abundance by number of log$_{10}$(Si/H)$=$$-$4.8
(these values also provide a reasonable fit to the higher resolution MagE detections of these lines). 
Limits for iron and oxygen are not restrictive (oxygen especially being confounded by emission)
so it is not possible to comment further on the composition of the parent body polluting
Gaia\,J0611$-$6931. While it may be tempting to credit the presence of volatile emission
seen for Gaia\,J0611$-$6931 to a volatile-rich nature of the parent body it is accreting,
it is prudent to wait for complete disk and/or atmospheric abundances before arriving at
such a conclusion (e.g., see the cautionary tale for over-interpreting gas emission strength
as abundance in \citealt{matlovic20}).

\subsection{Gaia\,J0644$-$0352}
\label{sec0644}

Only weak Ca~II IRT emission lines are seen in HIRES data for this star (Figure \ref{figcaiiirts}).
There may also possibly be a very weak hint of emission from Fe~II $\lambda$5316,
but with no other iron lines detected this is considered inconclusive.
While the measurements
have fairly high uncertainty (due to the low significance of the line detections),
there could be a slight hint of asymmetry in the maximum blue and red velocity gas seen
(Table \ref{tabline0644}).

Gaia\,J0644$-$0352 is notable for being one of two helium-dominated atmosphere
stars in the sample. The star also has hydrogen detected in low-resolution spectra
indicating the possibility of a parent body having some amount of water
(e.g., \citealt{gentilefusillo17}). A complete atmospheric abundance and oxygen
budget analysis (e.g., \citealt{klein10}; \citealt{farihi13}) 
will be necessary to determine how much water the parent body contained.

%Debes MIKE 0644 politics: agreed to the below before HIRESr was taken
%Our preference is instead for our team to mention in our paper about the three other white dwarf stars that 0644:
%(1) has been observed with MIKE;
%(2) like the three white dwarfs we otherwise found, also shows CaII IRT emission lines;
%(3) has parameters as measured from our low-resolution spectrum: Teff, logg, and H and Ca abundances (only pollution seen in low-res - this of course leaves you with Mg, Si, and O plus an Fe limit);
%(4) that details from the MIKE spectrum will be given in a separate paper by Debes et al. (or whomever will lead it).
%The paper from your team mentioned in point (4) would include our team (myself, Beth Klein, Ben Zuckerman, Siyi Xu, Allie Doyle, Patrick Dufour, and Alycia Weinberger of course) as co-authors. Our logic is that since your paper already has numerous co-authors, including several more isn't a big deal; the assumption being that the paper will only discuss 0644.

\subsection{WD0842+572}
\label{sec0842}

Initial detection and spectroscopy for WD0842+572 was conducted as a part of the
Second Byurakan Sky Survey where the star was found to have a spectral type of 
DA \citep{balayan97a}.
Until recently this star was not prominently featured in any published work. In
\citet{swan20}, {\it Spitzer} warm IRAC photometry is presented along with
an unsubstantiated note that the star is host to gas disk emission lines.

WD0842+572 is host to a selection of narrow emission lines with typically well-defined
double-peak morphology. We find contributions from silicon, magnesium, calcium, and iron
(Figures \ref{fig0842l1}-\ref{fig0842l8}).

Several characteristics make WD0842+572 stand out compared to the rest of the sample.
First is the highest atmospheric abundance of magnesium for 
any star in our sample (Table \ref{tabpars}).
Second is the clear detection of a semi-forbidden transition of magnesium,
Mg~I] $\lambda$4571 (Figure \ref{fig0842l3}),
and a bevy of semi-forbidden neutral iron lines between 4300-5200\,\AA\ 
(Figures \ref{fig0842l2} and \ref{fig0842l4}).
Additionally, semi-forbidden neutral calcium is seen in the red wing of the H$\alpha$ photospheric
absorption line (Figure \ref{fig0842l6}).
Last is a preponderance of emission lines from neutral atomic species like Mg~I, Si~I,
Ca~I, and Fe~I (Figures \ref{fig0842l1}-\ref{fig0842l6}). 
Notably absent is emission from the O~I triplet near 7772\,\AA\
which is seen at most other systems presented herein.

Strong Ca~II IRT lines are seen for this star (Figure  \ref{fig0842l8}), 
but do not quite rise to the level of the strongest emitters known (Table \ref{tabline0842}).
Possible variation is hinted at in the structure of lines seen between
the Kast and HIRES epochs (separated by roughly six months). Whether this
variation is due to the lower resolution in the Kast data or true changes in the
lines is not clear from the data available; in any case, Kast line measurements
differ from what is measured with HIRES
at the $\lesssim$3$\sigma$ level (see Table \ref{tabline0842}).

\subsection{WD1622+587}
\label{sec1622}

Spectroscopic characterization of WD1622+587 was performed through the
Second Byurakan Sky Survey where it was found to have a spectral type of 
DB \citep{balayan97b}.
The star received little attention following that initial observation and until now was
not singled out individually in any published study.

Oxygen, calcium, and iron emission lines are seen in WD1622+587
(Figures \ref{fig1622l1} to \ref{fig1622l3}). Most lines are only seen
in the HIRES data, although slight hints of Ca~II IRT emission are seen  in Kast spectra
(Figures \ref{fig1622l2} and \ref{fig1622l3}); 
we do not attempt measurements for these low-significance lines.
Oxygen emission in the 7772\,\AA\ complex is comparable in strength to the individual
Ca~II IRT lines (Table \ref{tabline1622}), 
although no oxygen emission from the 8446\,\AA\ complex is seen.

WD1622+587 is a helium-dominated atmosphere star, also with hydrogen present,
similar to Gaia\,J0644$-$0352 (Table \ref{tabpars}). 

\subsection{Gaia\,J2100+2122}
\label{sec2100}

Gaia\,J2100+2122 has an emission spectrum dominated by iron lines, a feature that
made it a clear emission line star despite discovery Kast spectra not covering
Ca~II IRT lines (Figures \ref{fig2100l1} and \ref{fig2100l2}). 
Emission is also seen from oxygen and calcium (Figures \ref{fig2100l3} and \ref{fig2100l4}).
Gaia\,J2100+211 also hosts unidentified emission in the 6200-6500\,\AA\ region like 
in the spectrum of Gaia\,J0510+2315 (Figure \ref{fig0510l4}).

While most iron and oxygen lines are relatively similar between epochs, variability
is present at the $\gtrsim$3$\sigma$ level for Ca~II IRT lines
between the HIRES and Kast epochs (Table \ref{tabline2100} and
Figure \ref{fig2100l4}). Gaia\,J2100+2122
thus stands out as the only star for which we saw variability beyond measurement uncertainties.
Dennihy et al.\ (2020, submitted) showcase the variability for this source across several
epochs.

% discussion/conclusion
\section{\large \bf Discussion and Conclusions}
\label{secdisc}

The nine stars presented herein significantly enlarge the known sample of single
white dwarfs that host gaseous debris disks. Previously known in the literature
were SDSS\,J0738+1835 (\citealt{dufour12}; \citealt{brinkworth12}), 
Ton 345$=$SDSS\,J0845+2257 (\citealt{gaensicke08}; \citealt{melis10}),  
WD\,J091405.30+191412.25 \citep{gaensicke19},
SDSS\,J0959$-$0200 \citep{farihi12a}, 
SDSS\,J1043+0855 (\citealt{gaensicke07}; \citealt{melis10}; \citealt{manser16b}), 
SDSS\,J1228+1040 (\citealt{gaensicke06}; \citealt{melis10}; \citealt{hartmann16};
\citealt{manser16}; \citealt{manser19}), 
HE\,1349$-$2305 (\citealt{melis12}; \citealt{dennihy18}),
and SDSS\,J1617+1620 \citep{brinkworth12, wilson14}. 
Two sources claimed in the literature to be single
white dwarfs hosting gas disks have been refuted (SDSS\,J1144+0529 reported first by
\citealt{guo15} and refuted by \citealt{swan20},
and SDSS\,J1344+0324 reported first by \citealt{li17} and refuted in \citealt{xu19}).
In sum, 17 gas disk-hosting white dwarfs are now known.

%Teff range for all known gas disk stars and those reported here, low Teff for WD0145+234
As an ensemble, the known gas disk-hosting white dwarf stars have an average effective 
temperature of $\approx$18,500\,K with a standard deviation of $\pm$4500\,K.
There are four sources in the sample with effective temperatures $\leq$ one standard deviation from
the mean: SDSS\,J0738+1835
(T$_{\rm eff}$$\approx$14,000\,K; \citealt{dufour12}),
SDSS\,J1617+1620 (T$_{\rm eff}$$\approx$13,500\,K; \citealt{brinkworth12}),
SDSS\,J0959$-$0200 (T$_{\rm eff}$$\approx$13,300\,K; \citealt{farihi12a}),
and WD0145+234 (T$_{\rm eff}$$\approx$12,720\,K, Table \ref{tabpars}). 
SDSS\,J0959$-$0200 and SDSS\,J1617+1620 are known to host
highly variable IRT emission line strengths.
On intra-night to $\sim$few month temporal baselines WD0145+234
shows no evidence for variability (Figure \ref{fig0145l1}).

%discuss total CaII IRT EW vs Teff and upper envelope 
WD0145+234 narrowly finds itself as the new ``coolest'' gas disk-hosting white dwarf known;
it is not clear what role (if any) the recent infrared ``outburst'' may have had in producing
the detected gas disk. All of the cool gas disk stars named in the previous paragraph 
have so far been found to host only Ca~II IRT emission lines.
An examination of summed Ca~II IRT emission line equivalent width
as a function of host white dwarf
T$_{\rm eff}$ reveals considerable scatter at all effective temperatures. However there 
appears to be an upper envelope of maximum observed Ca~II IRT equivalent widths 
of $\approx$50\,\AA\ for hotter white dwarfs that is
significantly lower ($\approx$10-20\,\AA ) for 
white dwarfs with T$_{\rm eff}$$<$16,000\,K. With only four sources known
to date in this temperature range it is not clear if this is the result of observational
bias or actual physical processes at play (e.g., disk heating from energetic photons
as described in \citealt{melis10}).

%kinematics... no UVW for SDSSJ0738, WDJ0914, SDSSJ0959, SDSSJ1617
Three-dimensional Galactic UVW space motions are computed and presented
for all systems discovered through this work (Table \ref{tabkins}). Additionally, space motions
were adopted for Ton 345, SDSS\,J1043+0855, and SDSS\,J1228+1040 from
\citet{melis10}. Kinematic measurements for HE\,1349$-$2305 from \citet{melis12}
are used to calculate UVW:  U$=$+3\,km\,s$^{-1}$, 
V$=$$-$11\,km\,s$^{-1}$, W$=$$-$4\,km\,s$^{-1}$.
As a whole for the class of gas disk single white dwarf stars, 
we see consistently negative Galactic V velocities, similar to V velocities for
white dwarfs with and without atmospheric pollution (see Table 6 of \citealt{zuckerman03}
and Table 4 of \citealt{zuckerman10}). As such, gas disk-hosting white dwarfs do not
appear to be kinematically ``special'' compared to other white dwarf stars
and join them in having space motions characteristic of an old population trailing the 
local standard of rest as they orbit around the Galactic Center. 

%DAs vs DBs
Within the sample of new emission line stars presented herein we see a distribution
of host star primary atmospheric types (hydrogen or DA, helium or DB) of
$\approx$78\% DA and 22\% DB. For the 17 total gas disk-hosting systems now
known, this distribution is 71\% DA and 29\% DB. Both are a reasonable match
to what is observed in the field population of white dwarf stars (e.g., \citealt{kilic20} and 
references therein).
As such, it is safe to conclude there is no bias for gas disks to 
be present around DB or DA white dwarfs.

%types of gas disk lines seen and comparison to past gas disk WDs
Previously identified gas disk white dwarfs have, with the exception of
SDSS\,J1228+1040 and WD\,J091405.30+191412.25, 
typically only been seen to host emission from calcium
and sometimes iron. SDSS\,J1228+1040 saw the additional identification of oxygen and magnesium
\citep{manser16} while WD\,J091405.30+191412.25 hosts more unusual emission dominated
by hydrogen, oxygen, and sulfur (the likes of which are due to accretion from a giant planet-like
companion; \citealt{gaensicke19}). Two-thirds of the sample presented herein
show at least oxygen, calcium, and iron
while a small subset additionally show magnesium and silicon (Table \ref{tablines}). 
Neutral silicon emission from the 3905\,\AA\ transition has never been documented
before; we detect it unambiguously in the spectra of Gaia\,J0611$-$6931 and
WD0842+572.
Magnesium may be more prevalent than indicated by Table \ref{tablines} as its neutral triplet
transitions near 5170\,\AA\ are typically confused with iron emission in that region.
The Mg~I $\lambda$8806 line appears to be comparably strong from Gaia\,J0611$-$6931
(Figure \ref{fig0611l5}) and SDSS\,J1228+1040 \citep{manser16} and is not blended
with other lines. Unfortunately, the HIRES red setup used for this study did not cover the
8806\,\AA\ region (it fell in an order gap).
Gaia\,J0611$-$6931 is unique in the identification of sodium in its emission spectrum.
The lack of hydrogen and other volatiles, combined with the detection of elements typically
found in rocky minerals (both in emission and absorption), 
points to the gas originating in the destruction of rocky bodies.
Future atmospheric pollution analyses and emission line modeling will reveal the
elemental composition of the material producing the gas and being accreted by each host
white dwarf.

%brief discussion of line structure and inclination angles
% +WD0842+527 semi-forbidden speculative interpretation: face-on disk?
As noted in \citet{melis10}, emission line profiles observed at high spectral resolution
exhibit a range of complex structures (e.g., Figure \ref{figcaiiirts}).
It is tempting to attribute the range of observed line widths for gas disk systems to different
disk inclination angles. For example, WD0842+572 could be viewed nearly face-on while
SDSS\,J0006+2858 could be viewed closer to edge-on (Tables \ref{tabline0842} and
\ref{tabline0006}). However, without performing detailed
physical modeling for the gas and/or dust disks in these systems, it is not possible to conclusively
assign viewing geometries and thus arrive at gas disk parameters like inner and outer
radii (e.g., as in \citealt{melis10}). Emission line modeling that reproduces
the observed line intensities and shapes will be necessary to arrive at robust gas disk
physical parameters (e..g, \citealt{gaensicke19}).
Despite this, a speculative interpretation for the unusual characteristics seen in
WD0842+572 is worth mentioning. The combination
of its narrow line profiles, abundance of neutral emission line species, and presence of
several semi-forbidden transitions could be seen as evidence for 
a face-on viewing angle. The nearly zero sin$i$
reduces the maximum velocity emission seen to its small (relative to other gas disk systems)
value, while the presence of neutrals and semi-forbidden emission lines come from
primarily seeing the lower-density region of an optically thick disk's upper atmosphere.

%var summary; include IR variability for gas disk systems
Variability in gas disk emission lines has been noted for several systems in the literature.
\citet{manser16} thoroughly characterize long term changes in Ca~II IRT emission features 
for SDSS\,J1228+1040, deciphering with doppler imaging what appears to be a precessing 
eccentric gas ring (see more below). In a follow-up study dedicated to shorter timescales,
\citet{manser19} find repeating changes with a period of $\sim$2~hours in the Ca~II IRT
line strengths which they interpret as evidence for an orbiting dense planetesimal.
\citet{manser16b} document variability over long timescales for SDSS\,J1043+0855.
\citet{wilson14} demonstrate that SDSS\,J1617+1620 is highly variable, with Ca~II IRT
emission lines strengthening dramatically between 2006 and 2008, then monotonically
declining thereafter.
\citet{dennihy18} conducted a monitoring campaign for HE1349$-$2305, finding
a cycling through having a red-dominated
peak structure as found in \citet{melis12} to a blue-dominated peak structure and
back to red-dominated again; a 
period of 1.4~years is estimated.
\citet{wilson15} and \citet{manser16b} show variability in line profiles for Ton\,345.

The above, typically more exhaustive, studies showcase a wide range of variability
behaviors for gas disk emission systems. Variability in Gaia\,J2100+2122 (Table \ref{tabline2100}
and Section \ref{sec2100}) and possibly in WD0842+572 (Table \ref{tabline0842} and Section
\ref{sec0842}) are presented herein while Dennihy et al.\ (2020, submitted) showcase
more extensive observations of SDSS\,J0347+1624 and Gaia\,J2100+2122.
An interesting tie-in to gas disk line variability comes from the recent explosion in
dust disk variability studies. \citet{xujura14} showed that SDSS\,J0959$-$0200 experienced
a dramatic drop in its infrared dust emission levels while \citet{farihi18} found similar
behavior combined with re-brightening episodes for GD\,56 (a star that does not host a known
gaseous debris disk component).
\citet{wang19} documented the ``outburst'' captured by the {\it WISE} satellite for WD0145+234.
\citet{swan19,swan20} provide a more systematic study of infrared variability in white dwarf
disk systems (those with dust only and both dust and gas)  finding that gas disk-hosting
systems as a class show more variability in their dust continuum emission than
non-gas disk-hosting stars. They interpret this finding within the context of circumstellar
material having a range of collisional activities, gas disk systems
being the more collisionally active cousins of the dust-only systems.

%discussion of line profiles seen at higher resolution
% asymmetric lines
\citet{manser16} demonstrate how emission line structures can change with time,
in their case through the long time baseline study of SDSS\,J1228+1040.
\citet{gaensicke06} and \citet{melis10}
find maximum gas velocities for SDSS\,J1228+1040 of $\approx$560\,km\,s$^{-1}$
with observations conducted between 2003 and 2008. 
%SDSS1228 in Melis+'10 looks asymmetric and SDSS1043 looks kind of asymmetric too
However, \citet{melis10}
find an asymmetry in their 2008 measurements with the maximum blue
wing emission being at $-$380\,km\,s$^{-1}$ and maximum red wing emission
at +550\,km\,s$^{-1}$. During monitoring between 2011 and 2015, 
\citet{manser16} show that this can evolve into a maximum blue extent of $-$390\,km\,s$^{-1}$
and maximum red extent of +780\,km\,s$^{-1}$ (it is not clear if these are ever achieved
at the same time).
SDSS\,J0006+2858 simultaneously hosts a maximum blue extent for its gas emission
of $-$800\,km\,s$^{-1}$ and maximum red extent of +400\,km\,s$^{-1}$ over the time
observations were collected for it in this work (Figure \ref{fig0006lv} and Table \ref{tabline0006}).
Similar to that found here for SDSS\,J0006+2858, elements other than calcium
in the spectrum of SDSS\,J1228+1040 also show the same asymmetry indicating well-mixed atomic
species \citep{manser16}.

The only other system to show pronounced red/blue asymmetry similar to SDSS\,J0006+2858
and SDSS\,J1228+1040 is HE1349$-$2305. This star had in 2011
red and blue extents similar to that measured for SDSS\,J0006+2858 here
\citep{melis12}. \citet{dennihy18} do not measure maximum red and blue gas extents
in their monitoring study of  HE1349$-$2305.
Beyond these stars, possible asymmetries in maximum gas velocities may 
also be seen in Gaia\,J0510+2315 (Table \ref{tabline0510}),
SDSS\,J1043+0855 \citep{melis10}, and Gaia\,J0644$-$0352 (Table \ref{tabline0644}).
Thus, 6 out of 17 total gas disk systems show evidence for asymmetric line profiles.
It is not clear at this time if all such systems cycle through phases of 
exhibiting these asymmetries or if an era 
of asymmetric line extents is part of an evolutionary sequence that ends in fully
symmetric and stable gas emission lines.

Starting with the discovery of single white dwarf stars with gaseous debris disks,
it was recognized that these systems are likely host to eccentric gas rings 
\citep{gaensicke06,gaensicke07,gaensicke08,melis10}. This interpretation continued
with the addition of new data sets and systems (e.g., \citealt{wilson15}; \citealt{manser16};
\citealt{manser16b}; \citealt{manser19}). Indeed, an eccentric gaseous ring component
has also been inferred for the transiting planetesimal system WD1145+147 
\citep{cauley18,fortinarchambault20}.
Asymmetric gas emission line structures bolster such interpretations.
Disk evolutionary models suggest that
eccentric debris systems may reflect an early evolutionary state and that the material
will eventually settle into a more circularized orbit \citep[e.g.,][]{veras14,nixon20}.

%something for Bonsor's student to investigate:
%mass accretion rates for all objects based on Ca and/or Mg, point out huge ones like WD0842
% comment on range of Mdot_acc for gas disk vs non gas-disk WDs? if interesting

%gas disk temps from CaII IRT and CaII H+K intensity ratio for different inclinations (FWZI)???
% maybe check at least WD0842 and compare to Melis+'10

%northern hem vs southern hem
In conclusion, we present a set of nine new gas disk-hosting single white dwarf stars.
Some properties of these systems are similar to previously discovered gas disk systems,
while some display new and exotic features. Of special note is that
only four of the known gas disk systems are found south of the celestial equator, and of those only
two are below a Decl.\ of $-$5$^{\circ}$. This 
likely implies a significant number of southern-hemisphere gas disk-hosting white dwarf
stars remain to be discovered (at least 10 to balance the current population).


\acknowledgments

C.M.\ and B.Z.\ acknowledge support from NSF grants SPG-1826583 and SPG-1826550.
B.K.\ acknowledges support from the APS M.\ Hildred Blewett Fellowship.
We thank E.\ Dennihy for useful discussion.
Research at Lick Observatory is partially supported by a generous gift from Google.
Some of the data presented herein were obtained at the W.M.\ Keck Observatory, which is operated as a scientific partnership among the California Institute of Technology, the University of California and the National Aeronautics and Space Administration. The Observatory was made possible by the generous financial support of the W.M.\ Keck Foundation.
We acknowledge that Keck Observatory rests on land that is important and 
significant to the Hawaiian Natives in many ways. Similarly, we acknowledge that
Lick Observatory resides on land traditionally inhabited by the Muwekma
Ohlone Tribe of Native Americans.
This research has made use of NASA's Astrophysics Data System, the National Institute
of Standards and Technology Atomic Spectra Database Lines Form, the SIMBAD database,
and the VizieR service.


%% for the paper.  Each keyword is check against the master list during
%% copy editing.  Individual instruments can be provided in parentheses,
%% after the keyword, but they are not verified.

\facilities{Shane(Kast), Keck~I(HIRES), Gemini-South(GMOS), Magellan(MagE)}

%\software{AIPS}




}



\clearpage

\begin{figure}
 %\centering
  \begin{minipage}[!t]{53mm}
   \includegraphics[width=80mm,trim={0cm 5.5cm 0cm 0cm},clip]{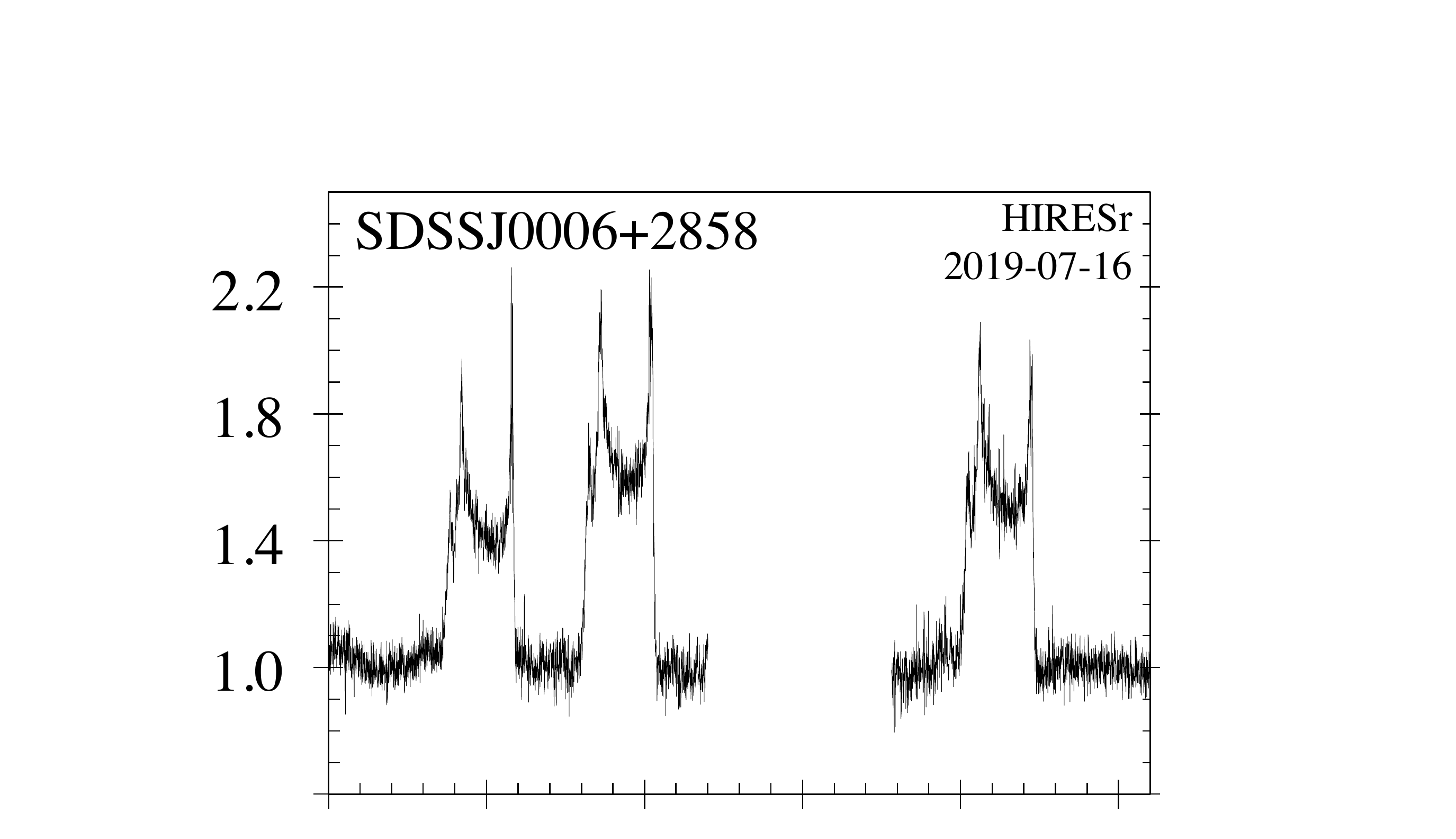}
  \end{minipage}
  \begin{minipage}[!t]{53mm}
   \includegraphics[width=80mm,trim={0cm 5.5cm 0cm 0cm},clip]{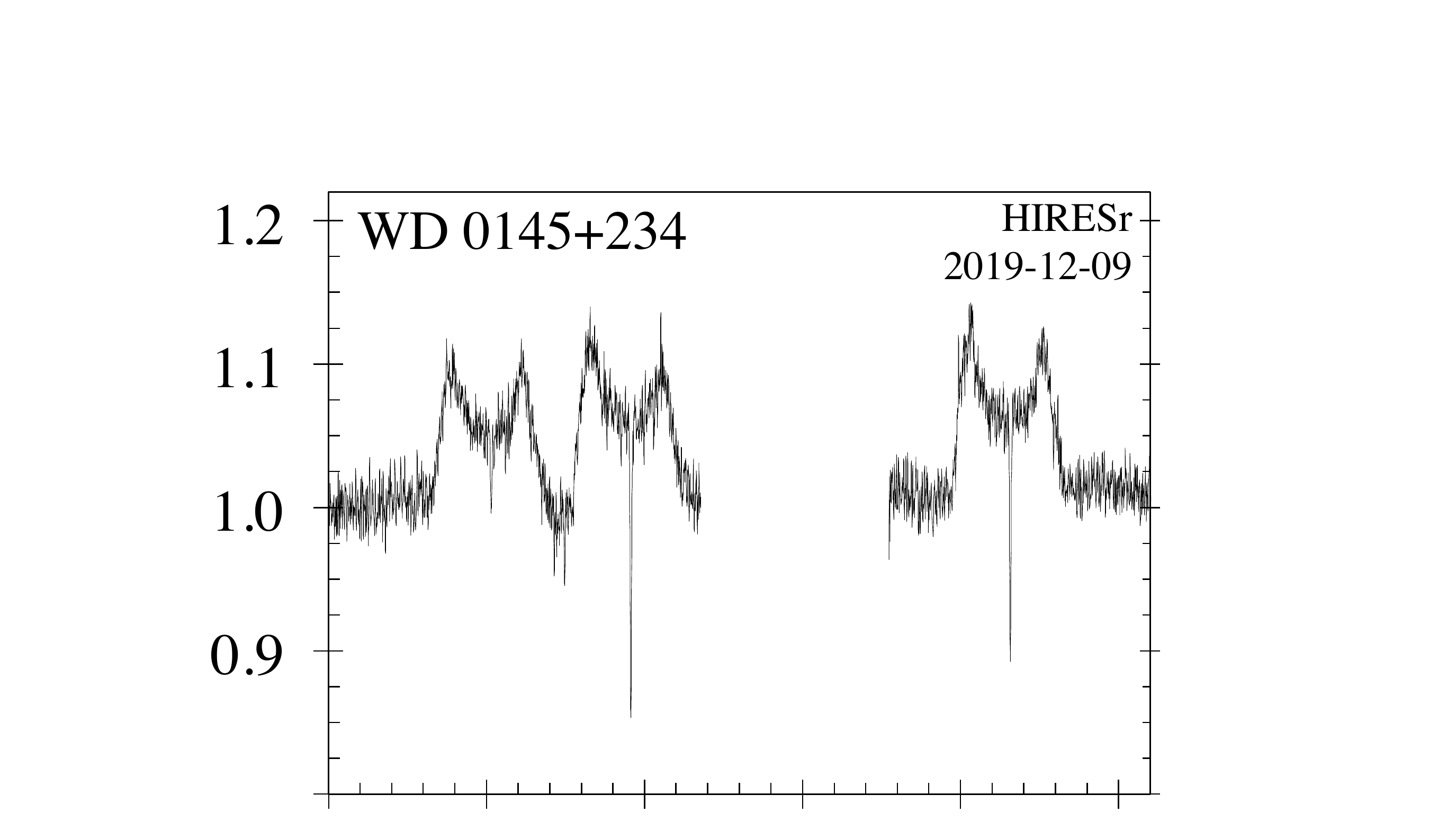}
  \end{minipage}
  \begin{minipage}[!t]{53mm}
   \includegraphics[width=80mm,trim={0cm 5.5cm 0cm 0cm},clip]{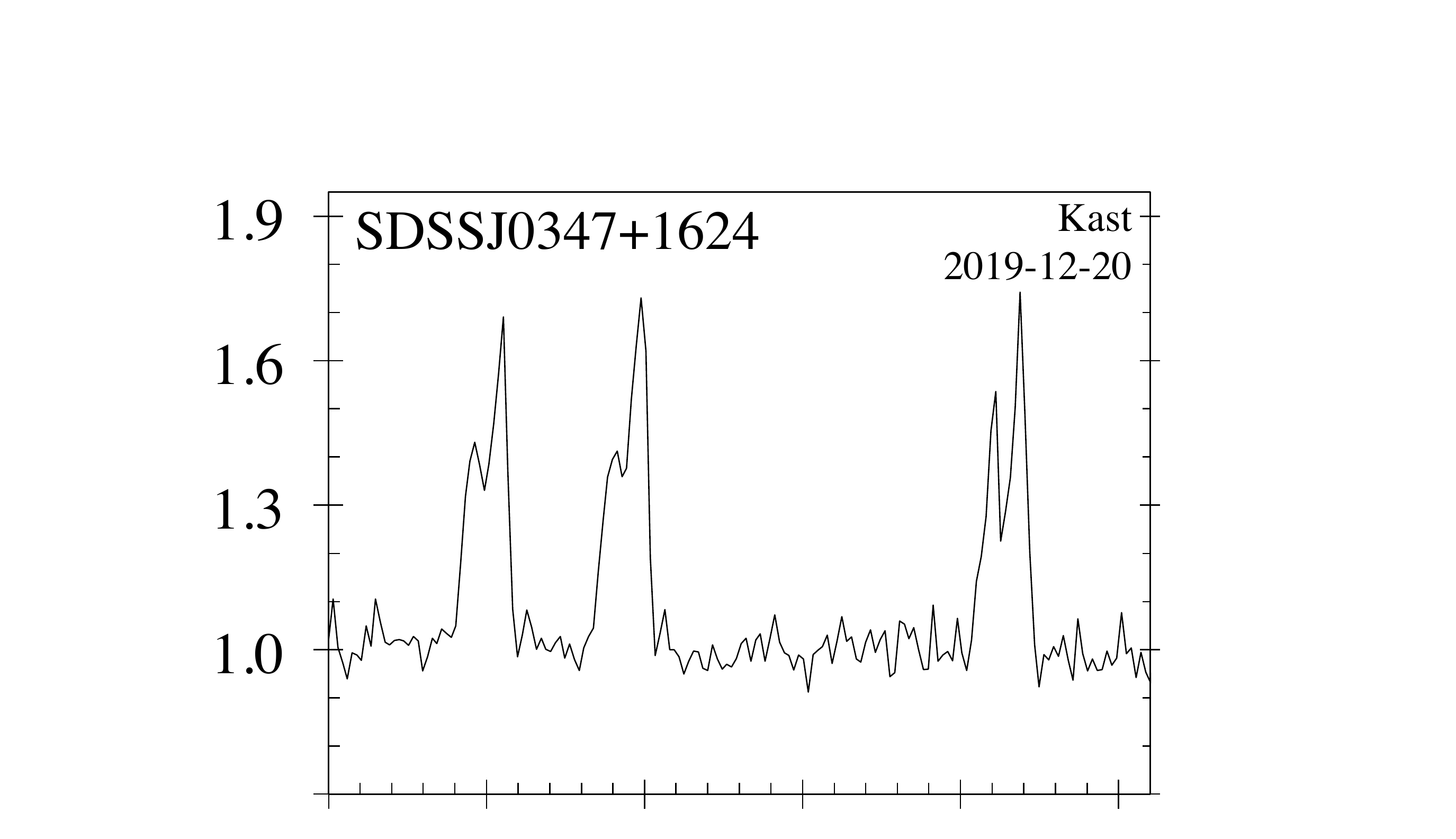}
  \end{minipage}
  \\*[-3.3mm]
  \begin{minipage}[!h]{53mm}
   \includegraphics[width=80mm,trim={0cm 5.5cm 0cm 3.5cm},clip]{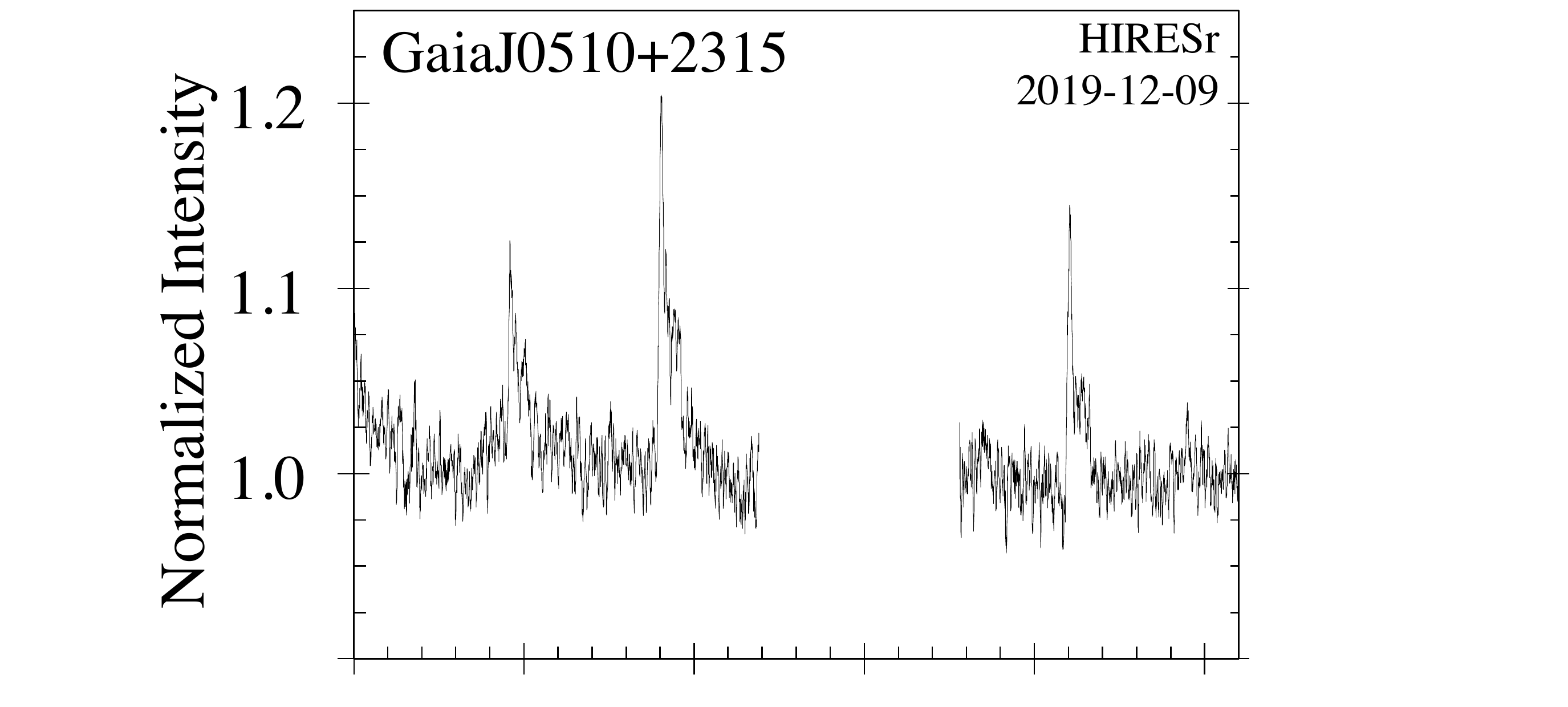}
  \end{minipage}
  \begin{minipage}[!h]{53mm}
   \includegraphics[width=80mm,trim={0cm 5.5cm 0cm 3.5cm},clip]{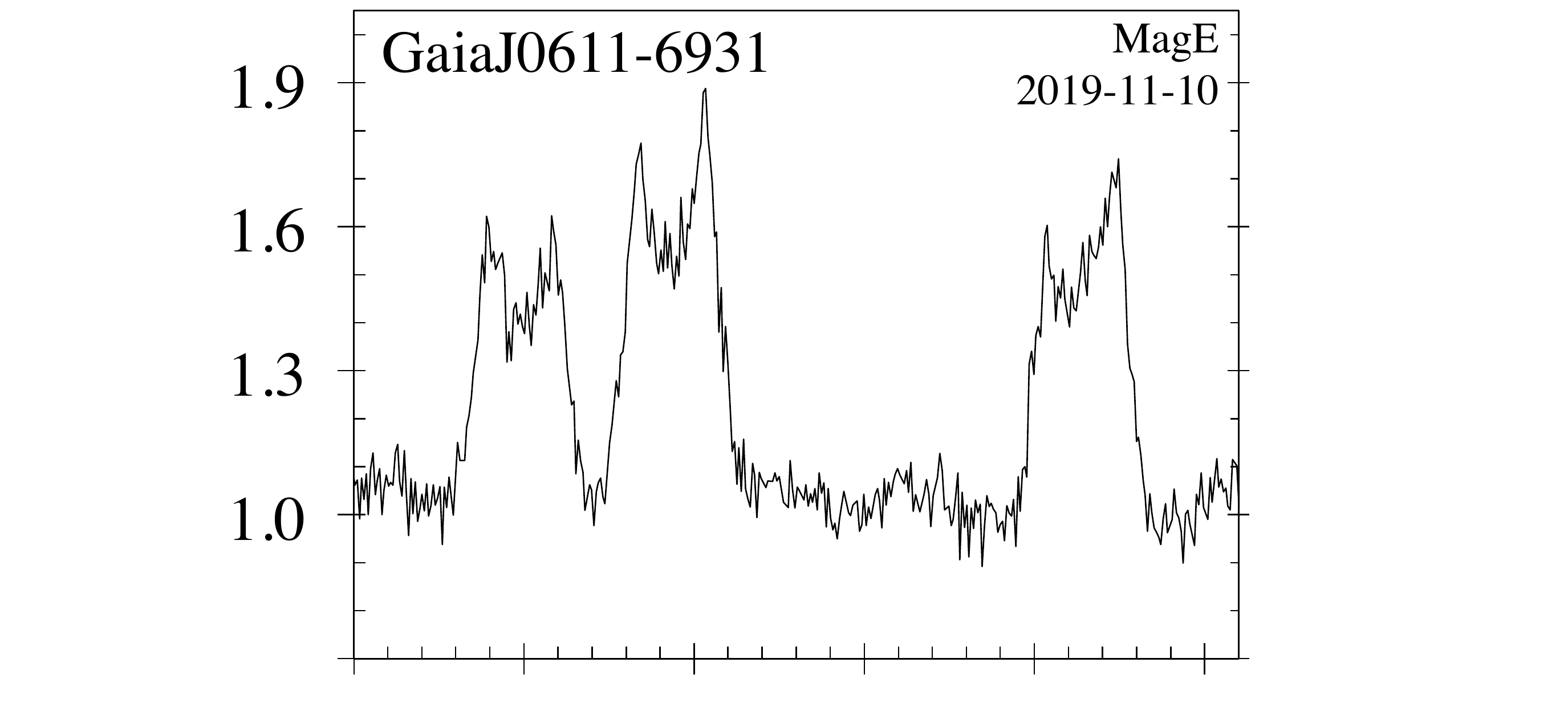}
  \end{minipage}
  \begin{minipage}[!h]{53mm}
   \includegraphics[width=80mm,trim={0cm 5.5cm 0cm 3.5cm},clip]{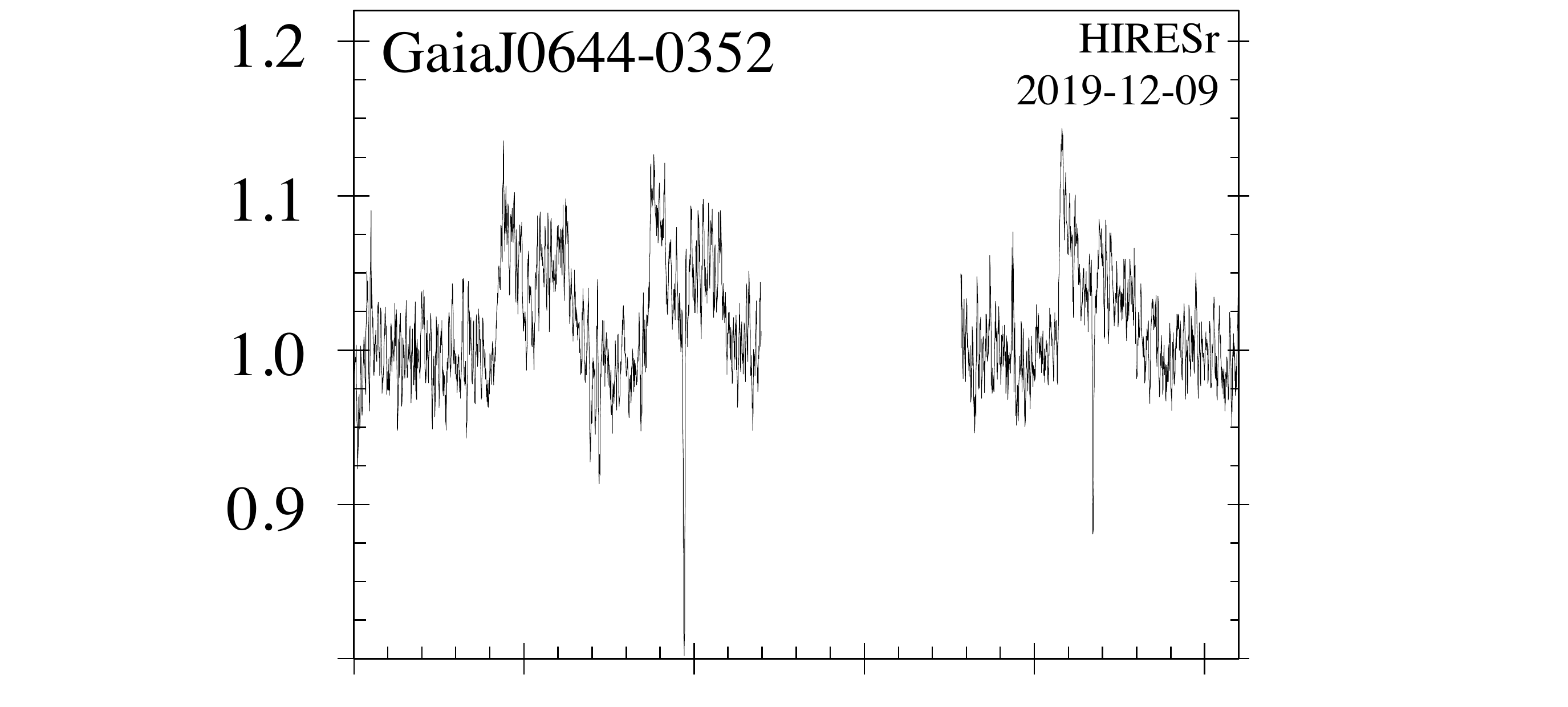}
  \end{minipage}
  \\*[-3.3mm]
  \begin{minipage}[!b]{53mm}
   \includegraphics[width=80mm,trim={0cm 0 0cm 3.5cm},clip]{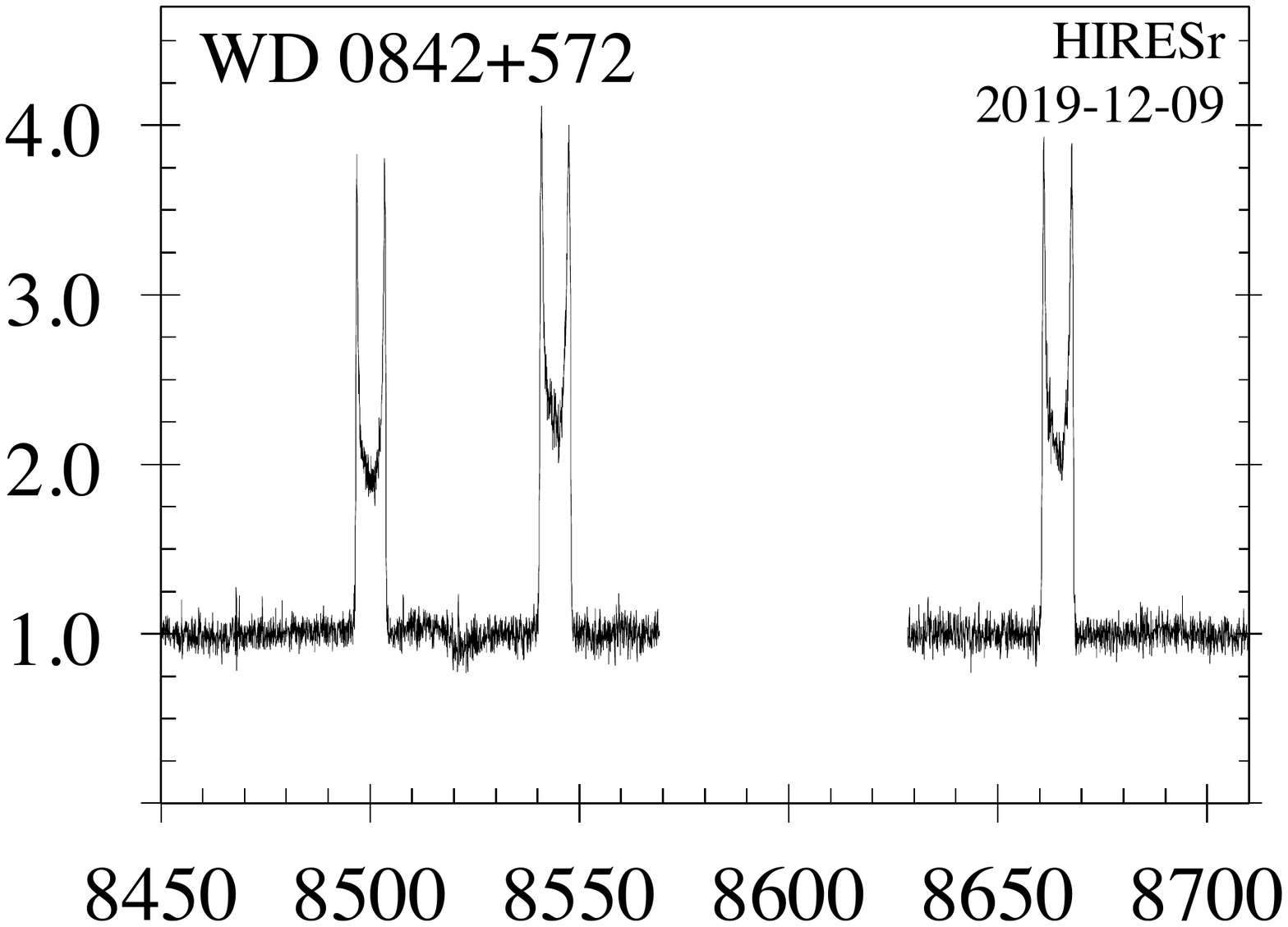}
  \end{minipage}
  \begin{minipage}[!b]{53mm}
   \includegraphics[width=80mm,trim={0cm 0 0cm 3.5cm},clip]{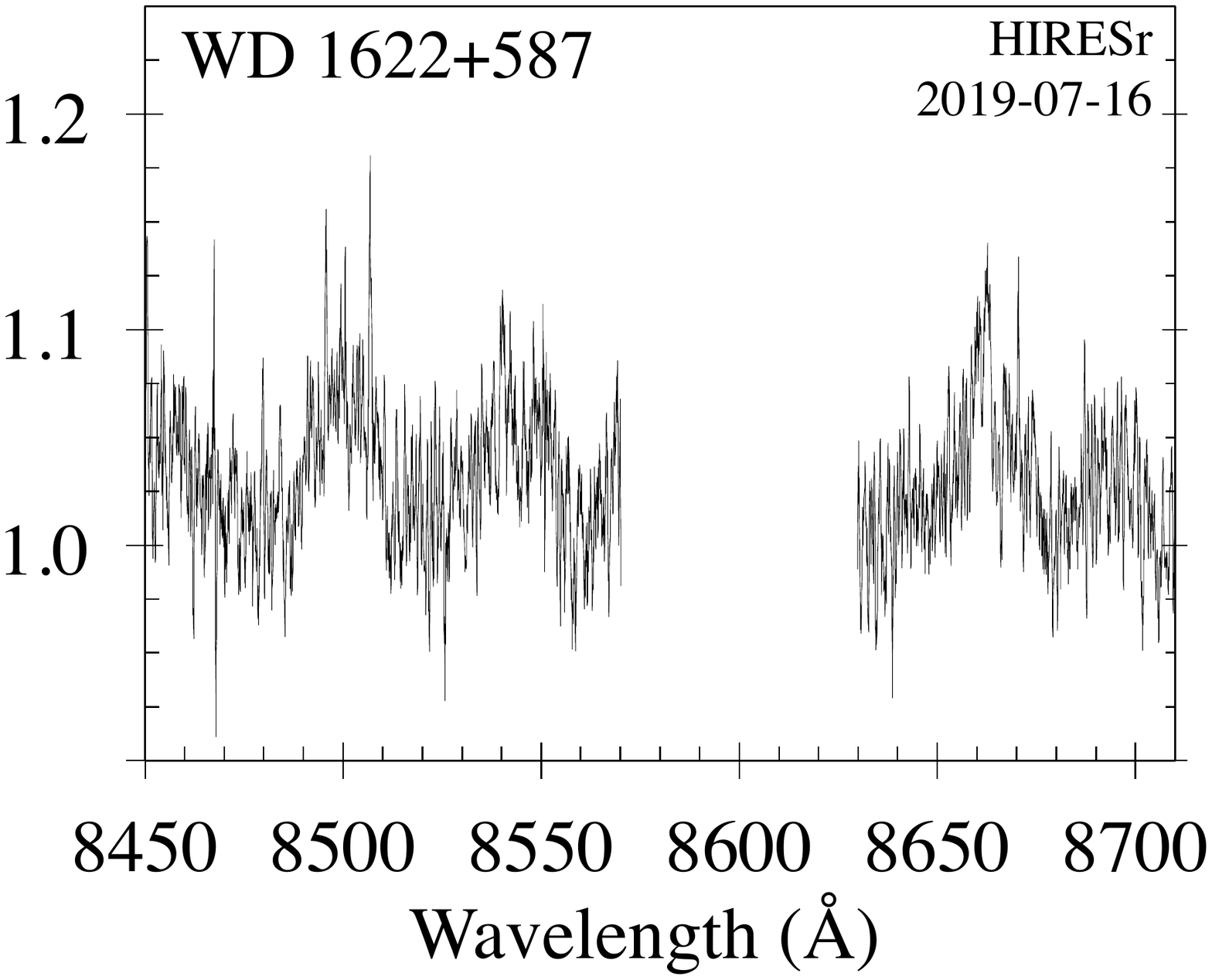}
  \end{minipage}
  \begin{minipage}[!b]{53mm}
   \includegraphics[width=80mm,trim={0cm 0 0cm 3.5cm},clip]{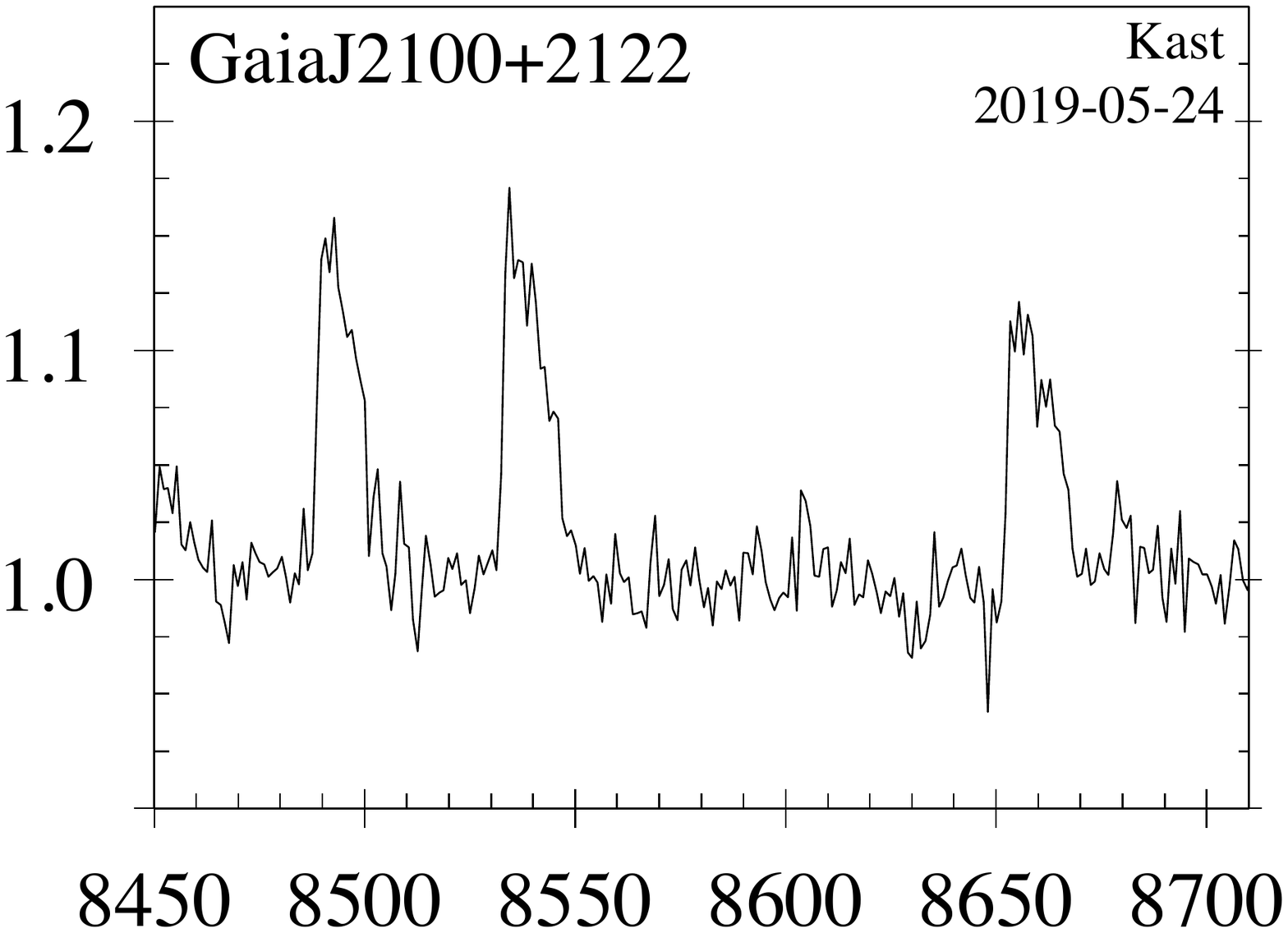}
  \end{minipage}
  \caption{\label{figcaiiirts} {\large An individual epoch of Ca~II IRT emission for each of the nine 
               white dwarfs discovered in this work to host a gas disk. 
               In all cases, spectra are continuum normalized.
               Wavelengths for HIRES data are in vacuum and shifted to the heliocentric reference frame.
               Kast spectra have wavelengths presented in air and are not shifted to the heliocentric
               reference frame. MagE spectra have wavelengths in air and are shifted to the heliocentric
               reference frame. In some cases the HIRES spectra are smoothed with a 5- or 11-pixel boxcar
               for display purposes.
               More complete spectra and multiple epochs for each star are presented in
               Appendix \ref{appfigtab}.} }
\end{figure}

\clearpage

\begin{figure}
 \centering
 \begin{minipage}[!h]{80mm}
  \includegraphics[width=87mm,trim={2cm 2cm 2cm 2cm},clip]{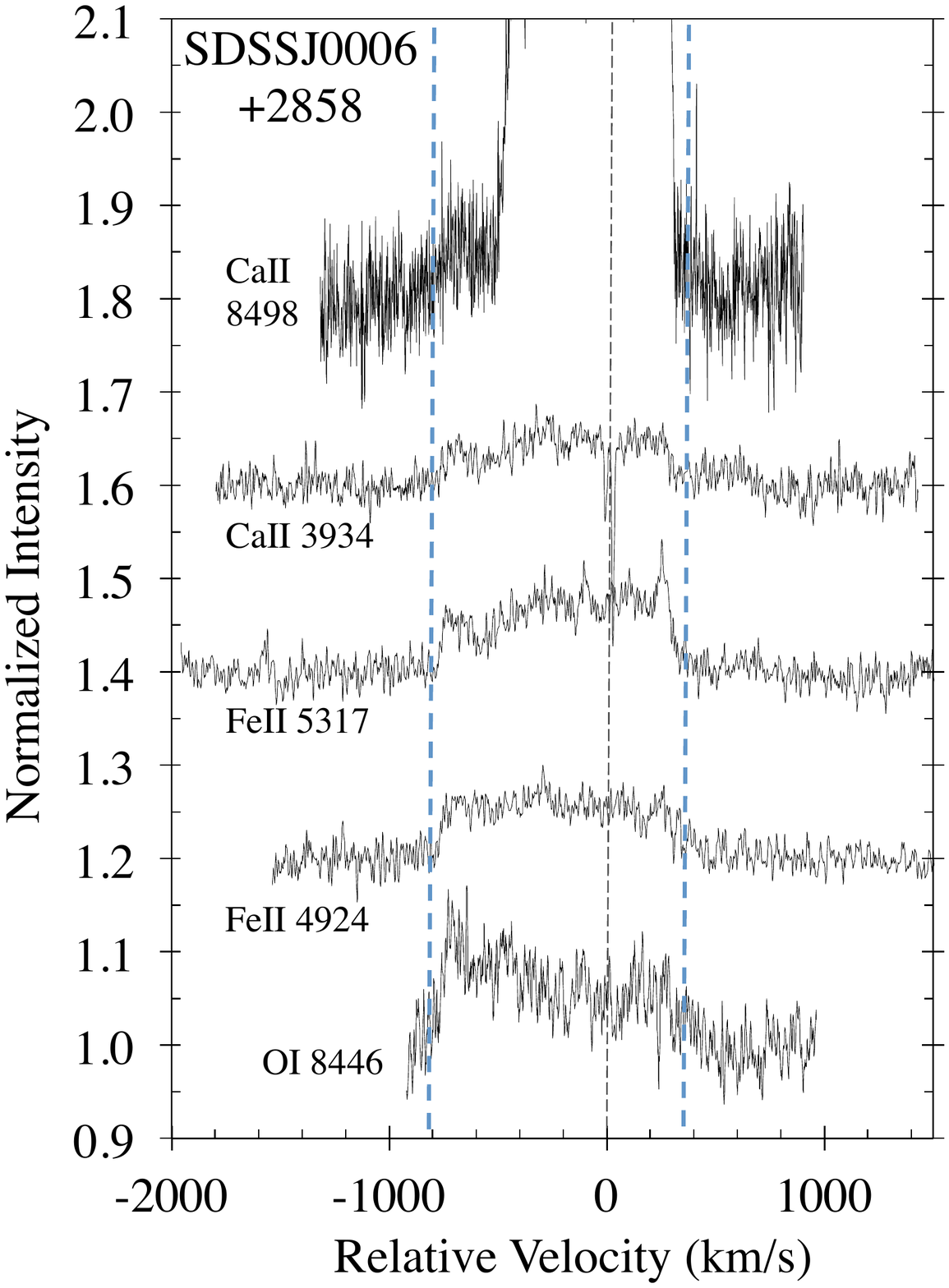}
 \end{minipage}
 \begin{minipage}[!h]{80mm}
  \includegraphics[width=87mm,trim={2cm 2cm 2cm 2cm},clip]{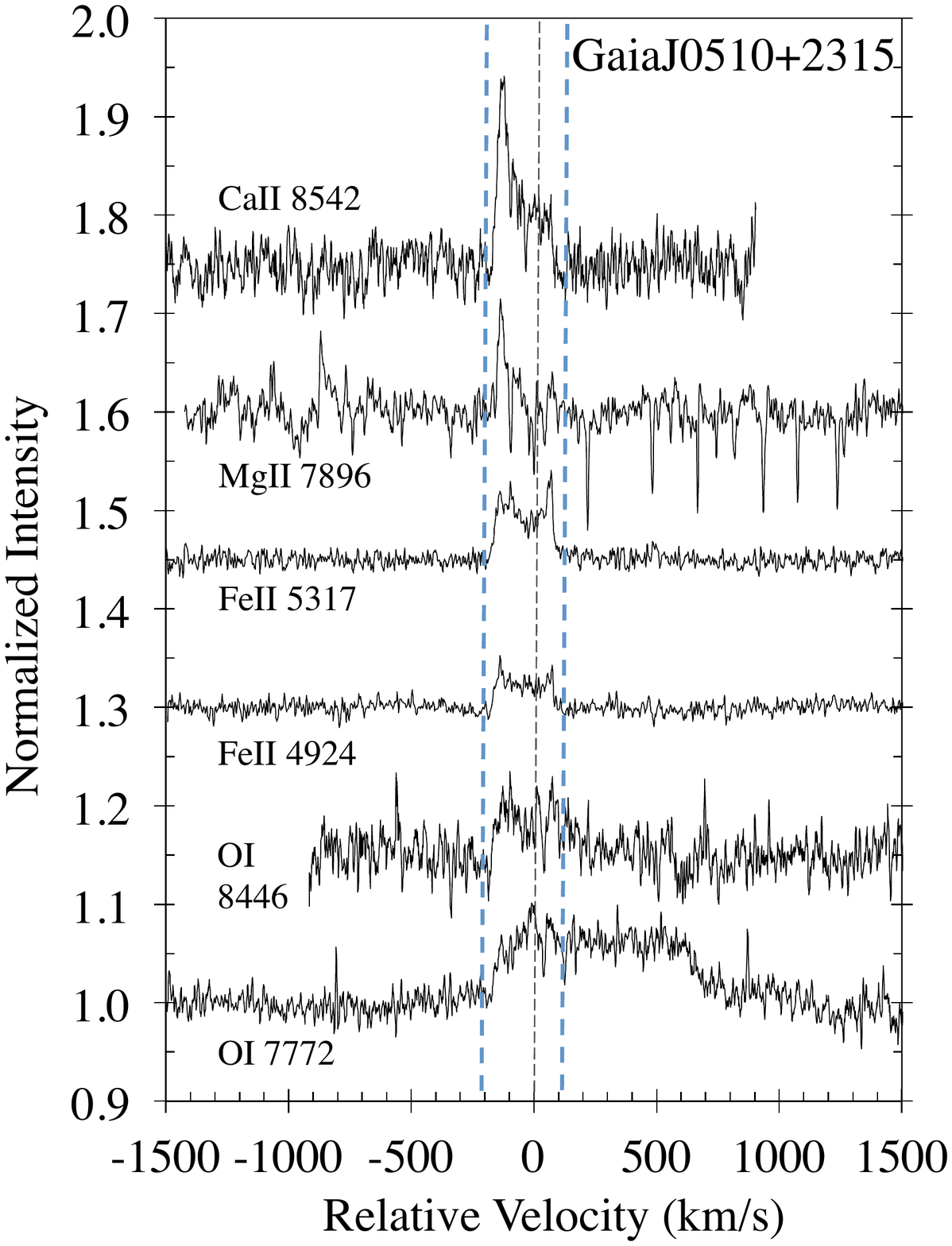}
 \end{minipage}
 \caption{\label{fig0006lv} \large{Velocity-space plots with 
             representative emission lines for each element
             detected for SDSS\,J0006+2858 (left) and Gaia\,J0510+2315 (right).
             Continuum levels have been fit and
             divided into each HIRES 
             spectral segment shown and individual segments offset by an additive constant
             for purposes of clarity.
             A velocity of 0 corresponds to the calculated systemic velocity for each star
             (Section \ref{secgasres} and
             Table \ref{tabkins}). Blue vertical dashed lines give the measured average maximum
             gas velocities seen in the blue and red wings of emission lines (Section \ref{secgasres} and
             Tables \ref{tabline0006} and \ref{tabline0510}).
             Clear velocity asymmetry is seen in the red and blue wings of all elements
             for SDSS\,J0006+2858 while more subtle velocity asymmetry is seen for
             Gaia\,J0510+2315.
             Extended red-wing emission in the O~I $\lambda$8446 complex for SDSS\,J0006+2858
             is due to a blend
             of multiple lines, the range delineated by the vertical dashed lines is centered on 
             the lowest wavelength
             transition in the multiplet.
             For Gaia\,J0510+2315, O~I lines near 7772 and 8446\,\AA\ are significantly
             more extended than other emission lines (see Section \ref{sec0510}).}  }
\end{figure}

\clearpage

\floattable

\begin{deluxetable}{ccccccccccc}
\rotate
\tabletypesize{\normalsize}
\tablecolumns{11}
\tablewidth{0pt}
\tablecaption{\large{White Dwarf Parameters \label{tabpars}} }
\tablehead{ 
  \colhead{Object} & 
  \colhead{R.A.} &
  \colhead{Decl.} &
  \colhead{$G$$_{\rm mag}$} &
  \colhead{Primary} &
  \colhead{T$_{\rm eff}$} & 
  \colhead{log$g$} &
  \colhead{Mass} &
  \colhead{Radius} &
  \colhead{[Ca/H(e)]} &
  \colhead{[Mg/H(e)]} \\
  \colhead{} & 
  \multicolumn{2}{c}{(J2000)} &
  \colhead{(mag)} &
  \colhead{Atm.\ Const.} &
  \colhead{(K)} & 
  \colhead{(cgs)} &
  \colhead{(M$_{\odot}$)} &
  \colhead{(R$_{\odot}$)} &
  \multicolumn{2}{c}{(log abundance by \#)}
}
\startdata
SDSSJ0006+2858 & 00 06 34.713 & +28 58 46.54 & 16.37 & H & 26,000 & 8.0 & 0.64 & 0.013 & $-$ & $-$ \\
WD0145+234         & 01 47 54.818 & +23 39 43.60 & 14.05 & H & 12,720 & 8.1 & 0.67 & 0.012 & $-$6.6 & $-$5.9 \\ %from HIRES data
SDSSJ0347+1624 & 03 47 36.692  & +16 24 09.73 & 16.65 & H & 21,815 & 8.1 & 0.69 & 0.012 & $-$ & $-$  \\
GaiaJ0510+2315 &  05 10 02.146 & +23 15 41.42  & 15.06  & H & 21,700 & 8.2 & 0.75 & 0.011 & $-$ & $-$5.5  \\
GaiaJ0611$-$6931 & 06 11 31.699 & $-$69 31 02.15 & 16.80 & H & 17,900 & 8.2 & 0.74 & 0.011 & $-$6.2 & $-$4.6  \\
GaiaJ0644$-$0352 & 06 44 05.231 & $-$03 52 06.42 & 16.19 & He\tablenotemark{a} & 18,350 & 8.2 & 0.72 & 0.011 & $-$6.7 & $-$ \\
WD0842+572         & 08 46 02.473 & +57 03 28.64 & 16.81 & H & 16,225 & 8.0 & 0.62 & 0.013 & $-$ & $-$3.9 \\
WD1622+587         & 16 22 59.645 & +58 40 30.89 & 16.87 & He\tablenotemark{a} & 18,850 & 7.8 & 0.49 & 0.015 & $-$ & $-$ \\
GaiaJ2100+2122 & 21 00 34.648 & +21 22 56.89 & 15.19 & H & 26,550 & 8.1 & 0.70 & 0.012 & $-$ & $-$5.0 \\
\enddata
\tablecomments{Abundances are quoted relative to the primary atmospheric constituent, hydrogen (H) or helium (He). The primary atmospheric constituent is indicated in the ``Primary Atm.\ Const.'' column. A ``$-$'' in the abundance column means no metal absorption lines are seen in the Kast spectra.}
\tablenotetext{a}{Hydrogen is also detected in the atmospheres of GaiaJ0644$-$0352 
and WD1622+587; we obtain log$_{10}$(H/He)$=$$-$5.1 and log$_{10}$(H/He)$=$$-$4.8 
respectively.}
\end{deluxetable}

\clearpage

\floattable

\begin{deluxetable}{lcccccccccc}
\rotate
\tabletypesize{\normalsize}
\tablecolumns{11}
\tablewidth{0pt}
\tablecaption{\large{Observed Emission Lines \label{tablines}} }
\tablehead{ 
  \colhead{Element} & 
  \colhead{rest $\lambda$$_{\rm air}$} &
  \colhead{SDSSJ0006} &
  \colhead{WD0145} &
  \colhead{SDSSJ0347} &
  \colhead{GaiaJ0510} &
  \colhead{GaiaJ0611} &
  \colhead{GaiaJ0644} &
  \colhead{WD0842} &
  \colhead{WD1622} & 
  \colhead{GaiaJ2100}
}
\startdata
\multicolumn{11}{l}{\bf Oxygen} \\
O~I  & 7772        &  Y   & $-$ &  w  &  Y  &  w   & $-$ & $-$  &  Y   &  w  \\
O~I  & 8446        &  w  & $-$ & $-$ &  w   &  w   & $-$ & $-$ & $-$ &  w  \\
\multicolumn{11}{l}{\bf Sodium} \\
Na~I & 5890       & $-$ & $-$ & $-$ & $-$ & Y & $-$ & $-$ & $-$ & $-$  \\
\multicolumn{11}{l}{\bf Magnesium} \\
Mg~I] & 4571.095 & $-$ & $-$ & $-$ & $-$ & $-$ & $-$ &  Y  & $-$ & $-$  \\
Mg~I  & 5167.321 &  ?  & $-$ &   ?  &   ?   & Y & $-$ &  Y  &  ?  & $-$ \\
Mg~I  & 5172.684 &  ?  & $-$ &   ?  &   ?   & Y & $-$ &  Y  &  ?  & $-$ \\
Mg~I  & 5183.604 &  ?  & $-$ &   ?  &   w   & Y & $-$ &  Y  &  ?  & $-$ \\
Mg~I  & 8806.76   &  ?   &  ?   & $-$  &  ?  &  Y   &   ?  &   ?  &   ?  &   ?  \\
Mg~II & 7877.05   & $-$ & $-$  & $-$ &  w  & $-$ & $-$ & $-$ & $-$ & $-$ \\
Mg~II & 7896.37   & $-$ & $-$ & $-$ &  w  & $-$ & $-$ & $-$ & $-$ & $-$ \\
\multicolumn{11}{l}{\bf Silicon} \\
Si~I  & 3905.523 & $-$ & $-$ & $-$ & $-$ &  Y  & $-$ &   w  & $-$ & $-$ \\
\enddata
\tablecomments{For each star and transition, a ``$-$''
means no line was seen in any epoch, ``w'' means a
weak transition is seen, ``Y'' means a clear emission line is
seen, and ``?'' means this wavelength was not covered by our observations
or lies within a strong transition that prevents conclusive analysis.
SDSS\,J0347+1624 was only observed with the Kast, all other stars have
higher resolution data products.}
\end{deluxetable}

\clearpage

\addtocounter{table}{-1}

\floattable

\begin{deluxetable}{lcccccccccc}
\rotate
\tabletypesize{\normalsize}
\tablecolumns{11}
\tablewidth{0pt}
\tablecaption{\large{(continued) Observed Emission Lines \label{tablines}} }
\tablehead{ 
  \colhead{Element} & 
  \colhead{rest $\lambda$$_{\rm air}$} &
  \colhead{SDSSJ0006} &
  \colhead{WD0145} &
  \colhead{SDSSJ0347} &
  \colhead{GaiaJ0510} &
  \colhead{GaiaJ0611} &
  \colhead{GaiaJ0644} &
  \colhead{WD0842} &
  \colhead{WD1622} & 
  \colhead{GaiaJ2100}
}
\startdata
\multicolumn{11}{l}{\bf Calcium} \\
Ca~I]  & 6572.78   & $-$ & $-$ & $-$ & $-$ & $-$ & $-$ &  w & $-$ & $-$ \\
Ca~II & 3933.663 &  w  & $-$ & $-$ & $-$ &  Y  & $-$ &  w  & $-$ & $-$ \\
Ca~II & 3968.469 &  w  & $-$ & $-$ & $-$ &  w  & $-$ &  w  & $-$ & $-$ \\
Ca~II & 8498.023 &  Y  &   Y  &  Y  & w  &   Y   &  Y  &  Y  &   w  &   Y  \\
Ca~II & 8542.091 &  Y  &   Y  &  Y  &  Y  &   Y   &  Y  &  Y  &   w  &   Y  \\
Ca~II & 8662.141 &  Y  &   Y  &  Y  &  w  &   Y   &  Y  &  Y  &   w  &   Y  \\
\multicolumn{11}{l}{\bf Neutral Iron} \\
Fe~I] & 4375.929 &  $-$ & $-$ & $-$ & $-$ & $-$  & $-$ &  Y   & $-$ & $-$  \\ 
Fe~I] & 4427.310 & $-$ & $-$ & $-$ & $-$ & $-$  & $-$ &   Y   & $-$ & $-$  \\ 
Fe~I] & 4461.652 &  $-$ & $-$ & $-$ & $-$ & $-$  & $-$ &  w   & $-$ & $-$  \\ 
Fe~I  & 5012.068 &   ?   &  $-$ & ?  &  $-$  &  $-$ &   ?  &  Y    &  ?   &  ?    \\
%Fe~I] & 5204.583 &  $-$ & $-$ & $-$ & $-$ & $-$  & $-$ &  w   & $-$ & $-$  \\ 
Fe~I] & 5110.413 & $-$ & $-$ & $-$ & $-$ & $-$ & $-$ & Y  & $-$ & $-$ \\
\enddata
%\tablecomments{For each star and transition, a ``$-$''
%means no line was seen in any epoch, ``w'' means a
%weak transition is seen, ``Y'' means a clear emission line is
%seen, and ``?'' means this wavelength was not covered by our observations
%or lies within a strong transition that prevents conclusive analysis.}
\end{deluxetable}

\clearpage

\addtocounter{table}{-1}

\floattable

\begin{deluxetable}{lcccccccccc}
\rotate
\tabletypesize{\normalsize}
\tablecolumns{11}
\tablewidth{0pt}
\tablecaption{\large{(continued) Observed Emission Lines \label{tablines}} }
\tablehead{
  \colhead{Element} & 
  \colhead{rest $\lambda$$_{\rm air}$} &
  \colhead{SDSSJ0006} &
  \colhead{WD0145} &
  \colhead{SDSSJ0347} &
  \colhead{GaiaJ0510} &
  \colhead{GaiaJ0611} &
  \colhead{GaiaJ0644} &
  \colhead{WD0842} &
  \colhead{WD1622} & 
  \colhead{GaiaJ2100}
}
\startdata
\multicolumn{11}{l}{\bf Ionized Iron} \\
Fe~II & 4178.853 &   w  & $-$ & $-$ & $-$ & $-$  & $-$ &  w   & $-$ &  w  \\
Fe~II & 4233.162 &   w  & $-$ & $-$ & $-$ & $-$  & $-$ &  w   & $-$ &  w  \\
Fe~II & 4351.762 &   w  & $-$ & $-$ & $-$ & $-$  & $-$ &  w   & $-$ &  w  \\
Fe~II & 4416.818 &   w  & $-$ & $-$ & $-$ & $-$  & $-$ &  $-$  & $-$ &  w  \\
Fe~II & 4515.333 &   w  & $-$ & $-$ & $-$ & $-$  & $-$ &  w   & $-$ &  w  \\
Fe~II & 4520.218 &   w  & $-$ & $-$ & $-$ & $-$  & $-$ & $-$ & $-$ &  w  \\
Fe~II & 4522.627 &   w  & $-$ & $-$ & $-$ & $-$  & $-$ & $-$ & $-$ &  w  \\
Fe~II & 4549.466 &   w  & $-$ &  w  &  $-$ &  $-$ & $-$ &  w  &  $-$  &  w    \\
Fe~II & 4583.829 &   w  & $-$ &  w   &  w   &  $-$ & $-$ &  w    &  $-$  &  Y    \\
Fe~II & 4629.332 &  w  & $-$ & w  &   w  & $-$ & $-$ &  w & $-$ &  Y  \\
Fe~II & 4923.921 &  Y  & $-$ &  w  &  Y  & $-$ &   ?   &  Y  &   ?   &  w  \\
Fe~II & 5018.436 &  Y  & $-$ &  Y  &   Y &  $-$ &   ?   &  Y  &   ?   &  Y  \\
Fe~II & 5169.028 &  Y  & $-$ &  Y   &  Y  &  ?  & $-$ &  Y  &  ?   &  Y \\
Fe~II & 5197.57   &  w  & $-$ &  w   &  w  &  ? & $-$ & $-$ & $-$ &   w \\
Fe~II & 5234.62   &  w  & $-$ &  w   & w  & $-$ & $-$ &  w  & $-$ &  w \\
Fe~II & 5276.00   &  w  & $-$ &  w  &  w   & $-$ & $-$ &  w &  w  &  w \\
Fe~II & 5316.609 &  Y  & $-$ &  Y   &  Y   & $-$ & $-$ &  Y  &  w  &  Y \\
Fe~II & 6458.164 &  w  & $-$ &  w   &  w    & $-$  & $-$ &  w  & $-$ &  w  \\
Fe~II & 7711.720 & $-$ & $-$ &  w  &  w  & $-$ & $-$ &  w  & $-$ &  w  \\
\enddata
%\tablecomments{For each star and transition, a ``$-$''
%means no line was seen in any epoch, ``w'' means a
%weak transition is seen, ``Y'' means a clear emission line is
%seen, and ``?'' means this wavelength was not covered by our observations
%or lies within a strong transition that prevents conclusive analysis.}
\end{deluxetable}

\clearpage

\addtocounter{table}{-1}

\floattable

\begin{deluxetable}{lcccccccccc}
\rotate
\tabletypesize{\normalsize}
\tablecolumns{11}
\tablewidth{0pt}
\tablecaption{\large{(continued) Observed Emission Lines \label{tablines}} }
\tablehead{
  \colhead{Element} & 
  \colhead{rest $\lambda$$_{\rm air}$} &
  \colhead{SDSSJ0006} &
  \colhead{WD0145} &
  \colhead{SDSSJ0347} &
  \colhead{GaiaJ0510} &
  \colhead{GaiaJ0611} &
  \colhead{GaiaJ0644} &
  \colhead{WD0842} &
  \colhead{WD1622} & 
  \colhead{GaiaJ2100}
}
\startdata
\multicolumn{11}{l}{\bf Ambiguous or Unidentified} \\
?        & 4202       &  w  & $-$ & $-$ & $-$ & $-$  & $-$ & $-$ & $-$ &  w  \\ %FeI?
?        & 4488       & $-$ & $-$  & $-$ &  w  & $-$ & $-$ &  w & $-$ & $-$ \\
Fe~I? & 5052       &   w  & $-$ & $-$ & $-$ &  $-$ & $-$ & w  & $-$ & $-$ \\ %FeI?
%?        & 5365       &   w  & $-$ & $-$  & $-$  & $-$  & $-$  & $-$ &  $-$  & $-$   \\ %not real per Beth?
?        & 6160       & $-$  & $-$ & $-$ & $-$ &   w  & $-$ & $-$  & $-$ & $-$ \\ %
?        & 6240       &   w  & $-$ &  w   &  w   &   w   &  $-$ & $-$  &   $-$ & $-$    \\ %Fe?
?        & 6318       &   ?  & $-$  &  w   &  w   &  $-$  & $-$  & $-$  &  $-$ & $-$  \\ %Fe?
?        & 6347        & $-$ & $-$ & $-$ &  w   & $-$ & $-$ & $-$ & $-$ & $-$ \\ %maybe SiII? no good red edge to confirm..
?        & 6382       &   w  & $-$ &  w   &  w   &  $-$  & $-$  & $-$  & $-$  &  w    \\ %Fe?
?        & 6515       &  w & $-$ &   w  &  w   & $-$ & $-$ &  Y  & $-$ &  w   \\ %FeII 6516.076?
?        & 7506           & $-$ & $-$ & $-$ &  w & $-$ & $-$ & $-$ & $-$ & $-$ \\ %a doublet? ~19 Ang separation
\enddata
%\tablecomments{For each star and transition, a ``$-$''
%means no line was seen in any epoch, ``w'' means a
%weak transition is seen, ``Y'' means a clear emission line is
%seen, and ``?'' means this wavelength was not covered by our observations
%or lies within a strong transition that prevents conclusive analysis.}
\end{deluxetable}

\clearpage

\appendix
\counterwithin{figure}{section}

\section{Observations Summary and Target Kinematics Tables}

\begin{deluxetable}{cccccc}
\tabletypesize{\normalsize}
\tablecolumns{6}
\tablewidth{0pt}
\tablecaption{\large{Observations Summary} \label{tabobs}}
\tablehead{ 
  \colhead{Observation} & 
  \colhead{Instrument} &
  \colhead{Wavelength} &
  \colhead{Resolving} &
  \colhead{S/N} &
  \colhead{$\lambda$ of S/N} \\
  \colhead{Date (UT)} & 
  \colhead{Used} &
  \colhead{Range (\AA)} &
  \colhead{Power} &
  \colhead{} &
  \colhead{(\AA)} 
}
\startdata
\multicolumn{6}{l}{\bf SDSSJ0006+2858} \\ %fully updated!
\hline
 \multirow{2}{*}{29 Dec 2018} &  \multirow{2}{*}{Kast} &  3450-5475 & 900 & 36 & 5100 \\ 
                                               &                               & 5600-7850 & 2,300 & 28 & 7450 \\ 
 \multirow{2}{*}{25 Jan 2019} &  \multirow{2}{*}{Kast} &  3450-5475 & 1,300 & 38 & 5100 \\
                                               &                               & 6520-8785 & 3,500 & 37 & 7450 \\ 
07 Jul 2019 &  HIRES &  3139-5950 & 40,000 & 38 & 5100 \\
16 Jul 2019 &  HIRES &  4720-8995 & 40,000 & 32 & 7450 \\
 \multirow{2}{*}{26 Jul 2019} &  \multirow{2}{*}{Kast}  &  3450-5480 & 1,300 & 48 & 5100 \\
                                               &                               & 5580-7850 & 3,500 & 29 & 7450 \\ 
\hline
\multicolumn{6}{l}{\bf WD0145+234} \\ %fully updated!
\hline
\multirow{2}{*}{14 Nov 2018} & \multirow{2}{*}{Kast} & 3420-5480 & 1,300 & 55 & 5100 \\
                                               &                            & 5710-7820 & 3,500 & 30 & 7450 \\
21 Sep 2019 &  HIRES &  3130-5950 & 40,000 & 53 & 5100 \\
 \multirow{2}{*}{12 Oct 2019} &  \multirow{2}{*}{Kast} &  3450-5480 & 1,300 & 173 & 5100 \\
                                               &                              &   6440-8790 & 3,500 & 165 & 7450 \\
 \multirow{2}{*}{04 Nov 2019} &  \multirow{2}{*}{Kast} &  3450-5465 & 1,300 & 215 & 5100 \\
                                               &                              &   6440-8790 & 3,500 & 195 & 7450 \\
 \multirow{2}{*}{18 Nov 2019} &  \multirow{2}{*}{Kast} &  3450-5460 & 1,300 & 250 & 5100 \\ %three exposures throughout night, all same as best I can tell
                                               &                              &   6465-8795 & 3,500 & 220 & 7450 \\ %three sets of exposures throughout night, all same as best I can tell
09 Dec 2019 &  HIRES &  4720-8995 & 40,000 & 85 & 7450 \\
\hline
\multicolumn{6}{l}{\bf SDSSJ0347+1624} \\ %fully updated!
\hline
 \multirow{2}{*}{21 Oct 2016} &  \multirow{2}{*}{Kast} &  3450-5430 & 1,300 & 50 & 5100 \\
                                               &                               & 5580-7690 & 3,500 & 30 & 7450 \\ 
 \multirow{2}{*}{20 Dec 2019} &  \multirow{2}{*}{Kast} &  3450-5450 & 700 & 57 & 5100 \\
                                               &                               & 5750-8900 & 1,700 & 59 & 7450 \\ 
\enddata
%\tablecomments{Resolving powers are derived from FWHM measurements of arclines in comparison spectra.
%HIRES wavelength coverage is quoted as continuous, but there are gaps in coverage 
%between each of the three CCDs and sometimes between red orders.}
\end{deluxetable}

\clearpage

\addtocounter{table}{-1}

\begin{deluxetable}{cccccc}
\tabletypesize{\normalsize}
\tablecolumns{6}
\tablewidth{0pt}
\tablecaption{\large{(continued) Observations Summary} }
\tablehead{ 
  \colhead{Observation} & 
  \colhead{Instrument} &
  \colhead{Wavelength} &
  \colhead{Resolving} &
  \colhead{S/N} &
  \colhead{$\lambda$ of S/N} \\
  \colhead{Date (UT)} & 
  \colhead{Used} &
  \colhead{Range (\AA)} &
  \colhead{Power} &
  \colhead{} &
  \colhead{(\AA)} 
}
\startdata
\multicolumn{6}{l}{\bf GaiaJ0510+2315} \\ %fully updated!
\hline
 \multirow{2}{*}{18 Mar 2019} &  \multirow{2}{*}{Kast} &  3465-5450 & 900 & 62 & 5100 \\
                                               &                               & 6470-8760 & 2,300 & 47 & 7450 \\ 
 \multirow{2}{*}{16 Nov 2019} &  \multirow{2}{*}{Kast} &  3450-5450 & 700 & 97 & 5100 \\
                                               &                               & 6475-8760 & 1,700 & 86 & 7450 \\ 
 \multirow{2}{*}{17 Nov 2019} &  \multirow{2}{*}{Kast} &  3450-5400 & 900 & 113 & 5100 \\
                                               &                               & 5590-7830 & 2,300 & 91 & 7450 \\ 
 \multirow{2}{*}{18 Nov 2019} &  \multirow{2}{*}{Kast} &  3450-5450 & 1,300 & 180 & 5100 \\ %two exposures throughout night, same as best I can tell
                                               &                               & 6480-8770 & 3,500 & 145 & 7450 \\ %two exposures throughout night, same as best I can tell
09 Dec 2019 &  HIRES &  4720-8995 & 40,000 & 65 & 7450 \\
\hline
\multicolumn{6}{l}{\bf GaiaJ0611-6931} \\ %fully updated!
\hline
 \multirow{3}{*}{10 Oct 2019} &  \multirow{3}{*}{GMOS-S} &  3700-4690 & 1,750 & 65 & 4500 \\
                                               &                               & 4730-5750 & 2,200 & 73 & 5100 \\ 
                                               &                               & 5800-6820 & 2,500 & 60 & 6200 \\ 
10 Nov 2019 &  MagE &  3120-9800 & 7,500 & 46 & 7450 \\
\hline
\hline
\multicolumn{6}{l}{\bf GaiaJ0644$-$0352} \\ %fully updated!
%not using MIKE data for J0644-0352 (Debes' data)
\hline
 \multirow{2}{*}{16 Dec 2018} &  \multirow{2}{*}{Kast} &  3450-5450 & 1,300 & 40 & 5100 \\
                                               &                               & 5720-7820 & 3,500 & 25 & 7450 \\
 \multirow{2}{*}{28 Dec 2018} &  \multirow{2}{*}{Kast} &  3450-5450 & 700 & 42 & 5100 \\
                                               &                               & 5620-7840 & 1,700 & 30 & 7450 \\ 
 \multirow{2}{*}{18 Mar 2019} &  \multirow{2}{*}{Kast} &  3450-5450 & 900 & 9 & 5100 \\ 
                                               &                               & 6420-8780 & 2,300 & 3 & 7450 \\ 
09 Dec 2019 &  HIRES &  4720-8995 & 40,000 & 26 & 7450 \\
\enddata
\tablecomments{Resolving powers are derived from FWHM measurements of arclines in comparison spectra.
HIRES wavelength coverage is quoted as continuous, but there are gaps in coverage 
between each of the three CCDs and sometimes between red orders.}
\end{deluxetable}

\clearpage

\addtocounter{table}{-1}

\begin{deluxetable}{cccccc}
\tabletypesize{\normalsize}
\tablecolumns{6}
\tablewidth{0pt}
\tablecaption{\large{(continued) Observations Summary} \label{tabobs}}
\tablehead{ 
  \colhead{Observation} & 
  \colhead{Instrument} &
  \colhead{Wavelength} &
  \colhead{Resolving} &
  \colhead{S/N} &
  \colhead{$\lambda$ of S/N} \\
  \colhead{Date (UT)} & 
  \colhead{Used} &
  \colhead{Range (\AA)} &
  \colhead{Power} &
  \colhead{} &
  \colhead{(\AA)} 
}
\startdata
\multicolumn{6}{l}{\bf WD0842+572} \\ %fully updated!
\hline
 \multirow{2}{*}{18 Mar 2019} &  \multirow{2}{*}{Kast} &  3450-5470 & 900 & 30 & 5100 \\
                                               &                               & 6440-8775 & 2,300 & 20 & 7450 \\ 
 \multirow{2}{*}{24 Apr 2019} &  \multirow{2}{*}{Kast} &  3450-5440 & 900 & 59 & 5100 \\
                                               &                               & 6460-8720 & 2,300 & 47 & 7450 \\ 
 \multirow{2}{*}{29 May 2019} &  \multirow{2}{*}{Kast} &  3480-5450 & 900 & 16 & 5100 \\
                                               &                               & 6470-8760 & 2,300 & 21 & 7450 \\ 
09 Dec 2019 &  HIRES &  4265-8995 & 40,000 & 31 & 7450 \\
15 Jun 2020  &  HIRES & 3130-5950 & 40,000 & 11 & 5100 \\
\hline
\multicolumn{6}{l}{\bf WD1622+587} \\ %fully updated!
\hline
 \multirow{2}{*}{25 Jan 2019} &  \multirow{2}{*}{Kast} &  3450-5455 & 700 & 38 & 5100 \\
                                               &                               & 5700-7810 & 1,700 & 35 & 7450 \\
 \multirow{2}{*}{24 Apr 2019} &  \multirow{2}{*}{Kast} &  3450-5450 & 900 & 44 & 5100 \\
                                               &                               & 6420-8720 & 2,300 & 37 & 7450 \\ 
 \multirow{2}{*}{12 Jul 2019} &  \multirow{2}{*}{Kast} &  3460-5420 & 900 & 53 & 5100 \\
                                               &                               & 6470-8770 & 2,300 & 49 & 7450 \\ 
16 Jul 2019 &  HIRES &  4720-8995 & 40,000 & 21 & 7450 \\
\hline
\multicolumn{6}{l}{\bf GaiaJ2100+2122} \\ %fully updated!
\hline
 \multirow{2}{*}{16 Dec 2018} &  \multirow{2}{*}{Kast} &  3450-5450 & 700 & 21 & 5100 \\
                                               &                               & 5720-7820 & 1,700 & 12 & 7450 \\
 \multirow{2}{*}{28 Dec 2018} &  \multirow{2}{*}{Kast} &  3450-5470 & 700 & 79 & 5100 \\
                                               &                               & 5600-7830 & 1,700 & 63 & 7450 \\ 
 \multirow{2}{*}{30 Dec 2018} &  \multirow{2}{*}{Kast} &  3460-5450 & 700 & 91 & 5100 \\
                                               &                               & 5650-7830 & 1,700 & 68 & 7450 \\
 \multirow{2}{*}{24 May 2019} &  \multirow{2}{*}{Kast} &  3460-5460 & 900 & 139 & 5100 \\
                                               &                               & 6470-8775 & 2,300 & 122 & 7450 \\ 
07 Jul 2019 &  HIRES &  3130-5950 & 40,000 & 54 & 5100 \\
16 Jul 2019 &  HIRES &  4720-8995 & 40,000 & 48 & 7450 \\
\enddata
%\tablecomments{Resolving powers are derived from FWHM measurements of arclines in comparison spectra.}
\end{deluxetable}

\clearpage

\floattable

\begin{deluxetable}{cccccccccc}
\rotate
\tabletypesize{\normalsize}
\tablecolumns{10}
\tablewidth{0pt}
\tablecaption{\large{White Dwarf Kinematics \label{tabkins}} }
\tablehead{ 
  \colhead{Object} & 
  \colhead{$\pi$\tablenotemark{a}} &
  \colhead{$\mu$$_{\alpha}$cos$\delta$} &
  \colhead{$\mu$$_{\delta}$} &
  \colhead{RV meas} &
  \colhead{Grav.\ Redshift} & 
  \colhead{Syst.\ Vel.} &
  \colhead{U} &
  \colhead{V} &
  \colhead{W} \\
  \colhead{} &
  \colhead{(mas)} &
  \colhead{(mas\,yr$^{-1}$)} &
  \colhead{(mas\,yr$^{-1}$)} &
  \colhead{(km\,s$^{-1}$)} &
  \colhead{(km\,s$^{-1}$)} &
  \colhead{(km\,s$^{-1}$)} & 
  \colhead{(km\,s$^{-1}$)} & 
  \colhead{(km\,s$^{-1}$)} & 
  \colhead{(km\,s$^{-1}$)} 
}
\startdata
SDSSJ0006+2858 & 6.4$\pm$0.3 & +23.2$\pm$0.2 & +13.3$\pm$0.1 & +27$\pm$5 & 30 & $-$3$\pm$5 & $-$18 & $-$6 & +7 \\
WD0145+234         & 33.9$\pm$0.3 & $-$5.2$\pm$0.1 & $-$97.6$\pm$0.1 & +43$\pm$5 & 35 & +9$\pm$5 & $-$1 & $-$3 & $-$16 \\ 
SDSSJ0347+1624 & 6.9$\pm$0.2  & +5.7$\pm$0.2 & $-$62.9$\pm$0.1 & +27$\pm$5 & 36 & $-$9$\pm$5 & +17 & $-$36 & $-$20  \\
GaiaJ0510+2315   & 15.3$\pm$0.3 & $-$44.4$\pm$0.1 & $-$38.2$\pm$0.1 & +26$\pm$5 & 41 & $-$15$\pm$5 & +18 & $-$2 & $-$15  \\
GaiaJ0611$-$6931 & 6.99$\pm$0.25 & +22.7$\pm$0.1 & +100.4$\pm$0.1 & +60$\pm$5 & 41 & +19$\pm$5 & $-$63 & $-$36 & +5  \\
GaiaJ0644$-$0352 & 8.9$\pm$0.2 & +29.9$\pm$0.1 & $-$24.0$\pm$0.1 & +92$\pm$5 & 41 & +51$\pm$5 & $-$31 & $-$45 & +5 \\
WD0842+572         & 7.52$\pm$0.08 & $-$21.9$\pm$0.1 & $-$24.2$\pm$0.1 & +24$\pm$5 & 29 & $-$5$\pm$5 & $-9$ & $-$15 & $-$12 \\
WD1622+587         & 5.5$\pm$0.2 & +43.2$\pm$0.1 & $-$128.7$\pm$0.1 & $-$17$\pm$5& 21 & $-$38$\pm$5 & +118 & $-$29 & $-$27 \\
GaiaJ2100+2122  & 11.5$\pm$0.3 & +73.7$\pm$0.1 & +38.4$\pm$0.1 & +3$\pm$5 & 35 & $-$32$\pm$5 & $-$42 & $-$20 & $-$5 \\
\enddata
\tablecomments{``RV meas'' is the average velocity measured for individual metal and hydrogen lines in high resolution data (see Section \ref{secgasres}). ``Grav.\ Redshift'' is derived from the atmospheric
parameters produced from models reported in Table \ref{tabpars}. ``Syst.\ Vel.'' is the calculated
systemic velocity for each white dwarf star. We do not report nor propagate uncertainties for the
gravitational redshift which could be up to 15\,km\,s$^{-1}$. UVW space motions are reported relative 
to the Sun such that positive U is towards the Galactic center, positive V is in the direction of Galactic 
rotation, and positive W is toward the north Galactic pole \citep{johnson87}.}
\tablenotetext{a}{For the $Gaia$ DR2 parallax uncertainty we add in quadrature any contribution from excess astrometric noise.
}
\end{deluxetable}

%kinematics info
%SDSSJ0006: Hbeta velocity = 27.7 km/s, CaIIK velocity = 26.9 km/s (CS component at -8.2 km/s)
%   grav. redshift = 30.1 km/s
%    => systemic velocity = -2.8 +/- 5 km/s
%WD0145: Hbeta velocity = 43.2 km/s, CaIIK velocity = 43.3 km/s
%   grav. redshift = 34.6 km/s
%    => systemic velocity = +8.7 +/- 5 km/s
%SDSSJ0347:  Hbeta velocity = +27.1 km/s, CaIIK is interstellar? same velocity as Na D.. (18.3 km/s)
%   grav. redshift = 35.8 km/s
%    => systemic velocity = -8.7 +/- 5 km/s
%GaiaJ0510: Hbeta velocity = 27.1 km/s, SiII 6350 velocity = 26.0 km/s
%   grav. redshift = 40.9 km/s
%    => systemic velocity = -14.4 +/- 5 km/s
%GaiaJ0611: Hbeta velocity = 51.2 km/s, CaIIK velocity = 66.4 km/s (avg 60)
%   (heliocentric correction = -0.63 km/s)
%   grav. redshift = 41.0 km/s
%    => systemic velocity = +19.0 +/- 5 km/s
%GaiaJ0644: SiII 6350 velocity = +91.6 +/- 5 km/s, SiII 6370 velocity = 91.3 +/- 5 km/s, Halpha velocity = +94.52 +/- 5 km/s
%   grav. redshift = 41.2 km/s
%    => systemic velocity = +51.3 +/- 5 km/s
%WD0842: Hbeta velocity = 24.0 km/s, FeII 5018 velocity = 24.8 km/s
%   grav. redshift = 29.5 km/s
%    => systemic velocity = -5.1 +/- 5 km/s
%WD1622: SiII 6350 velocity = -17.9 +/- 5 km/s, SiII 6370 velocity = -21.6 +/- 5 km/s, Halpha velocity = -10.5 +/- 5 km/s
%   grav. redshift = 21.3 km/s
%    => systemic velocity = -37.9 +/- 5 km/s
%GaiaJ2100: Hbeta velocity = 4.3 km/s, CaIIK velocity = 2.5 km/s (CS component at -13.5 km/s)
%   grav. redshift = 35.1 km/s
%    => systemic velocity = -31.7 +/- 5 km/s

\clearpage

\section{Individual Target Spectra Figures and Tables of Line Measurements}
\label{appfigtab}

\begin{figure}[!h]
 \centering
 \includegraphics[width=120mm,trim={2cm 2cm 2cm 2cm},clip]{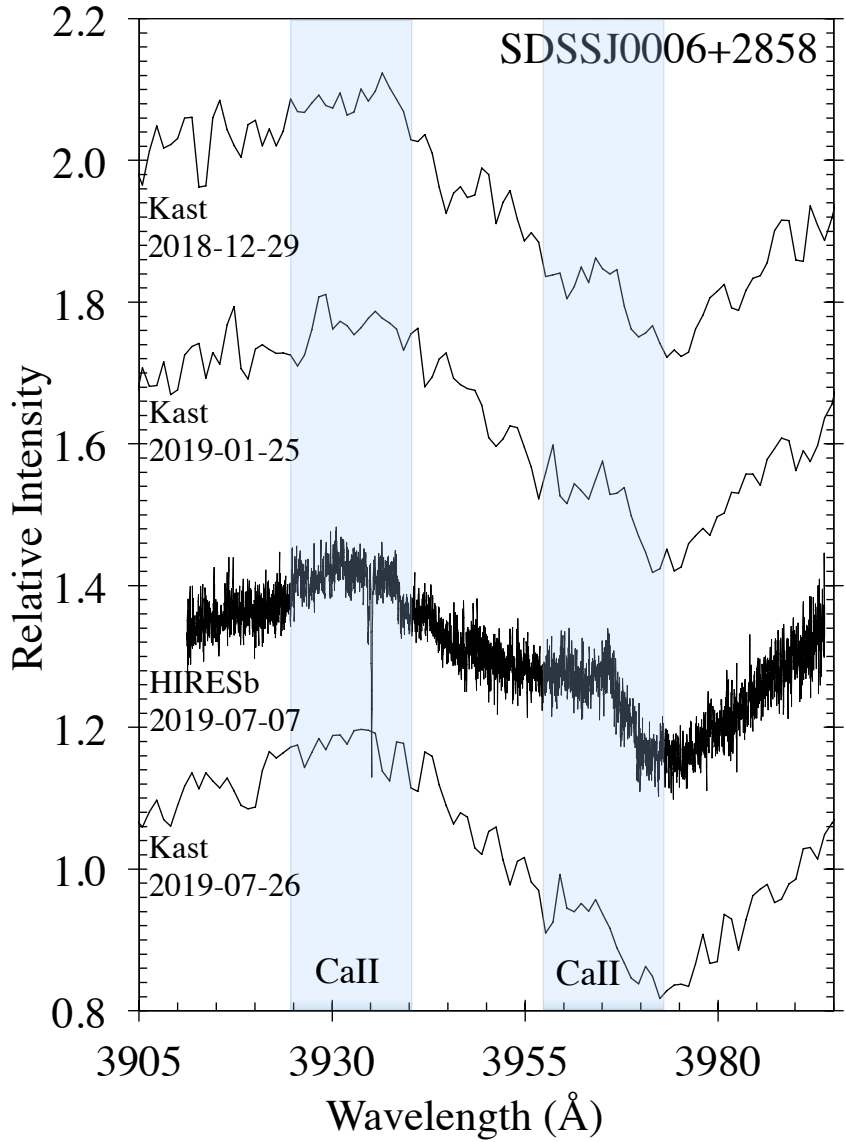}
 \caption{\label{fig0006l1} \large{Ca~II H+K spectra for SDSS\,J0006+2858.
             Likely emission is seen in Kast epochs while clear emission as well as photospheric
             and non-photospheric (interstellar or circumstellar) Ca II K-line 
             absorption is seen in the HIRES data.
             For this and all figures shown hereafter,
             when there are multiple epochs present in the figure the
             individual epochs have been offset by an additive constant
             for the purposes of clarity. 
             Wavelengths are in vacuum and shifted to the heliocentric reference frame for HIRES data,
             while they are in air and not shifted to the heliocentric frame for Kast data.
             Emission regions are marked with blue highlighted vertical bars.}  }
\end{figure}

\clearpage

\begin{figure}
 \centering
 \includegraphics[width=155mm,trim={2cm 5cm 2cm 2cm},clip]{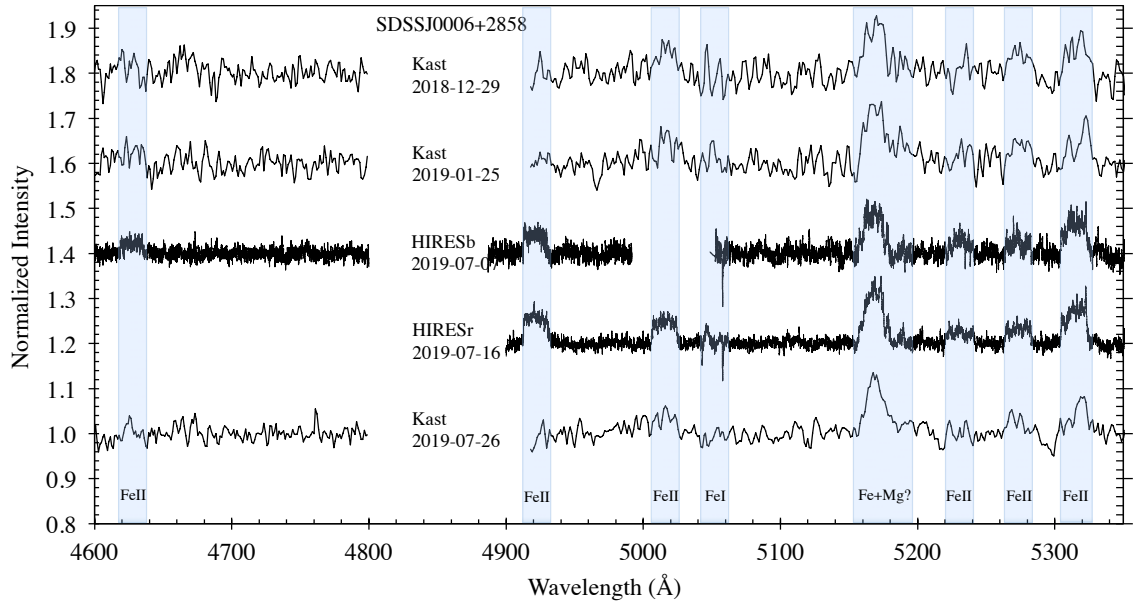}
 \caption{\label{fig0006l2} \large{Iron and possibly magnesium emission line region for
               SDSS\,J0006+2858. For this and all remaining figures for SDSS\,J0006+2858, the
               continuum levels have been fit and
               divided into each spectrum. 
               H$\beta$ at 4863\,\AA\  is cut out of each spectrum for plotting purposes.
               The HIRESb spectrum contains a gap
               between CCDs centered around 5025\,\AA .
               HIRES data are smoothed with an 11-pixel boxcar for display purposes.  
               Weak emission from an Fe~I line near 5052\,\AA\ may be present.
               It is possible that there is emission from Mg~I between 5150-5190\,\AA , but a
               clear identification is prevented due to dominating emission from Fe~II in the same region.}  }
\end{figure}

\clearpage

\begin{figure}
 \centering
 \includegraphics[width=160mm]{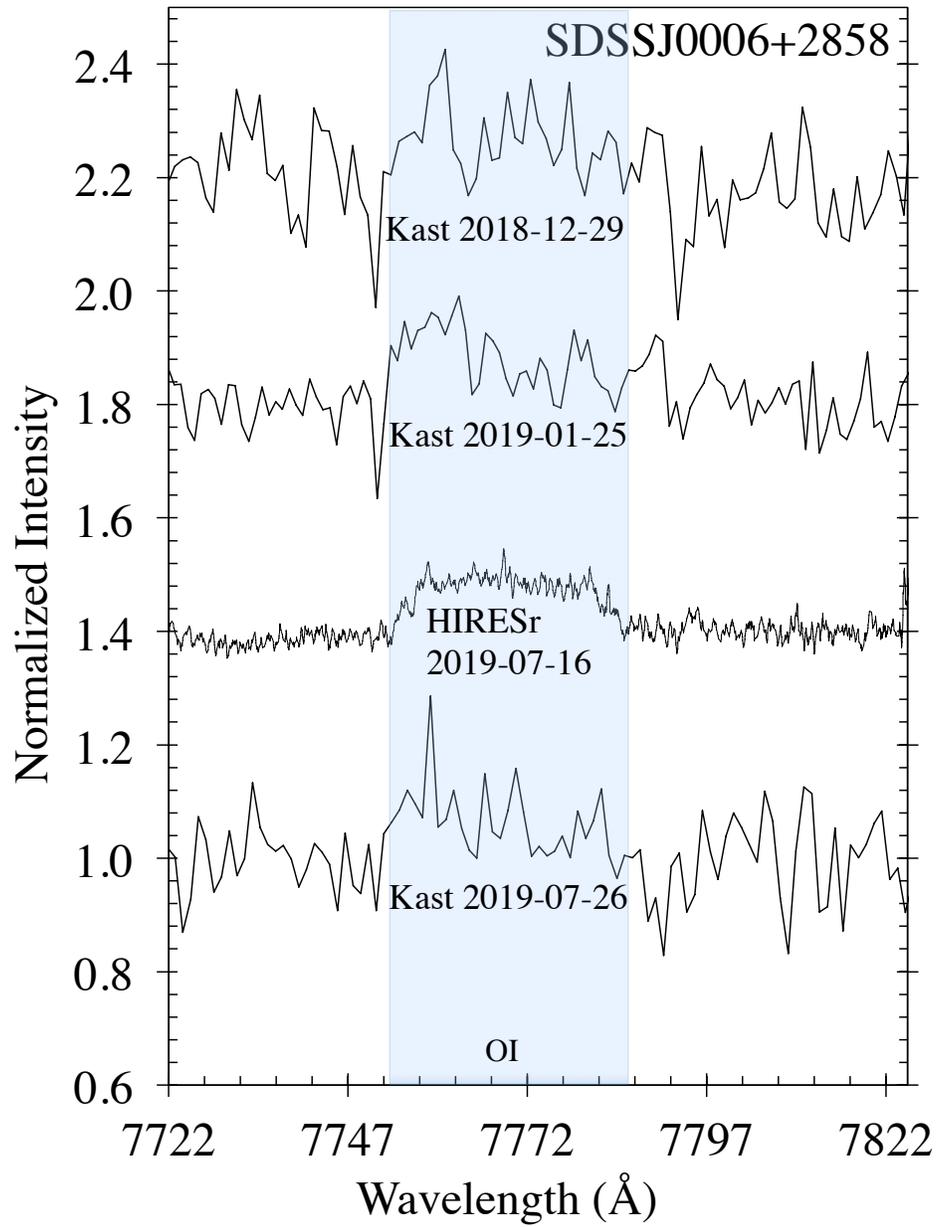}
 \caption{\label{fig0006l4} \large{O~I spectra for SDSS\,J0006+2858.
             HIRES spectra are smoothed with an 11-pixel boxcar.
             Probable emission is seen in Kast epochs while clear emission is
             seen in the HIRES data.}  }
\end{figure}

\clearpage

\begin{figure}
 \centering
 \includegraphics[width=155mm,trim={2cm 2cm 3cm 2cm},clip]{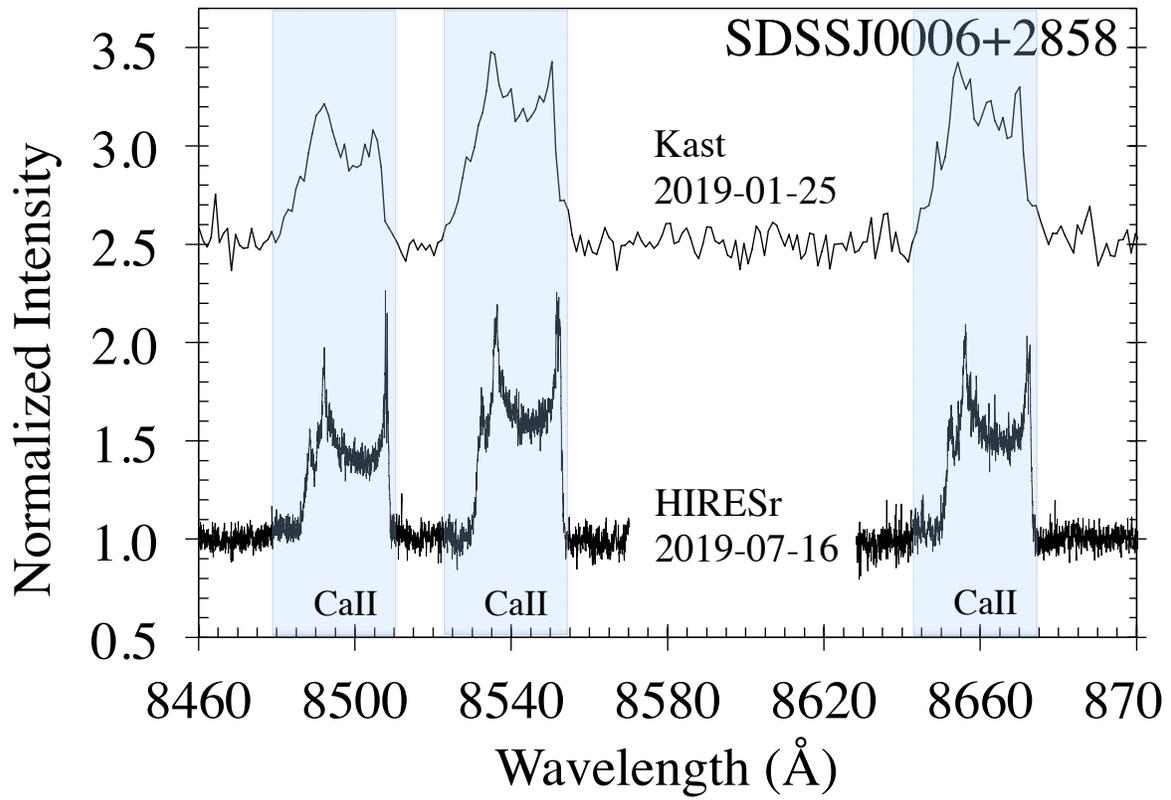}
 \caption{\label{fig0006l5} \large{Ca~II IRT portion of spectra for SDSS\,J0006+2858. 
             Highly structured
             emission features are seen in the HIRES spectra, including a ``third'' peak on the
             blue shoulder of the emission feature; this ``third'' peak appears to also be present
             in the Kast epoch.}  }
\end{figure}

\clearpage

\begin{figure}
 \centering
 \includegraphics[width=145mm,trim={2cm 2cm 2cm 2cm},clip]{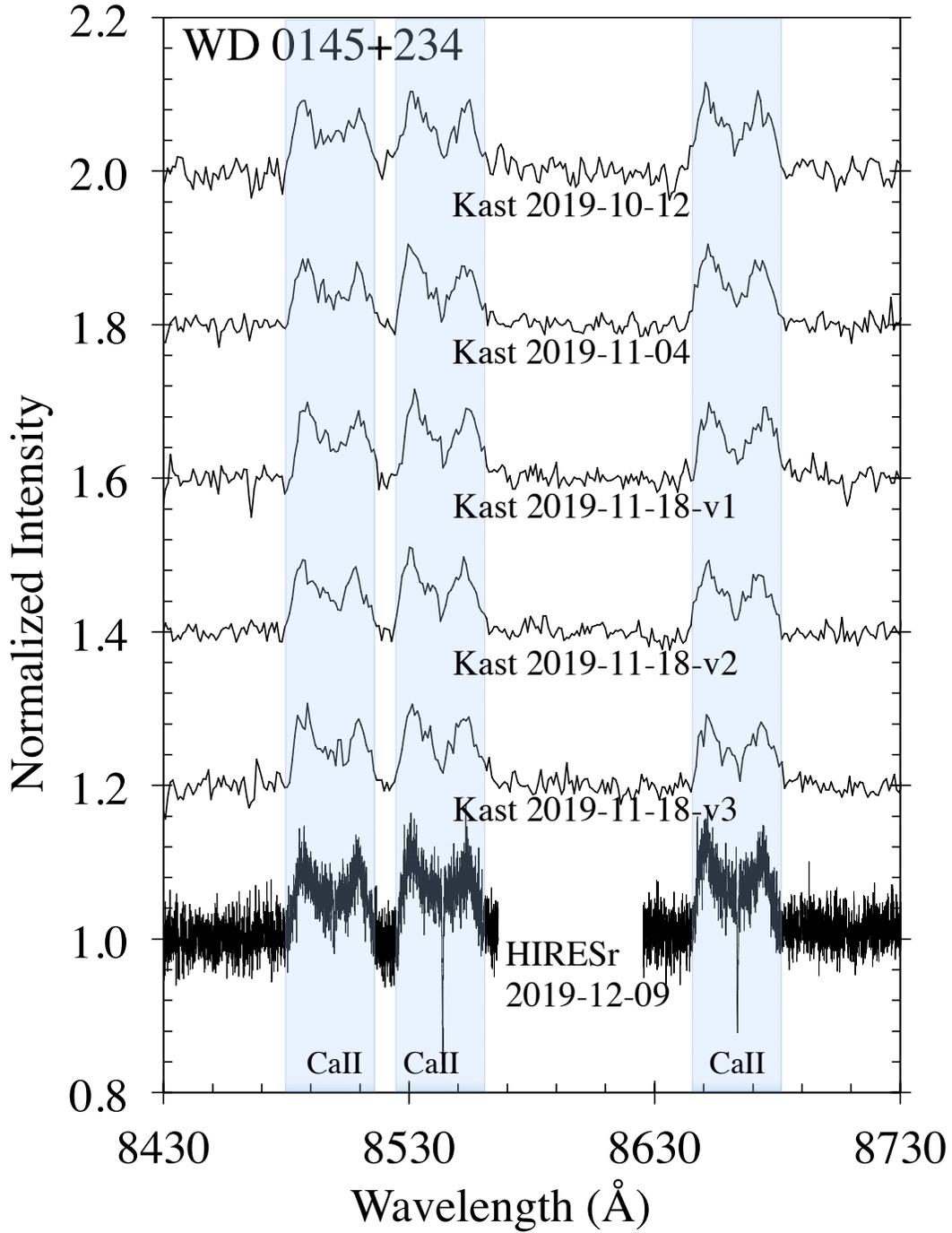}
 \caption{\label{fig0145l1} \large{Ca~II IRT spectra for WD0145+234. 
             Continuum levels have been fit and divided into each spectrum for this figure.
             Intra-night epochs taken with the Kast are marked as ``2019-11-18-v\#'' for three 
             observations taken throughout the night of UT 2019 November 18; no obvious variability 
             is seen between these or other epochs.}  }
\end{figure}

\clearpage

\begin{figure}
 \centering
 \includegraphics[width=155mm,trim={2cm 2cm 2cm 2cm},clip]{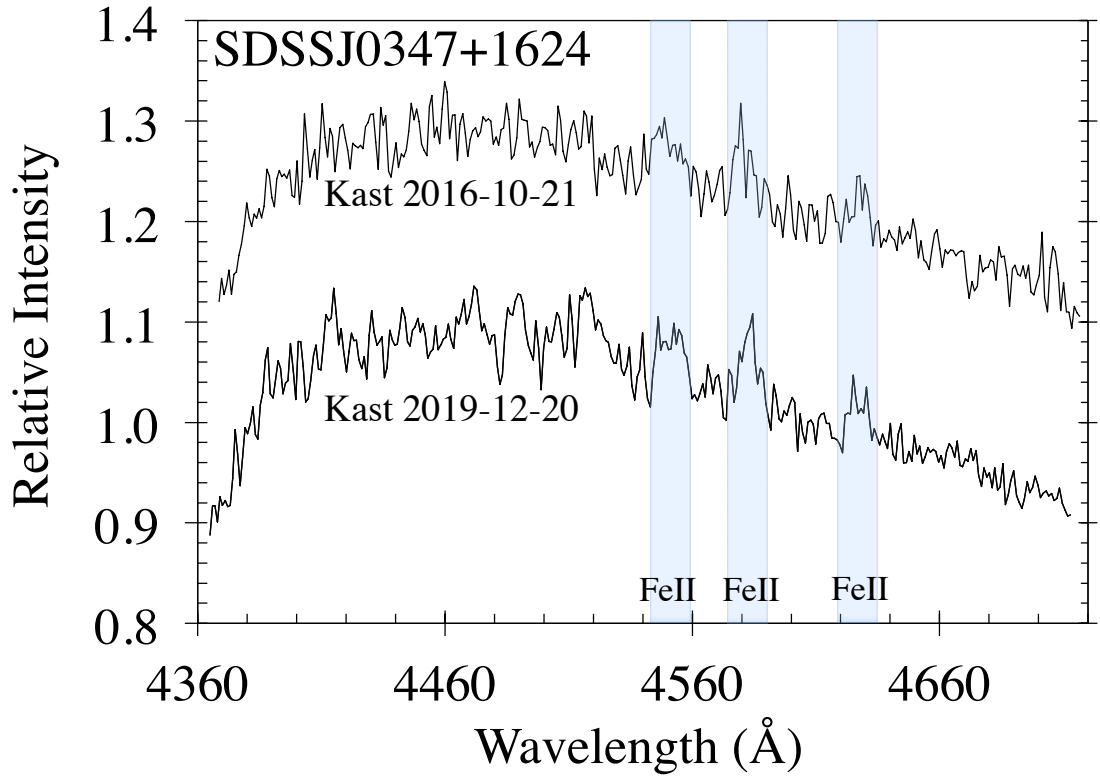}
 \caption{\label{fig0347l1} \large{Iron emission seen in one region of spectra for
               SDSS\,J0347+1624. Spectra are normalized to the median
               flux value in this plotted range.}  }
\end{figure}

\clearpage

\begin{figure}
 \centering
 \includegraphics[width=155mm,trim={2cm 2cm 2cm 2cm},clip]{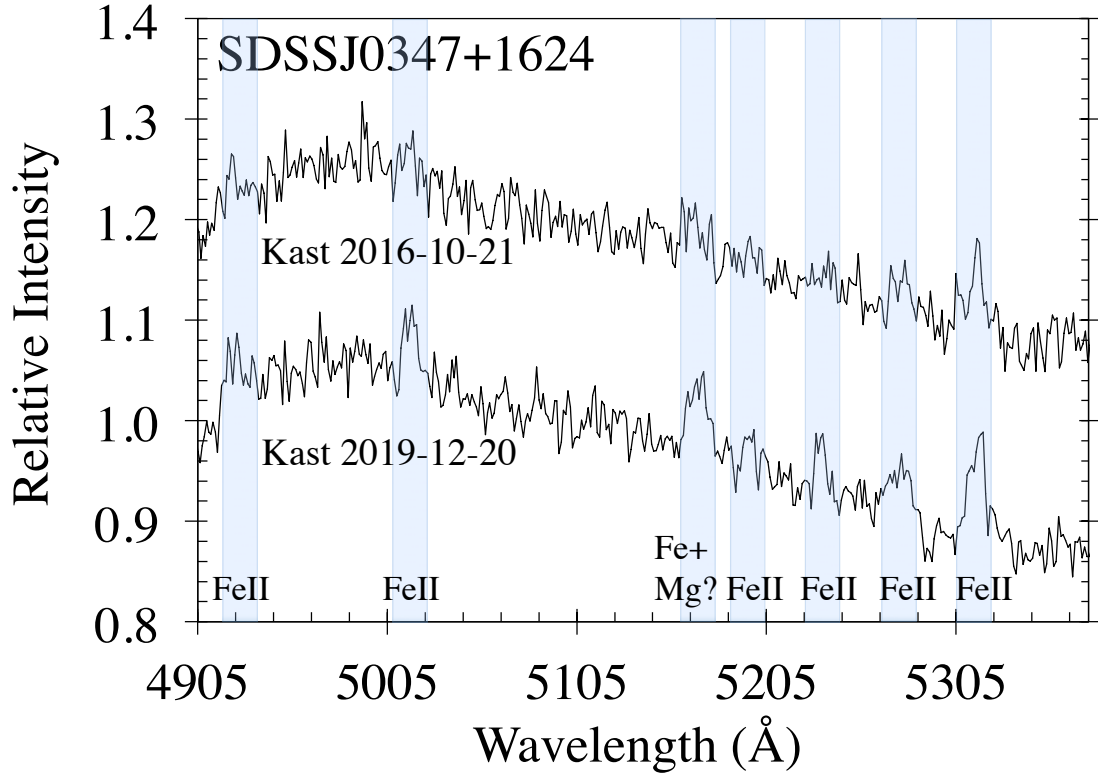}
 \caption{\label{fig0347l2} \large{Iron and possibly magnesium emission seen between
               4900-5350\,\AA\ for
               SDSS\,J0347+1624. Mg~I triplet emission, if present, would be blended with the stronger
               Fe~II $\lambda$5169 emission. Spectra are normalized to the median
               flux value in this plotted range.}  }
\end{figure}

%\clearpage
%Ben doesn't like this figure..
%\begin{figure}
% \centering
% \includegraphics[width=105mm,angle=90,trim={2cm 2cm 3cm 2cm},clip]{sdssj0347_Fe6300_plot.ps}
 %\caption{\label{fig0347l3} \large{Iron emission region around
 %              6300\,\AA\ for
 %              SDSS\,J0347+1624. Wavelengths are in air. Spectra are normalized to the median
 %              flux value in this plotted range and individual epochs offset by an additive constant.
 %              Emission regions where iron lines are seen in other systems with numerous iron
 %              emission lines are marked with blue highlighted vertical bars. It is not clear from
 %              these low-resolution data if such emission is present for SDSS\,J0347+1624.}  }
%\end{figure}

\clearpage

\begin{figure}
 \centering
 \includegraphics[width=155mm,trim={2cm 2cm 2cm 2cm},clip]{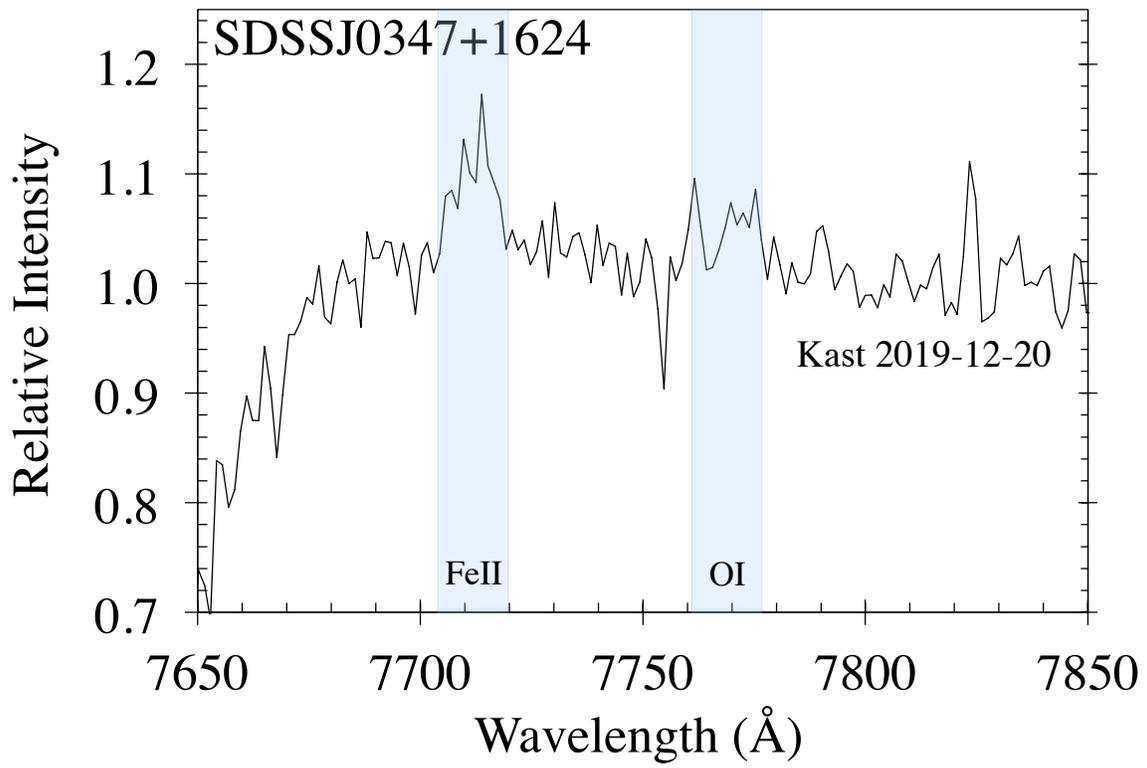}
 \caption{\label{fig0347l4} \large{Oxygen and iron emission for
               SDSS\,J0347+1624. The spectrum is normalized to the median
               flux value in this plotted range, no continuum normalization has been done.}  }
\end{figure}

\clearpage

%\begin{figure}
% \centering
% \includegraphics[width=155mm,trim={2cm 2cm 3cm 2cm},clip]{sdssj0347_CaIRT_plot.pdf}
% \caption{\label{fig0347l5} \large{Ca~II IRT emission region for
%               SDSS\,J0347+1624. The continuum level has been fit and
%               divided into the spectrum in this figure.}  }
%\end{figure}

%\clearpage

\begin{figure}
 \centering
 \includegraphics[width=155mm,trim={2cm 2cm 2cm 2cm},clip]{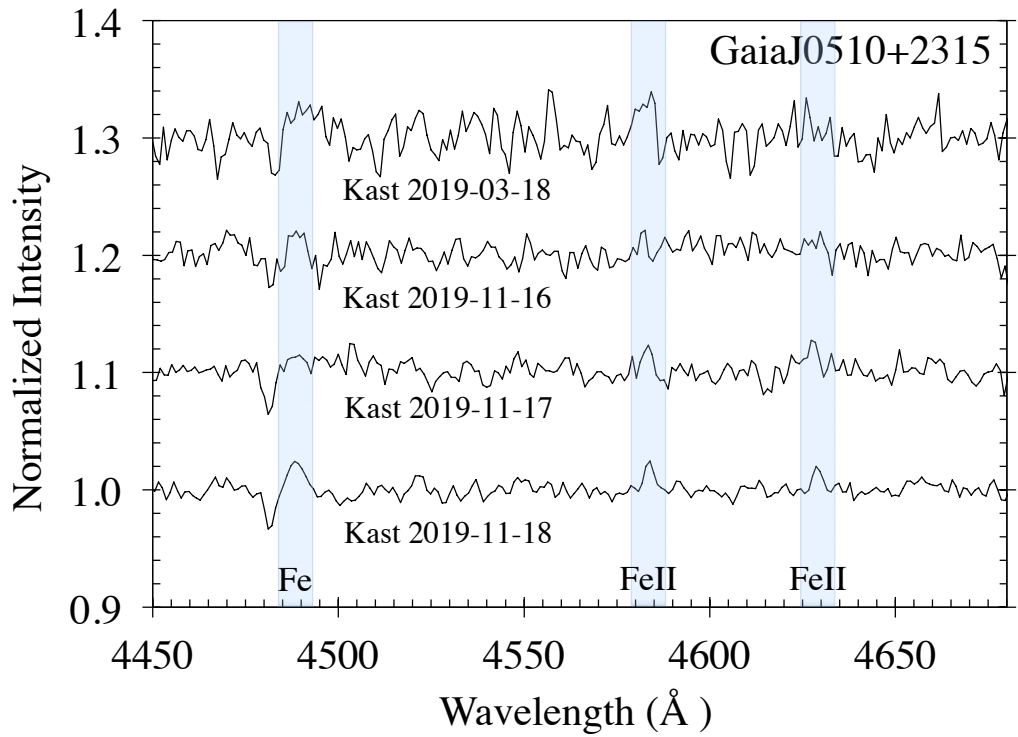}
 \caption{\label{fig0510l1} \large{Iron emission region for
               Gaia\,J0510+2315. An unidentified feature (likely some combination of Fe lines) 
               appears near 4488\,\AA .
               Spectra continuum levels have been fit and
               divided into each spectrum for this figure.}  }
\end{figure}

\clearpage

\begin{figure}
 \centering
 \includegraphics[width=155mm,trim={2cm 2cm 2cm 2cm},clip]{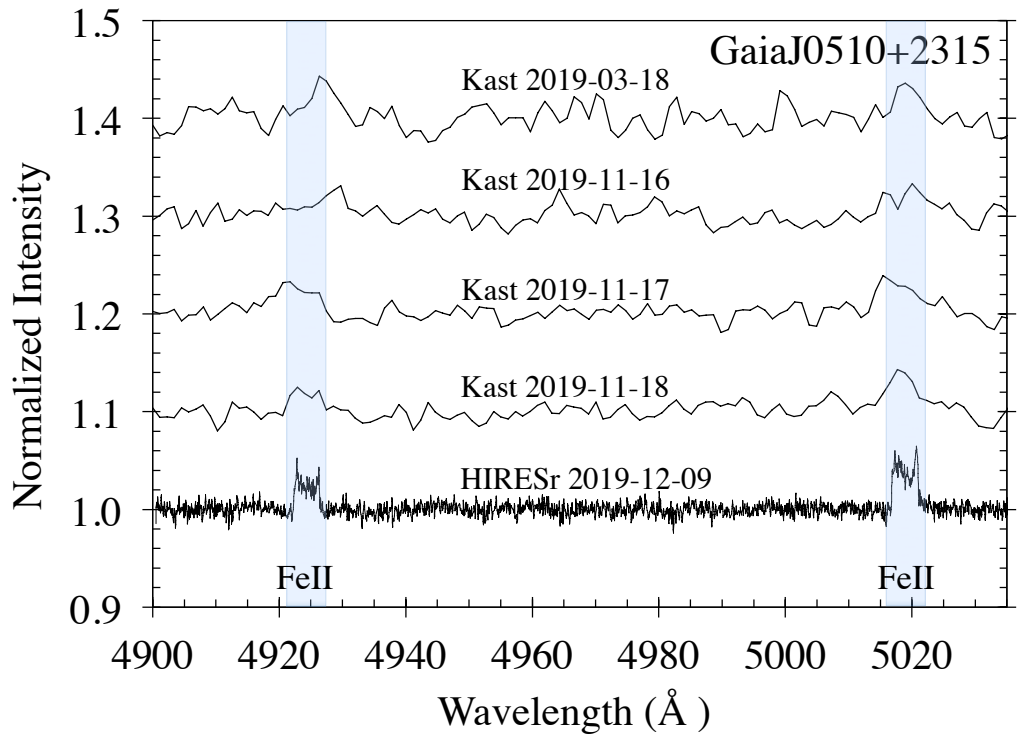}
 \caption{\label{fig0510l2} \large{Iron emission near 5000\,\AA\ for
               Gaia\,J0510+2315. For all remaining plots for Gaia\,J0510+2315 the
               spectra continuum levels have been fit and
               divided into each spectrum and
               HIRES data are smoothed with a 5-pixel boxcar.}  }
\end{figure}

\clearpage

\begin{figure}
 \centering
 \includegraphics[width=155mm,trim={2cm 2cm 2cm 2cm},clip]{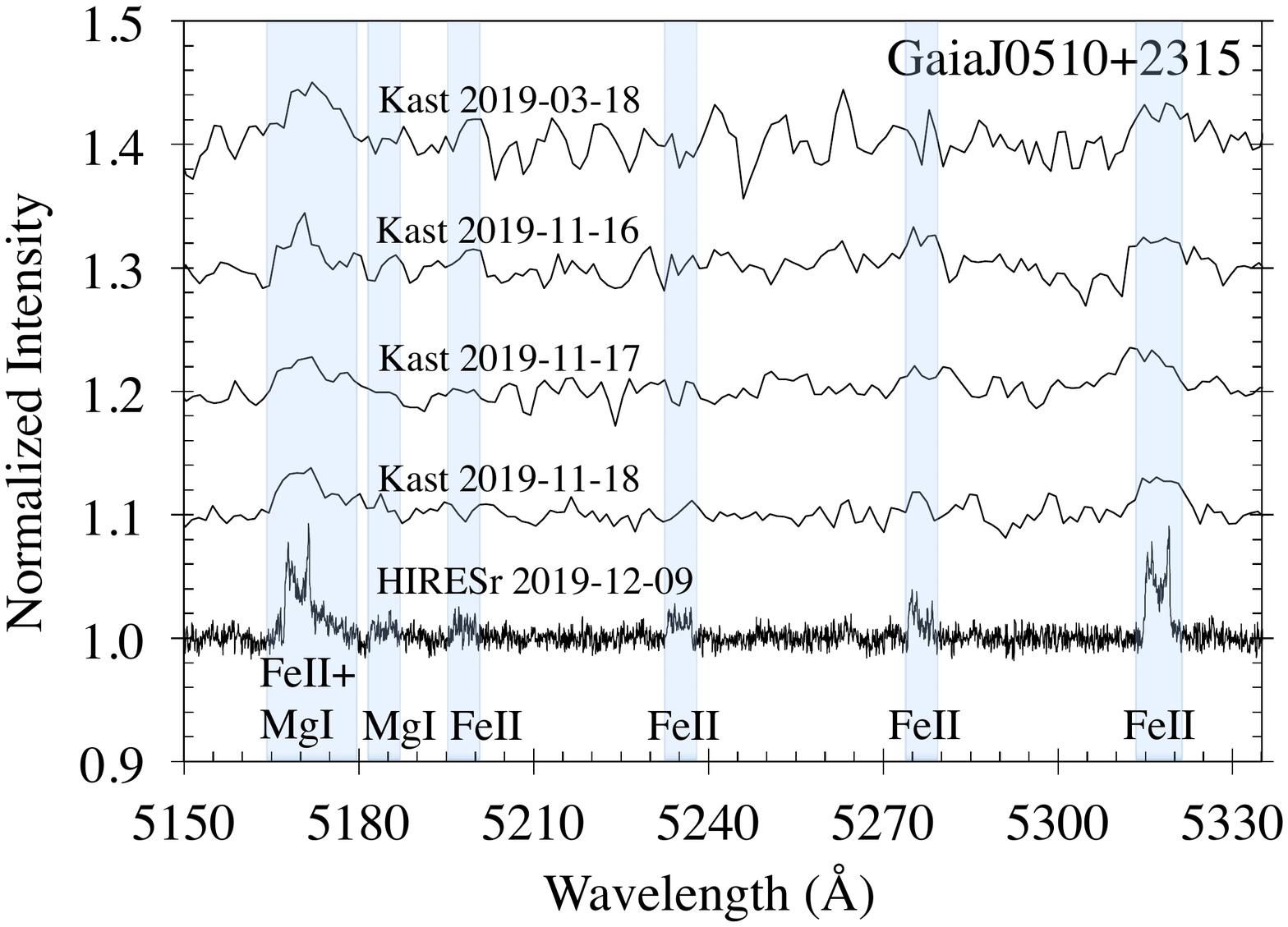}
 \caption{\label{fig0510l3} \large{Iron and magnesium emission for
               Gaia\,J0510+2315.}  }
\end{figure}

\clearpage

\begin{figure}
 \centering
 \includegraphics[width=155mm,trim={2cm 2cm 2cm 2cm},clip]{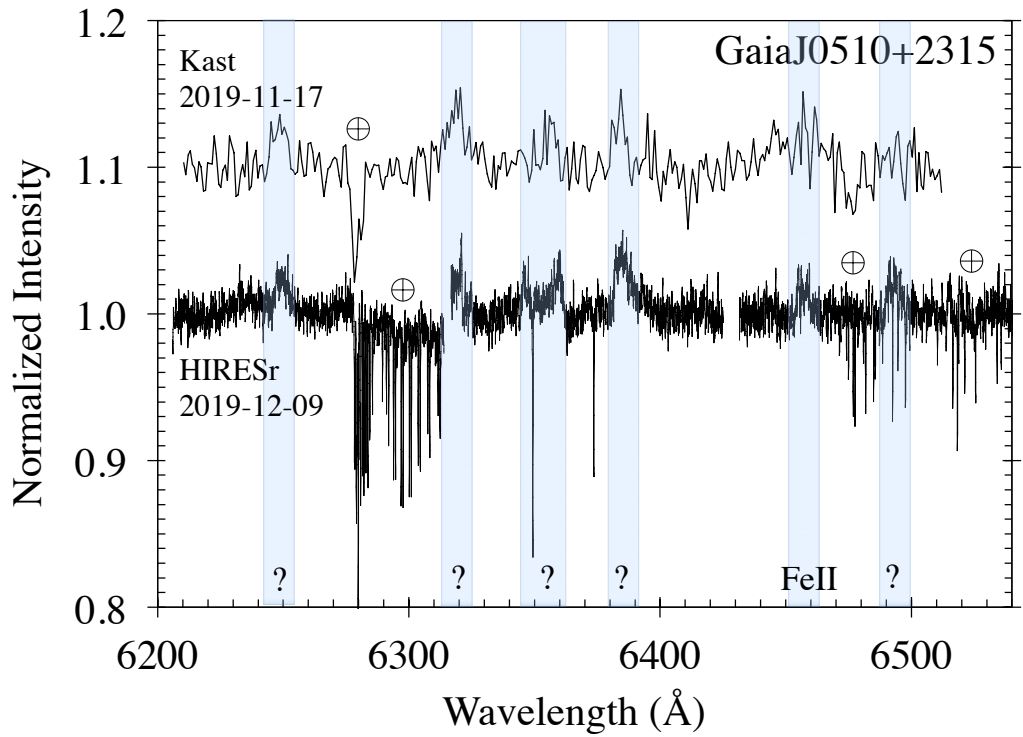}
 \caption{\label{fig0510l4} \large{Iron and unidentified emission region for
               Gaia\,J0510+2315. Earth atmospheric absorption features are labeled, while strong
               Si~II photospheric absorption lines are apparent near 6350 and 6370\,\AA .
               HIRES order gaps appear near 6315 and 6440\,\AA .}  }
\end{figure}

\clearpage

\begin{figure}
 \centering
 \includegraphics[width=155mm,trim={2cm 2cm 2cm 2cm},clip]{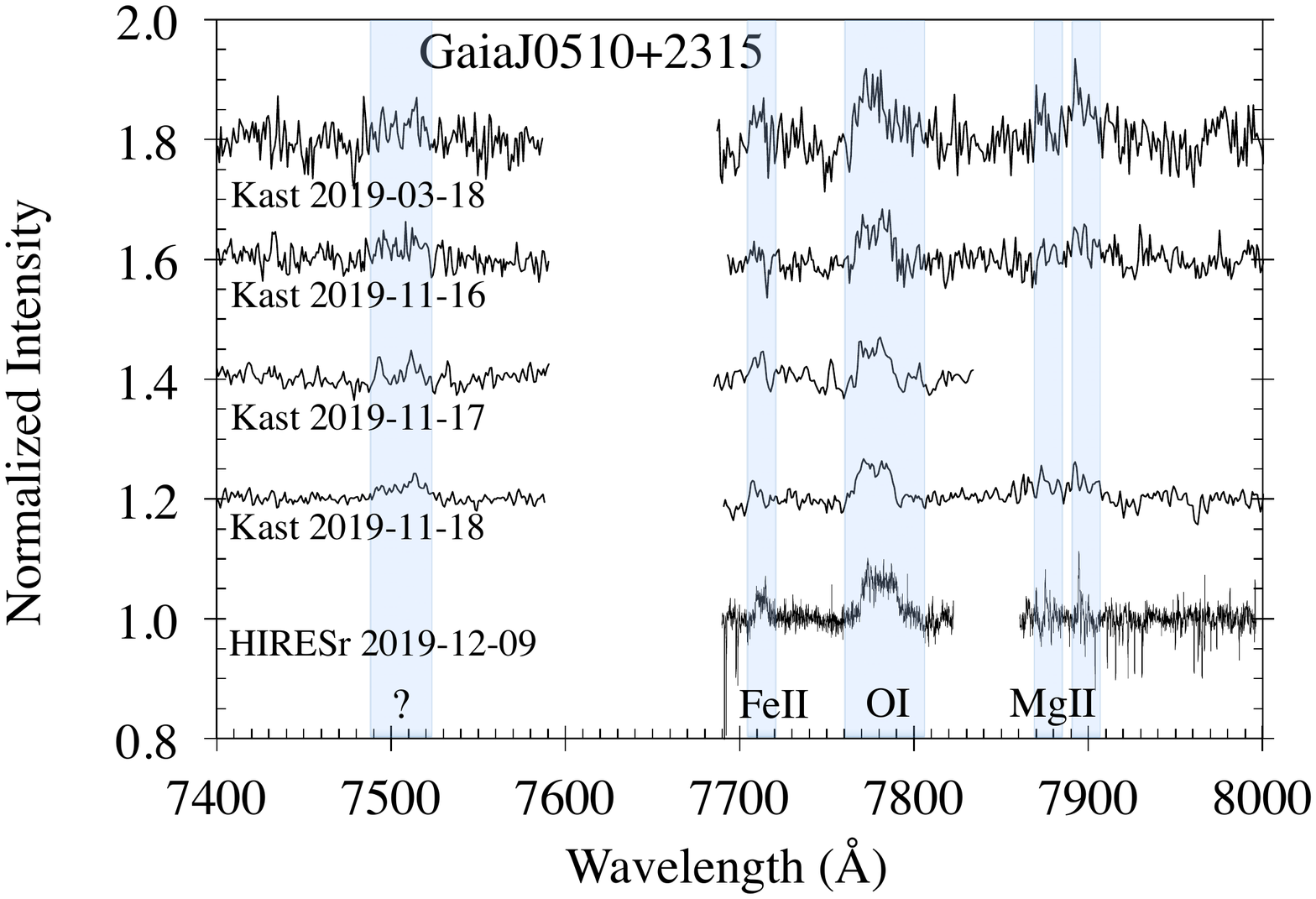}
 \caption{\label{fig0510l5} \large{Iron, oxygen, and magnesium emission region for
               Gaia\,J0510+2315. O~I emission is significantly broader than that measured
               for all other emission lines for this star 
               (Table \ref{tabline0510} and Figure \ref{fig0006lv}).
               Mg~II emission is contaminated by strong telluric absorption features.
               An unidentified emission feature is seen near 7506\,\AA ; this region
               of the spectrum is not covered by HIRES as it falls in between detectors.
               An inter-order gap also appears in the HIRES spectrum near 7850\,\AA .
               Strong telluric absorption near 7600\,\AA\ has been masked.}  }
\end{figure}

\clearpage

\begin{figure}
 \centering
 \includegraphics[width=155mm,trim={2cm 2cm 2cm 2cm},clip]{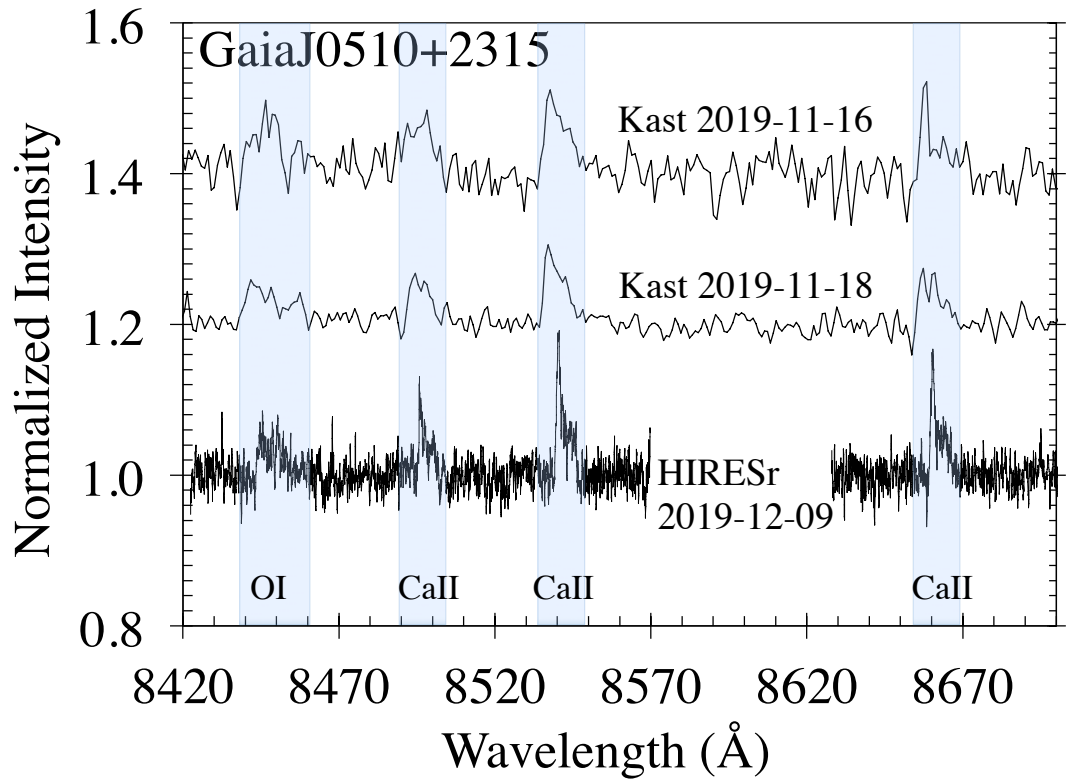}
 \caption{\label{fig0510l6} \large{Oxygen and calcium emission for
               Gaia\,J0510+2315.}  }
\end{figure}

\clearpage

\begin{figure}
 \centering
 \includegraphics[width=155mm,trim={2cm 2cm 2cm 2cm},clip]{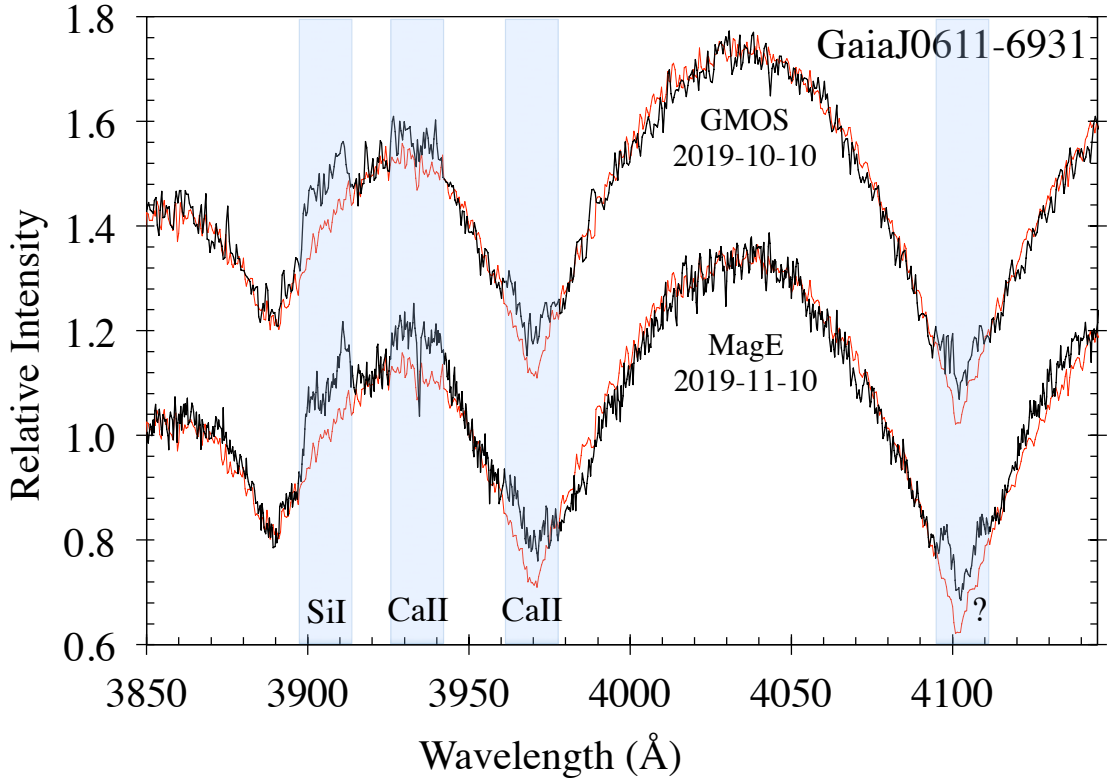}
 \caption{\label{fig0611l1} \large{Blue spectral region for
               Gaia\,J0611$-$6931 showing several emission structures blended with strong photospheric
               Balmer absorption lines. Wavelengths are in air for all GMOS-S and MagE spectra
               shown for Gaia\,J0611$-$6931. Spectra are normalized to the median
               flux value in this plotted range.
               The red underplotted spectrum for both epochs is the DA white 
               dwarf Gaia\,J0720$-$4250 that has atmospheric
               parameters similar to Gaia\,J0611$-$6931 (Table \ref{tabpars}). Gaia\,J0720$-$4250
               has T$_{\rm eff}$$=$17,700\,K and log$g$$=$8.06 from \citet{gentilefusillo19} 
               and was observed with GMOS-S on UT 2019-11-15.
               Ca~II H+K emission
               (with Ca~II H emission in the H$\epsilon$ absorption line core) is evident as well
               as Ca~II K absorption. Also apparent is emission from the 
               Si~I $\lambda$3905 line.
               Possible unidentified emission appears to straddle the H$\delta$ absorption line
               core.}  }
\end{figure}

\clearpage

\begin{figure}
 \centering
 \includegraphics[width=155mm,trim={2cm 2cm 2cm 2cm},clip]{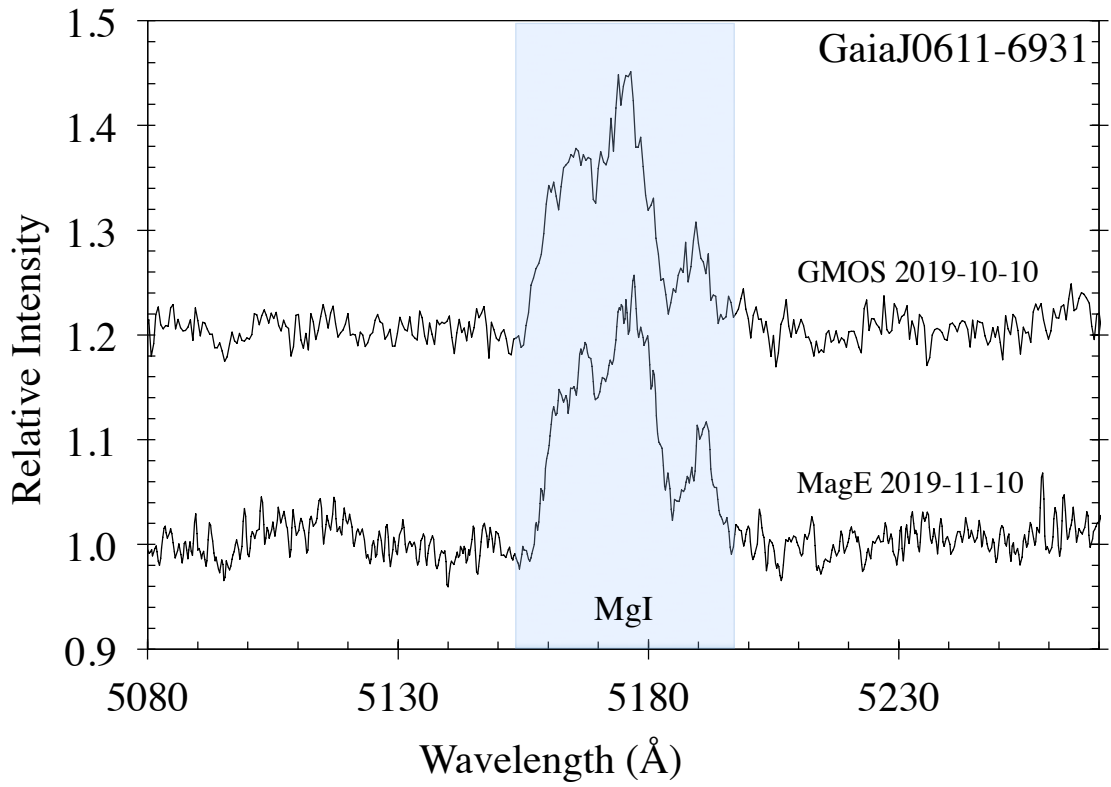}
 \caption{\label{fig0611l2} \large{Magnesium spectral region for
               Gaia\,J0611$-$6931. Continuum levels have been fit and
               divided into each spectrum.}  }
\end{figure}

\clearpage

\begin{figure}
 \centering
 \includegraphics[width=155mm,trim={2cm 2cm 2cm 2cm},clip]{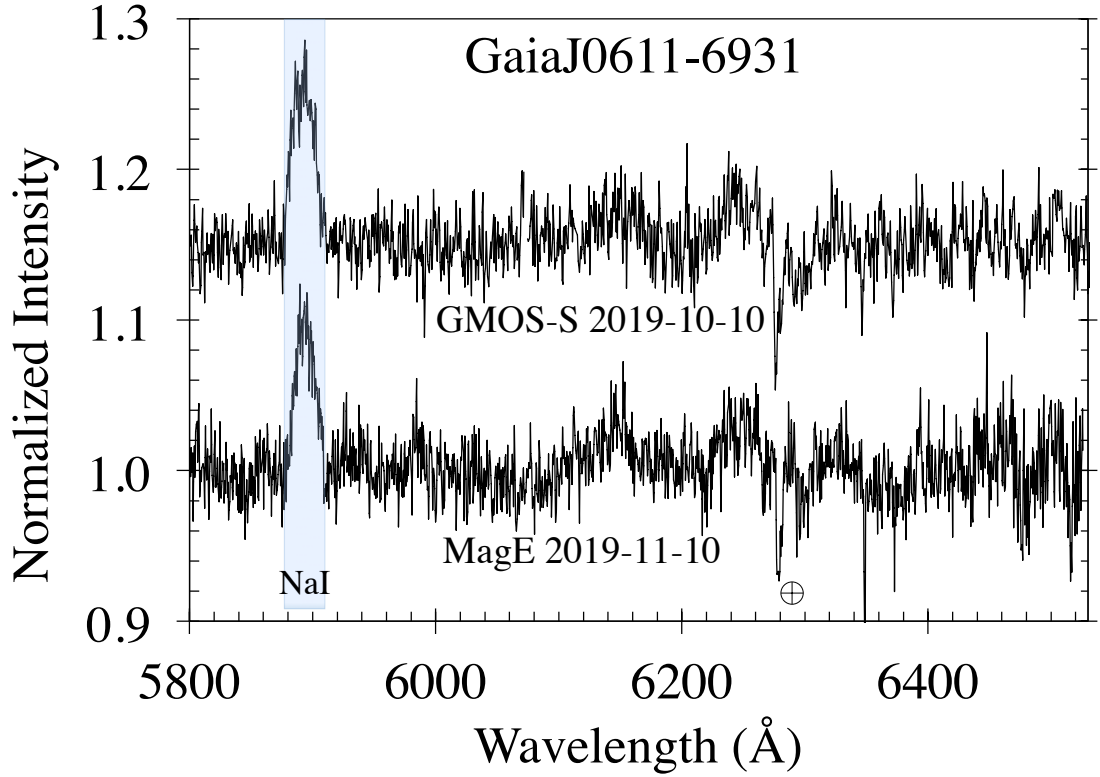}
 \caption{\label{fig0611l3} \large{Sodium and possible other emission lines in
               Gaia\,J0611$-$6931. Continuum levels have been fit and
               divided into each spectrum. Emission from the Na~D doublet is clearly evident.
               Possible emission from unidentified transitions may be present from 6100-6300\,\AA .
               Earth telluric absorption is marked near 6280\,\AA . Photospheric Si~II absorption lines
               are detected near 6350 and 6370\,\AA\ in both spectra.}  }
\end{figure}

\clearpage

\begin{figure}
 \centering
 \includegraphics[width=155mm,trim={2cm 2cm 2cm 2cm},clip]{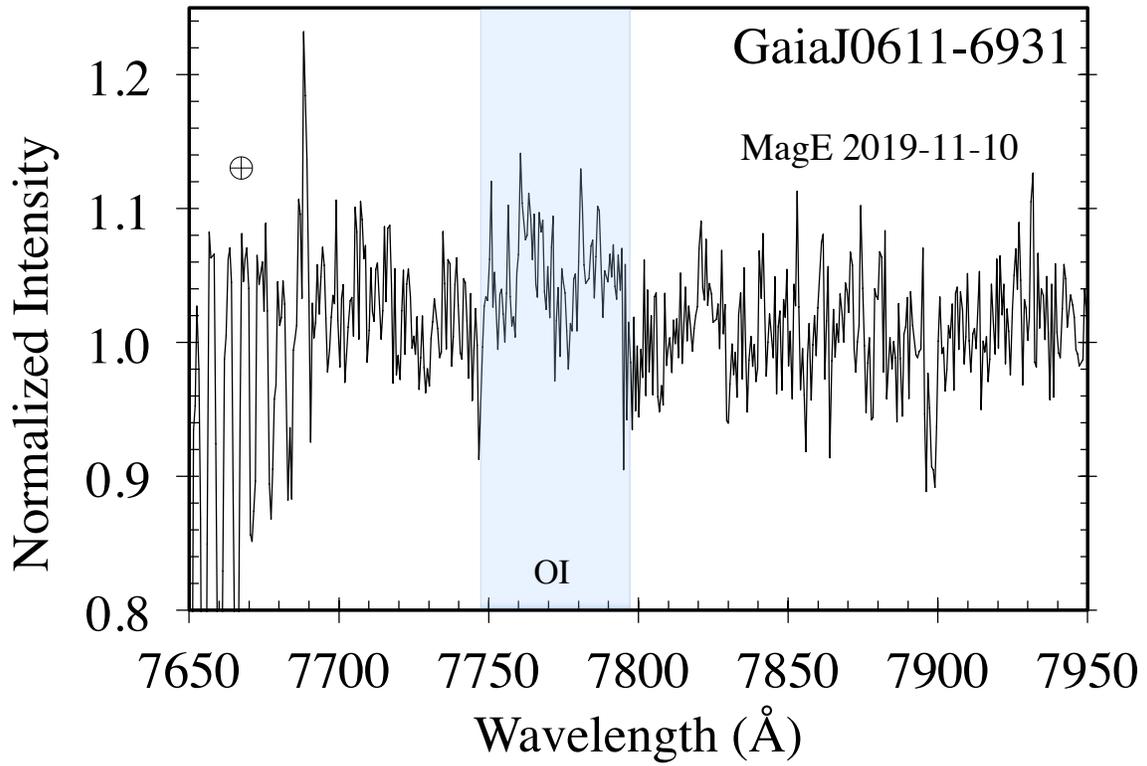}
 \caption{\label{fig0611l4} \large{Oxygen emission region in
               Gaia\,J0611$-$6931. The continuum level has been fit and
               divided into the spectrum. Earth atmospheric absorption is marked.}  }
\end{figure}

\clearpage

\begin{figure}
 \centering
 \includegraphics[width=155mm,trim={2cm 2cm 2cm 2cm},clip]{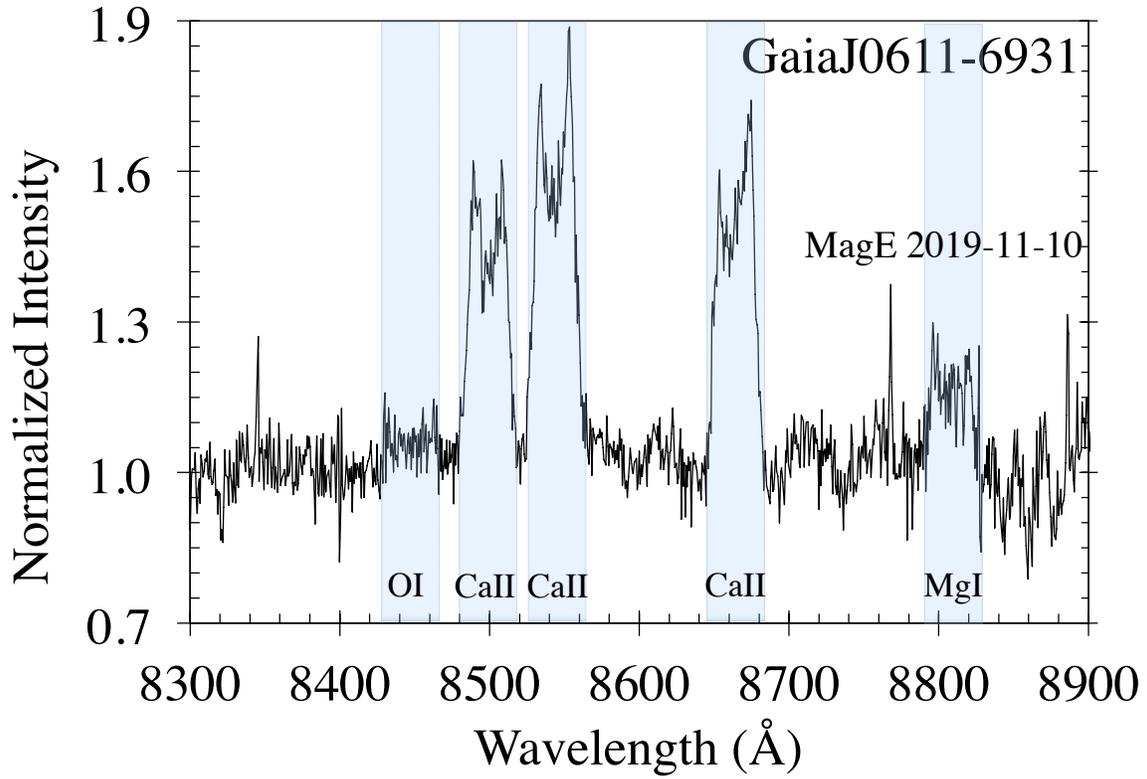}
 \caption{\label{fig0611l5} \large{Calcium, oxygen, and magnesium
               emission lines in
               Gaia\,J0611$-$6931. The continuum level has been fit and
               divided into the spectrum.}  }
\end{figure}

%\clearpage

%same thing as in 9-panel figure
%\begin{figure}
% \centering
 %\includegraphics[width=105mm,angle=90,trim={2cm 2cm 2cm 2cm},clip]{gaiaj0644_CaIRT_plot.ps}
% \caption{\label{fig0644l1} \large{Ca~II IRT emission region for
%               Gaia\,J0644$-$0352. Wavelengths are in air. The continuum level has been fit and
%               divided into the spectrum.
%               Emission regions are marked with blue highlighted vertical bars.}  }
%\end{figure}

\clearpage

\begin{figure}
 \centering
 \includegraphics[width=155mm,trim={2cm 2cm 2cm 2cm},clip]{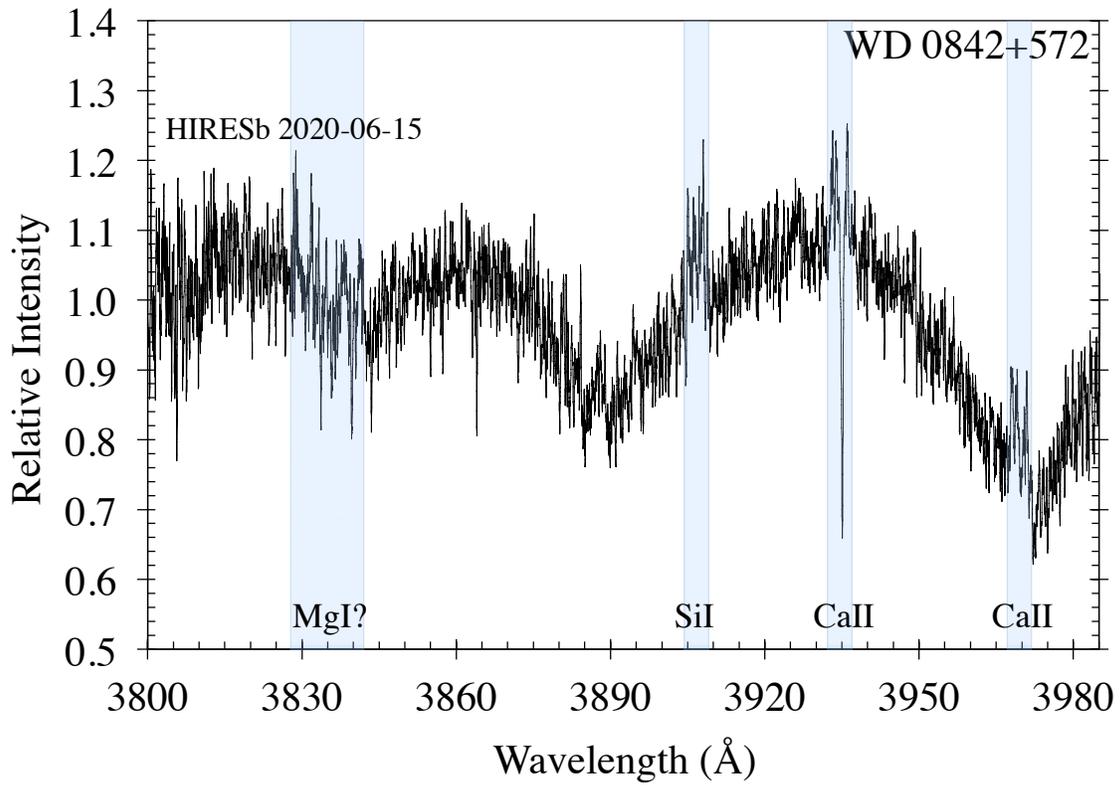}
 \caption{\label{fig0842l1} \large{Blue spectral region for
               WD0842+572. Spectra are normalized to the median
               flux value in this plotted range.
               Ca~II K emission and absorption are evident as well as
               Ca~II H emission in the H$\epsilon$ absorption line core. 
               Also present is Si~I emission and possibly two Mg~I features.}  }
\end{figure}

\clearpage

\begin{figure}
 \centering
 \includegraphics[width=155mm,trim={2cm 2cm 2cm 2cm},clip]{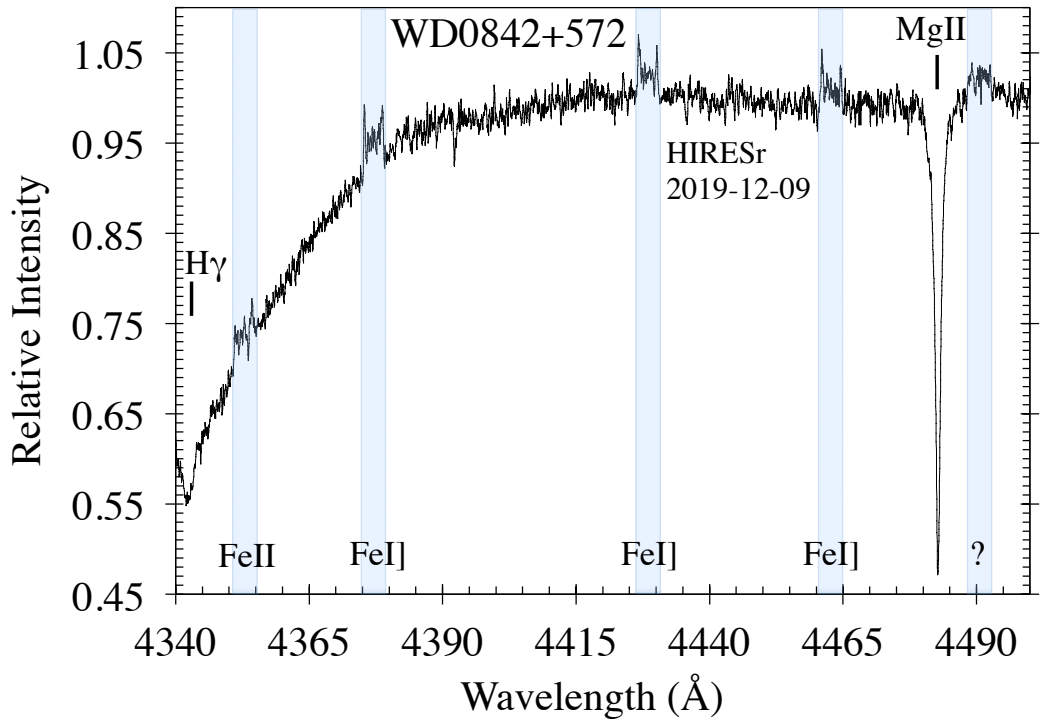}
 \caption{\label{fig0842l2} \large{Semi-forbidden neutral iron emission lines and an Fe~II
               line in
               WD0842+572. A weak unidentified feature appears on the red side of the strong Mg~II
               absorption line; given the propensity for weak iron lines seen throughout the spectrum of 
               WD0842+572, it could possibly be iron.
               The spectrum flux is normalized to the continuum emission level outside of the photospheric
               Balmer absorption.}  }
\end{figure}

\clearpage

\begin{figure}
 \centering
 \includegraphics[width=155mm,trim={2cm 2cm 2cm 2cm},clip]{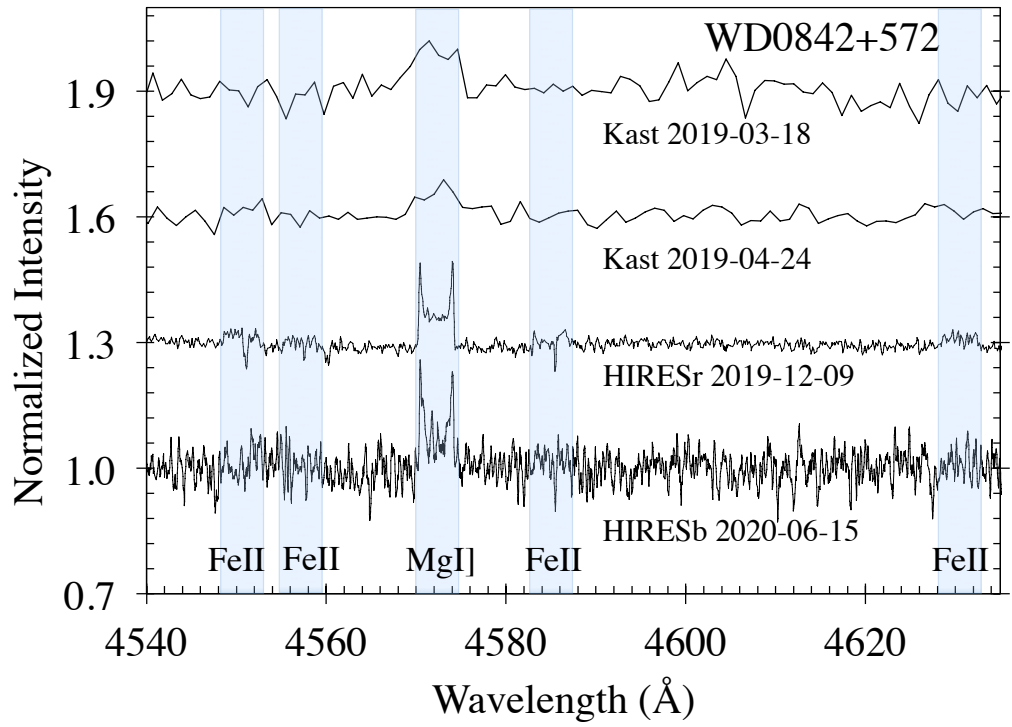}
 \caption{\label{fig0842l3} \large{Iron and magnesium emission line region
               for
               WD0842+572, which features the semi-forbidden 
               Mg~I] $\lambda$4571 line. Spectra continuum levels have been fit and
               divided into each spectrum. Absorption lines from Fe are present.}  }
\end{figure}

\clearpage

\begin{figure}
 \centering
 \includegraphics[width=155mm,trim={3cm 3cm 3cm 3cm},clip]{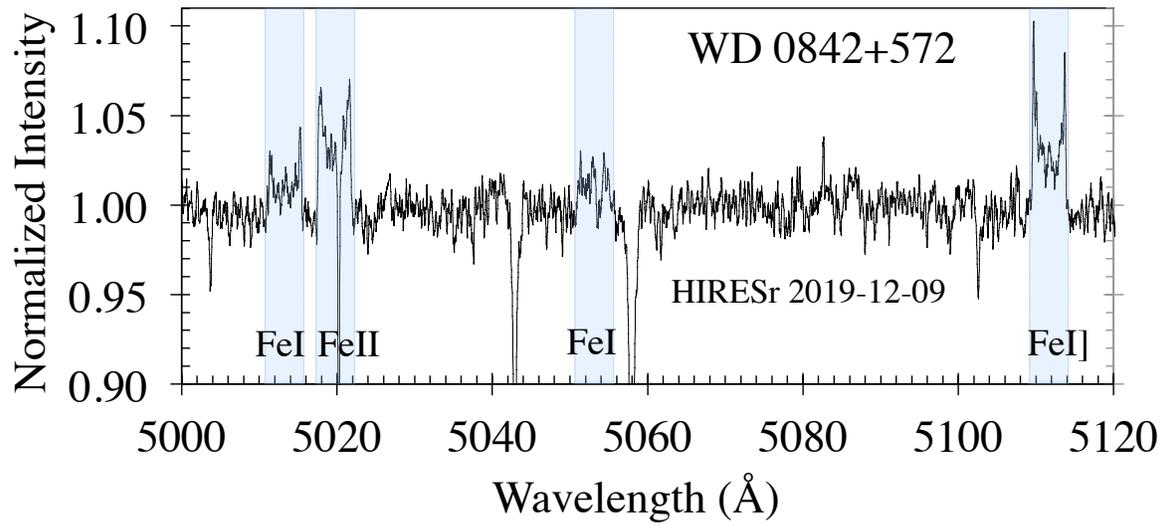}
 \caption{\label{fig0842l4} \large{Another iron emission line region
               for
               WD0842+572.
               The spectrum continuum level has been fit and
               divided into the spectrum. 
               Weak emission from an Fe~I line near 5052\,\AA\ may be present.
               Strong photospheric absorption lines from Si
               (near 5040 and 5055\,\AA ) and Fe are also seen.}  }
\end{figure}

\clearpage

\begin{figure}
 \centering
 \includegraphics[width=155mm,trim={2cm 4cm 2cm 2cm},clip]{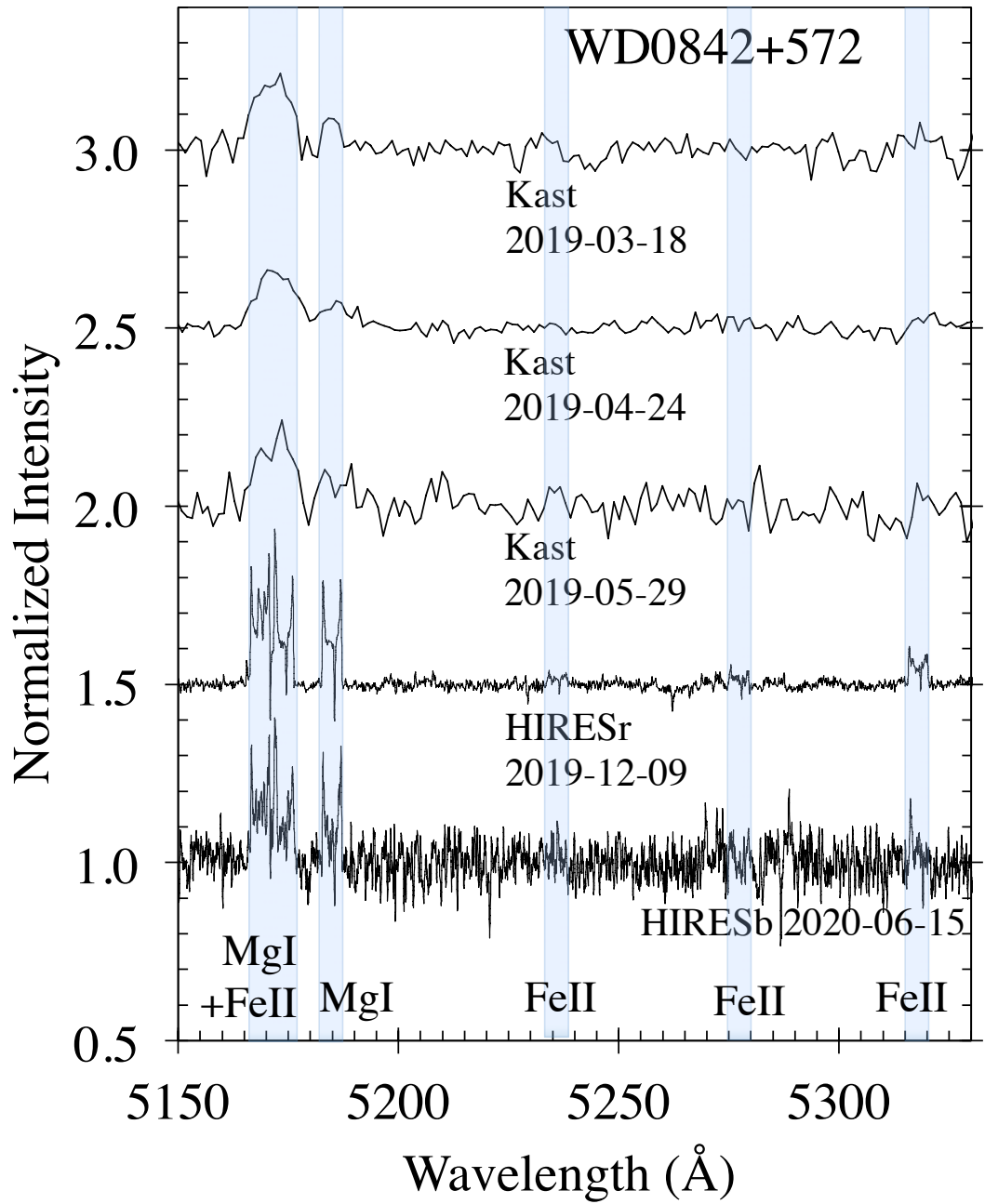}
 \caption{\label{fig0842l5} \large{Iron and magnesium emission line region
               for
               WD0842+572. Based on the HIRESr spectrum, there 
               may be numerous weak unidentified emission
               lines between 5200-5300\,\AA .
               Spectra continuum levels have been fit and
               divided into each spectrum. Absorption lines from Fe and the Mg~I
               triplet are apparent.}  }
\end{figure}

\clearpage

\begin{figure}
 \centering
 \includegraphics[width=155mm,trim={3cm 3cm 3cm 3cm},clip]{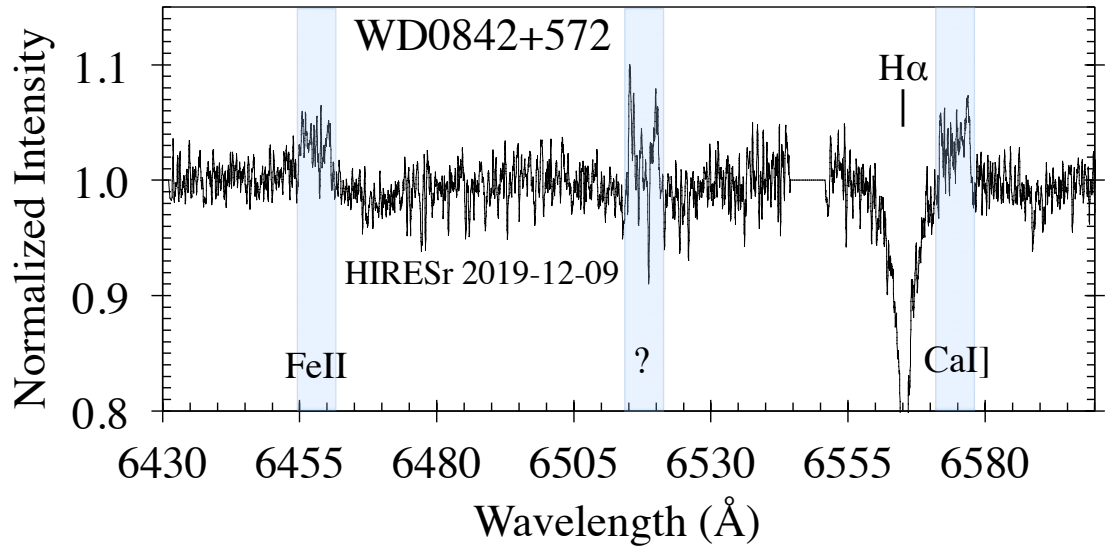}
 \caption{\label{fig0842l6} \large{Iron and semi-forbidden calcium emission line region
               for WD0842+572. An unidentified emission line appears near 6517\,\AA .
               The spectrum continuum level and wings of the H$\alpha$ line profile
               have been fit and divided into the spectrum; there is an order gap near 6550\,\AA .}  }
\end{figure}

\clearpage

%\begin{figure}
% \centering
 %\includegraphics[width=105mm,angle=90,trim={2cm 2cm 2cm 2cm},clip]{wd0842+572_7700-7800.ps}
% \caption{\label{fig0842l7} \large{Another iron emission line region
%               for WD0842+572.
%               The spectrum continuum level has been fit and divided 
%               into the spectrum.}  }
%\end{figure}

\clearpage

\begin{figure}
 \centering
 \includegraphics[width=155mm,trim={2cm 3cm 2cm 2cm},clip]{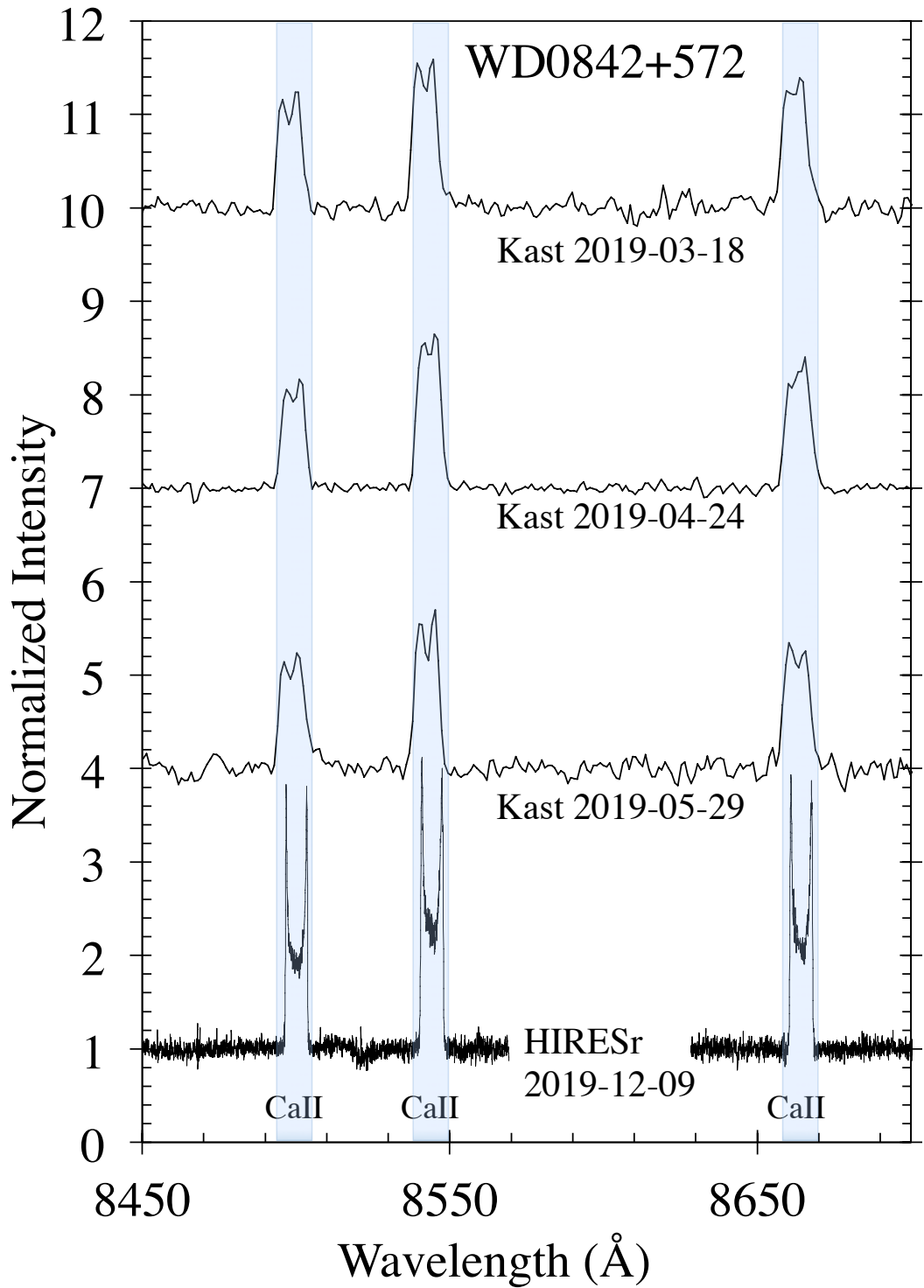}
 \caption{\label{fig0842l8} \large{Ca~II IRT emission line region
               for WD0842+572.
               Spectra continuum levels have been fit and
               divided into each spectrum.}  }
\end{figure}

\clearpage

\begin{figure}
 \centering
 \includegraphics[width=155mm,trim={2cm 2cm 3cm 2cm},clip]{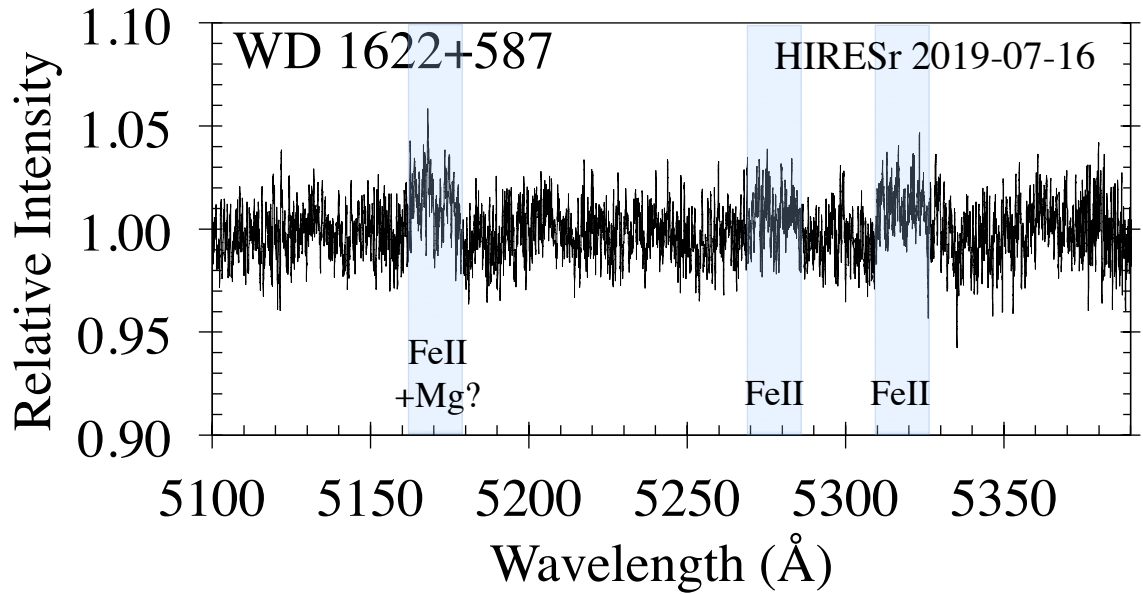}
 \caption{\label{fig1622l1} \large{Iron emission region for
               WD1622+587. For all  WD1622+587 figures the
               spectrum continuum levels have been fit and
               divided into each displayed spectrum.
               The spectrum shown here is smoothed with an 11-pixel boxcar for display purposes.}  }
\end{figure}

\clearpage

\begin{figure}
 \centering
 \includegraphics[width=155mm,trim={2cm 2cm 2cm 2cm},clip]{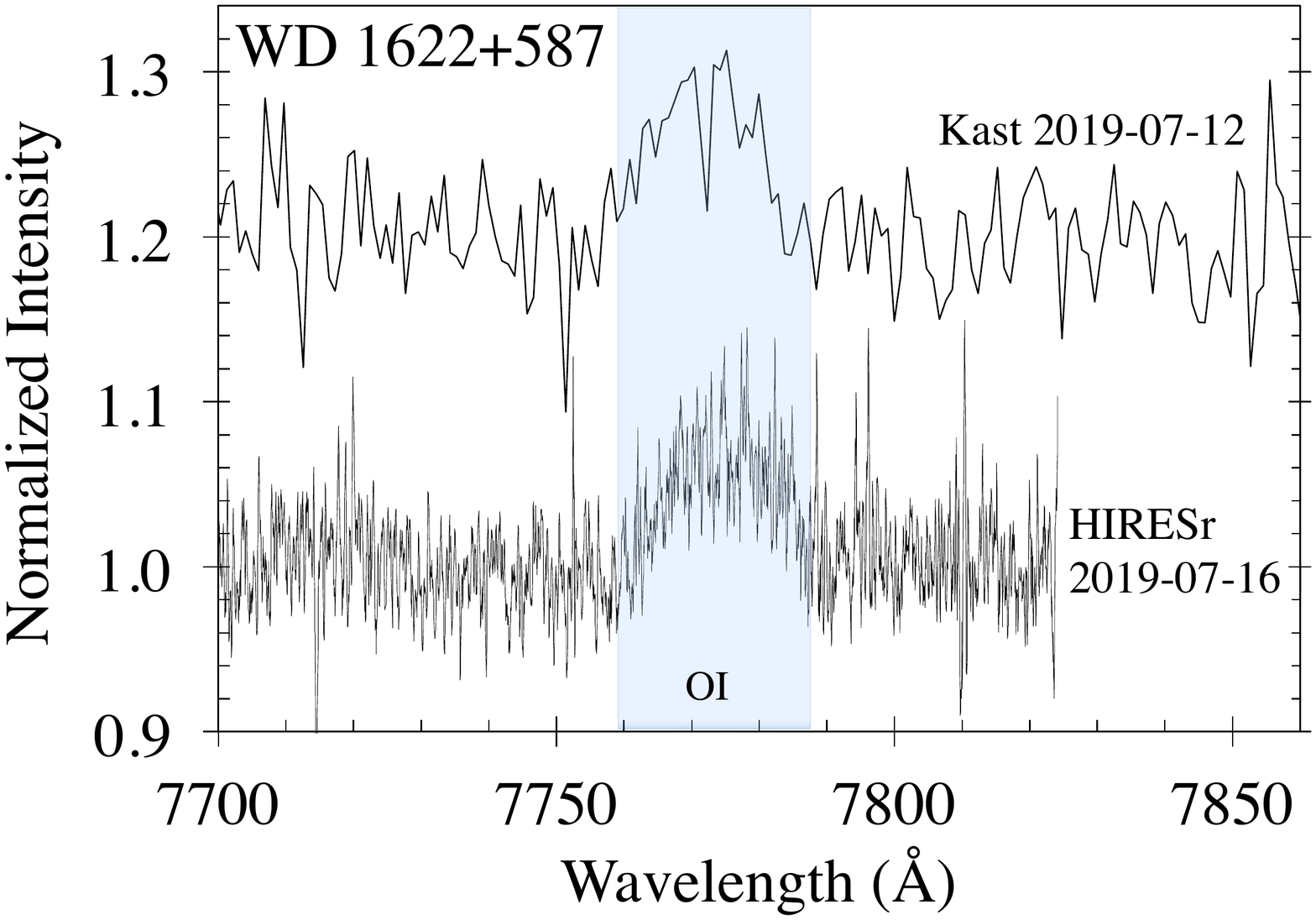}
 \caption{\label{fig1622l2} \large{Oxygen emission region for
               WD1622+587. The HIRES spectrum has been smoothed with a 7-pixel boxcar.}  }
\end{figure}

\clearpage

\begin{figure}
 \centering
 \includegraphics[width=155mm,trim={2cm 2cm 2cm 2cm},clip]{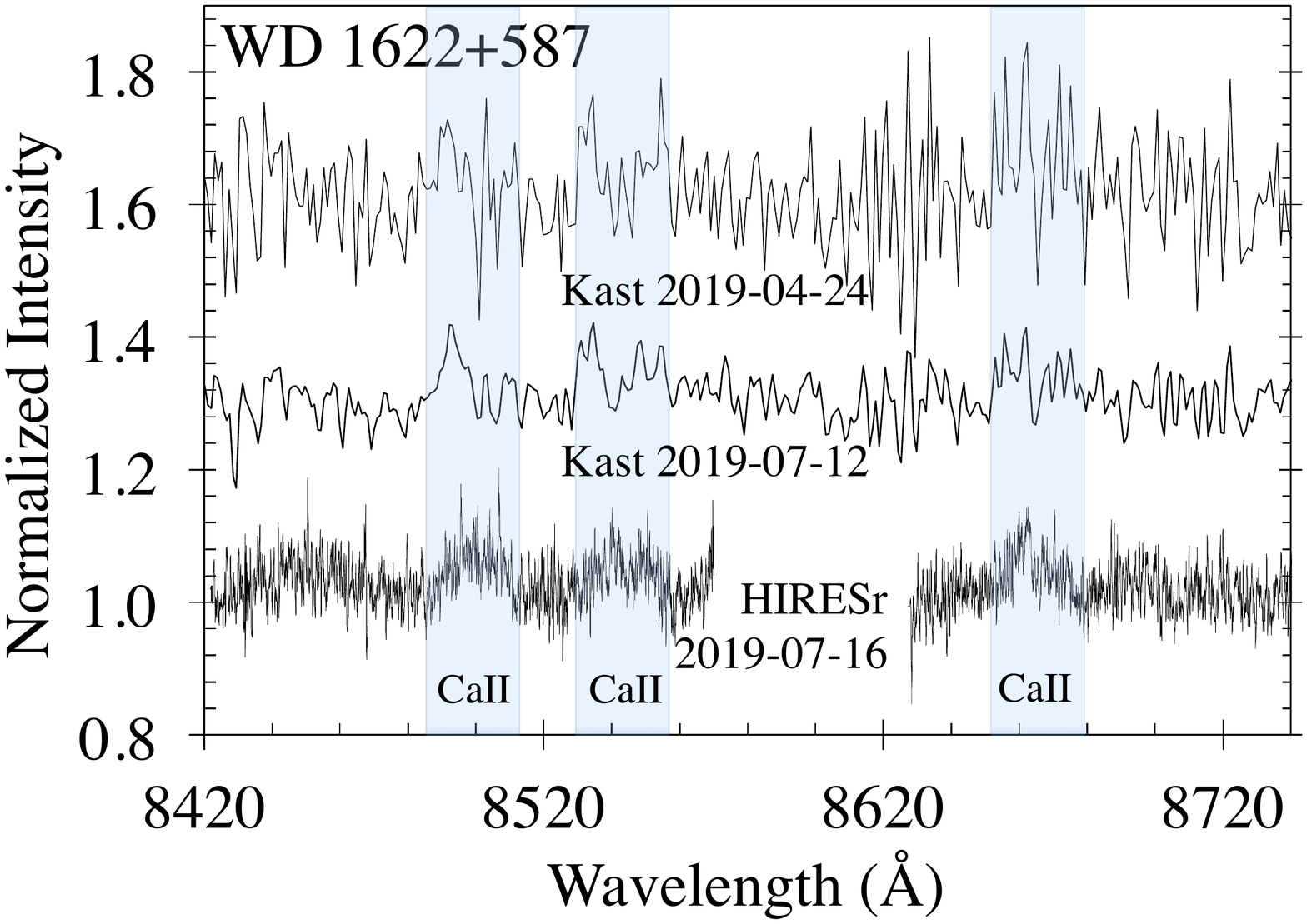}
 \caption{\label{fig1622l3} \large{Calcium emission region for
               WD1622+587. The HIRES spectrum has been smoothed with a 7-pixel boxcar.}  }
\end{figure}

\clearpage

\begin{figure}
 \centering
 \includegraphics[width=155mm,trim={2cm 5cm 2cm 2cm},clip]{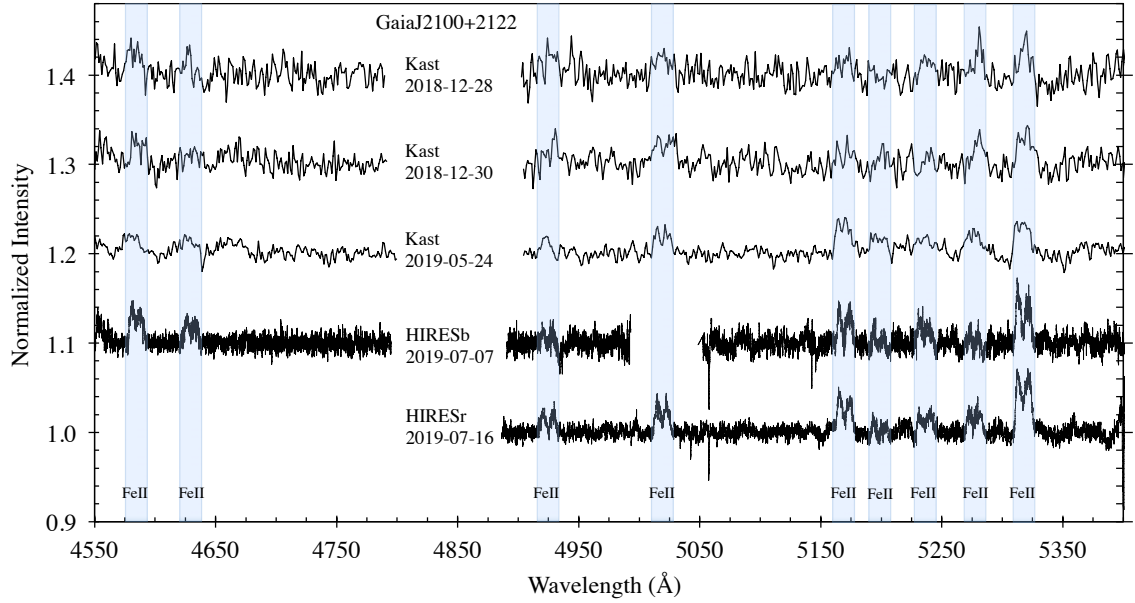}
 \caption{\label{fig2100l1} \large{Iron emission region for
               Gaia\,J2100+2122. 
               H$\beta$ is cut out of each spectrum for plotting purposes, HIRESb data also feature a gap
               between CCDs centered around 5025\,\AA .
               For this and all figures for Gaia\,J2100+2122 the spectra continuum levels have been fit and
               divided into each spectrum displayed. 
               The HIRES spectra in this figure are smoothed with an 11-pixel boxcar for 
               display purposes. Si~II photospheric absorption lines appear near 5040 and 5055\,\AA .}  }
\end{figure}

\clearpage

\begin{figure}
 \centering
 \includegraphics[width=155mm,trim={2cm 2.9cm 2.6cm 2.6cm},clip]{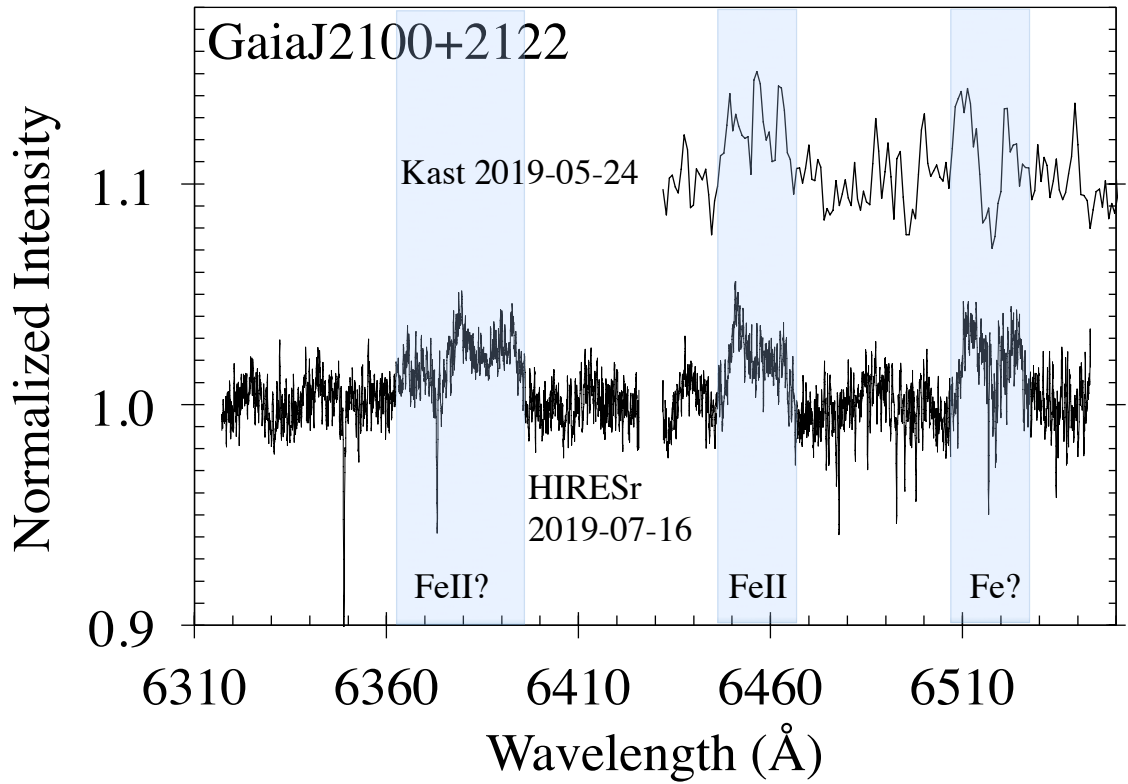}
 \caption{\label{fig2100l2} \large{Another iron emission region for
               Gaia\,J2100+2122. 
               Lines near 6382 and 6515\,\AA\ are unidentified; they may be due to Fe.
               The HIRES spectra are smoothed with a
               7-pixel boxcar and feature an order gap near 6420\,\AA . Strong photospheric
               Si~II absorption appears near 6350 and 6370\,\AA ; other
               absorption lines are from Earth's atmosphere.
               }  }
\end{figure}

\clearpage

\begin{figure}
 \centering
 \includegraphics[width=105mm,trim={2cm 2cm 3cm 2cm},clip]{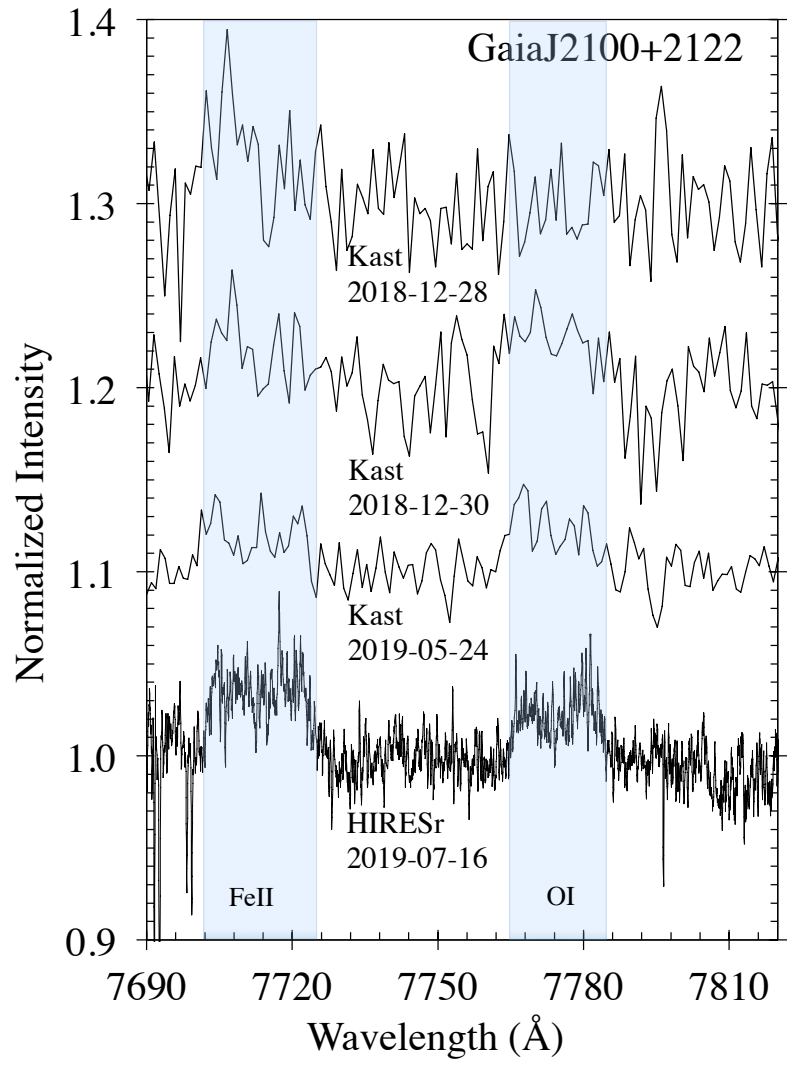}
 \caption{\label{fig2100l3} \large{Oxygen and iron emission for
               Gaia\,J2100+2122.}  }
\end{figure}

\clearpage

\begin{figure}
 \centering
 \includegraphics[width=155mm,trim={2cm 2cm 2cm 2cm},clip]{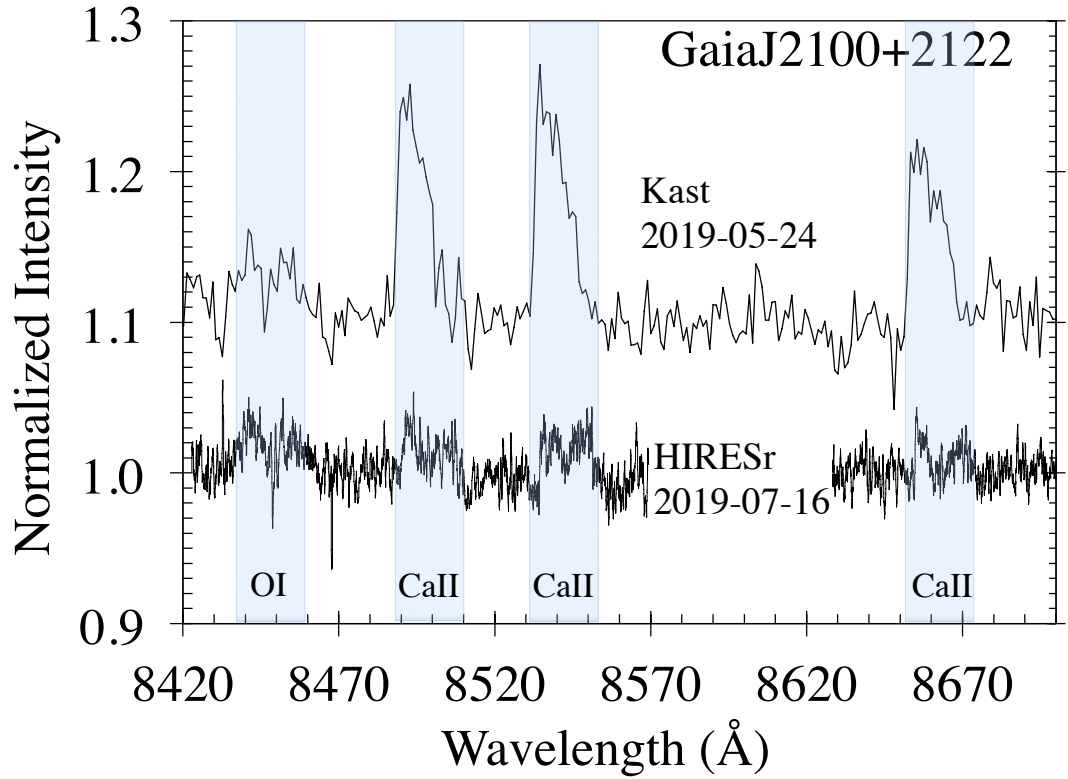}
 \caption{\label{fig2100l4} \large{Calcium and oxygen emission for
               Gaia\,J2100+2122. Variability in the Ca~II IRT emission lines is evident
               between the Kast and HIRES epochs.}  }
\end{figure}

\clearpage

\begin{deluxetable}{lccccc}
\tabletypesize{\small}
\tablecolumns{6}
\tablewidth{0pt}
\tablecaption{\large{SDSS\,J0006+2858 Emission Line Measurements \label{tabline0006}} }
\tablehead{ 
  \colhead{Transition} & 
  \colhead{UT Date} &
  \colhead{Eq.\ Width} &
  \colhead{Peak Separation} &
  \colhead{v$_{\rm max}$sin$i$} &
  \colhead{Full Width} \\
  \colhead{} & 
  \colhead{} &
  \colhead{(\AA )} &
  \colhead{(km\,s$^{-1}$)} &
  \colhead{(km\,s$^{-1}$)} &
  \colhead{(km\,s$^{-1}$)}
}
\startdata
%lines at 7771.94, 7774.17, and 7775.39 Ang in air; include 10 km/s systematic error component
%lines at 7774.08, 7776.31, and 7777.53 Ang in vacuum -> 7774.01, 7777.46 Ang in WD systemic frame; include 5 km/s systematic error component
OI $\lambda$7772        & 29 Dec 2018 & 1.6$\pm$1.2 & $-$ & $-$ & $-$ \\ %left edge is 7750.5 Ang, right edge is 7792.20 Ang
OI $\lambda$7772        & 25 Jan 2019 &  2.4$\pm$0.6 & $-$ & $-$ & $-$ \\ %left edge is 7751.1 Ang, right edge is 7792.65 Ang
OI $\lambda$7772        & 16 Jul 2019 & 2.6$\pm$0.2 & $-$ & $-$825$\pm$25/300$\pm$30 & $-$ \\ %left edge is 7752.63+/-0.6 Ang, right edge is 7785.25+/-0.8 Ang
OI $\lambda$7772        & 26 Jul 2019 & 2.5$\pm$1.1 & $-$ & $-$ & $-$ \\ %left edge is 7750.2 Ang, right edge is 7786.81 Ang
%
%lines at 8448.57, 8448.68, and 8449.08 Ang in vacuum -> 8448.49, 8449.00 in WD systemic frame; include 5 km/s systematic error component
OI $\lambda$8446        & 16 Jul 2019 & 0.9$\pm$0.4 & $-$ & $-$780$\pm$25/290$\pm$25 & $-$ \\
\enddata
\tablecomments{See Section \ref{secgasres} for descriptions of Peak Separation, v$_{\rm max}$sin$i$, and Full Width and when these measurements can be made. The two different values reported for v$_{\rm max}$sin$i$ correspond to the maximum velocity gas seen in the blue and red wings of the emission features, respectively. Full velocity width of the emission feature is the velocity extent from the blue to the red wings and is only measured when a line is not blended with another transition.}
\end{deluxetable}

\clearpage

\begin{deluxetable}{lccccc}
\tabletypesize{\small}
\tablecolumns{6}
\tablewidth{0pt}
\tablecaption{\large{WD0145+234 Emission Line Measurements \label{tabline0145}} }
\tablehead{ 
  \colhead{Transition} & 
  \colhead{UT Date} &
  \colhead{Eq.\ Width} &
  \colhead{Peak Separation} &
  \colhead{v$_{\rm max}$sin$i$} &
  \colhead{Full Width} \\
  \colhead{} & 
  \colhead{} &
  \colhead{(\AA )} &
  \colhead{(km\,s$^{-1}$)} &
  \colhead{(km\,s$^{-1}$)} &
  \colhead{(km\,s$^{-1}$)}
}
\startdata
%line at 8498.02 rest in air -> 8498.27 in WD systemic frame; include 10 km/s systematic error component
%line at 8500.35 rest in vacuum -> 8500.60 in WD systemic frame; include 5 km/s systematic error component
CaII $\lambda$8498     & 12 Oct 2019 & 2.1$\pm$0.5 & 800$\pm$15 & $-$ & 1410$\pm$115 \\  %left edge is 8479.0+/-2.0 Ang, right edge is 8519.0+/-2.5 Ang -- left peak is 8486.6+/-0.1 Ang, right peak is 8509.3+/-0.1 Ang
CaII $\lambda$8498     & 04 Nov 2019 & 1.8$\pm$0.2 & 770$\pm$15 & $-$ & 1450$\pm$90 \\  %left edge is 8478.8+/-1.5 Ang, right edge is 8520.0+/-2.0 Ang -- left peak is 8487.33+/-0.13 Ang, right peak is 8509.1+/-0.3 Ang
CaII $\lambda$8498     & 18 Nov 2019\tablenotemark{a} & 2.0$\pm$0.2 & 750$\pm$10 & $-$ & 1380$\pm$60 \\  %v1: EW=2.1+/-0.3 Ang; left edge is 8480.3+/-2.3 Ang, right edge is 8519.5+/-2.0 Ang -- left peak is 8488.6+/-0.1 Ang, right peak is 8509.3+/-0.1 Ang
%v2: EW=2.0+/-0.3 Ang; left edge is 8478.5+/-2.0 Ang, right edge is 8518.0+/-1.5 Ang -- left peak is 8486.5+/-0.1 Ang, right peak is 8507.6+/-0.5 Ang
%v3: EW=1.95+/-0.3 Ang; left edge is 8479.0+/-2.0 Ang, right edge is 8518.0+/-2.0 Ang -- left peak is 8487.3+/-0.3 Ang, right peak is 8509.5+/-0.1 Ang
CaII $\lambda$8498     & 09 Dec 2019 & 2.0$\pm$0.2 & 770$\pm$6 & $-$650$\pm$35/700$\pm$70 & 1355$\pm$79 \\ % left edge is 8482.1+/-1.0 Ang, right edge is 8520.5+/-2.0 Ang -- left peak is 8489.0+/-0.1 Ang, right peak is 8510.90+/-0.05 Ang
%
%line at 8542.09 rest in air -> 8542.34 in WD systemic frame; include 10 km/s systematic error component on any relative velocity measurement
%line at 8544.44 rest in vacuum -> 8544.69 in WD systemic frame; include 5 km/s systematic error component on any velocity measurement
CaII $\lambda$8542     & 12 Oct 2019 & 1.9$\pm$0.5 & 765$\pm$15 & $-$ & 1455$\pm$113 \\
\enddata
\tablecomments{See notes to Table \ref{tabline0006}.}
\tablenotetext{a}{Three exposures were taken over the course of the night. The values reported are the average of measurements made for each individual exposure with individual values for each separate exposure agreeing to within their uncertainties.}
\end{deluxetable}

\clearpage

\begin{deluxetable}{lccccc}
\tabletypesize{\small}
\tablecolumns{6}
\tablewidth{0pt}
\tablecaption{\large{SDSSJ0347+1624 Emission Line Measurements \label{tabline0347}} }
\tablehead{ 
  \colhead{Transition} & 
  \colhead{UT Date} &
  \colhead{Eq.\ Width} &
  \colhead{Peak Separation} &
  \colhead{v$_{\rm max}$sin$i$} &
  \colhead{Full Width} \\
  \colhead{} & 
  \colhead{} &
  \colhead{(\AA )} &
  \colhead{(km\,s$^{-1}$)} &
  \colhead{(km\,s$^{-1}$)} &
  \colhead{(km\,s$^{-1}$)}
}
\startdata
OI $\lambda$7772        & 20 Dec 2019 & 0.4$\pm$0.2 & $-$ & $-$ & $-$ \\
%
%line at 8498.02 rest in air -> 8495.185 in WD systemic frame; include 10 km/s systematic error component
CaII $\lambda$8498     & 20 Dec 2019 & 6.8$\pm$0.6 & 300$\pm$10 & $-$ & 760$\pm$75 \\  %left edge is 8488.5+/-1.5 Ang, right edge is 8510.0+/-1.5 Ang -- left peak is 8496.2+/-0.05 Ang, right peak is 8504.75+/-0.05 Ang
%
%line at 8542.09 rest in air -> 8539.24 in WD systemic frame; include 10 km/s systematic error component on any relative velocity measurement
CaII $\lambda$8542     & 20 Dec 2019 & 7.9$\pm$0.6 & 280$\pm$15 & $-$ & 790$\pm$80 \\  %left edge is 8531.5+/-2.0 Ang, right edge is 8554.0+/-1.0 Ang -- left peak is 8540.8+/-0.08 Ang, right peak is 8548.7+/-0.2 Ang
%
%line at 8662.14 rest in air -> 8659.25 in WD systemic frame; include 10 km/s systematic error component on any relative velocity measurement
CaII $\lambda$8662     & 20 Dec 2019 & 7.1$\pm$0.9 & 280$\pm$10 & $-$ & 780$\pm$75 \\  %left edge is 8652.0+/-1.5 Ang, right edge is 8674.5+/-1.5 Ang -- left peak is 8660.59+/-0.04 Ang, right peak is 8668.7+/-0.045 Ang
FeII $\lambda$4549     & 21 Oct 2016 & 0.6$\pm$0.4 & $-$ & $-$  & $-$  \\
FeII $\lambda$4549     & 20 Dec 2019 & 0.5$\pm$0.3 & $-$ & $-$  & $-$  \\
FeII $\lambda$4584     & 21 Oct 2016 & 0.5$\pm$0.2 & $-$ & $-$  & $-$  \\
FeII $\lambda$4584     & 20 Dec 2019 & 0.5$\pm$0.2 & $-$ & $-$  & $-$  \\
\enddata
\tablecomments{See notes to Table \ref{tabline0006}.}
\end{deluxetable}

\clearpage

\begin{deluxetable}{lccccc}
\tabletypesize{\small}
\tablecolumns{6}
\tablewidth{0pt}
\tablecaption{\large{GaiaJ0510+2315 Emission Line Measurements \label{tabline0510}} }
\tablehead{ 
  \colhead{Transition} & 
  \colhead{UT Date} &
  \colhead{Eq.\ Width} &
  \colhead{Peak Separation} &
  \colhead{v$_{\rm max}$sin$i$} &
  \colhead{Full Width} \\
  \colhead{} & 
  \colhead{} &
  \colhead{(\AA )} &
  \colhead{(km\,s$^{-1}$)} &
  \colhead{(km\,s$^{-1}$)} &
  \colhead{(km\,s$^{-1}$)}
}
\startdata
%lines at 7774.08, 7776.31, and 7777.53 Ang in vacuum -> 7773.71, 7777.16 Ang in WD systemic frame; include 5 km/s systematic error component
OI $\lambda$7772        & 18 Mar 2019 & 2.2$\pm$1.1 & $-$ & $-$ & $-$ \\
OI $\lambda$7772        & 16 Nov 2019 & 1.3$\pm$0.4 & $-$ & $-$ & $-$ \\
OI $\lambda$7772        & 17 Nov 2019 & 1.4$\pm$0.4 & $-$ & $-$ & $-$ \\
OI $\lambda$7772        & 18 Nov 2019\tablenotemark{a} & 1.3$\pm$0.3 & $-$ & $-$ & $-$ \\ %v1: EW=1.3+/-0.4
%v2: EW=1.3+/-0.4
OI $\lambda$7772        & 09 Dec 2019 & 1.0$\pm$0.2 & $-$ & $-$185$\pm$25/650$\pm$60 & $-$ \\ %left edge is 7768.9+/-0.6 Ang, right edge is 7792.6+/-1.5 Ang
% using red edge value taken from Figure 2!
%
%lines at 8448.57, 8448.68, and 8449.08 Ang in vacuum -> 8448.16, 8448.67 in WD systemic frame; include 5 km/s systematic error component
OI $\lambda$8446        & 16 Nov 2019 & 0.6$\pm$0.4 & $-$ & $-$ & $-$ \\
OI $\lambda$8446        & 18 Nov 2019\tablenotemark{a} & 0.6$\pm$0.2 & $-$ & $-$ & $-$ \\ %v1: EW=0.6+/-0.4
%v2: EW=0.55+/-0.2
OI $\lambda$8446        & 09 Dec 2019 & 0.4$\pm$0.1 & $-$ & $-$185$\pm$20/500$\pm$200 & $-$ \\
\enddata
\tablecomments{See notes to Table \ref{tabline0006}.}
\tablenotetext{a}{Two exposures were taken over the course of the night. The values reported are the average of measurements made for each individual exposure with individual values for each separate exposure agreeing to within their uncertainties.}
\end{deluxetable}

\clearpage

\begin{deluxetable}{lccccc}
\tabletypesize{\small}
\tablecolumns{6}
\tablewidth{0pt}
\tablecaption{\large{GaiaJ0611$-$6931 Emission Line Measurements \label{tabline0611}} }
\tablehead{ 
  \colhead{Transition} & 
  \colhead{UT Date} &
  \colhead{Eq.\ Width} &
  \colhead{Peak Separation} &
  \colhead{v$_{\rm max}$sin$i$} &
  \colhead{Full Width} \\
  \colhead{} & 
  \colhead{} &
  \colhead{(\AA )} &
  \colhead{(km\,s$^{-1}$)} &
  \colhead{(km\,s$^{-1}$)} &
  \colhead{(km\,s$^{-1}$)}
}
\startdata
%lines at 7771.94, 7774.17, and 7775.39 Ang in air -> 7772.43, 7775.88 in WD systemic frame; include 7 km/s systematic error component
OI $\lambda$7772        & 10 Nov 2019 & 2.7$\pm$1.1 & $-$ & $-$790$\pm$250/775$\pm$60 & $-$ \\ %left edge is 7752.0+/-6.5 Ang, right edge is 7796.0+/-1.5 Ang
%
%lines at 8446.25, 8446.36, and 8446.76 Ang in air -> 8446.785, 8447.295 in WD systemic frame; include 7 km/s systematic error component
OI $\lambda$8446        & 10 Nov 2019 & 1.7$\pm$0.8 & $-$ & $-$615$\pm$70/700$\pm$70 & $-$ \\ %left edge is 8429.5+/-2.0 Ang, right edge is 8467.0+/-2.0 Ang
%
%lines at 5889.95095 and 5895.92424 rest in air -> 5890.32 and 5896.30  in WD systemic frame; include 7 km/s systematic error component
NaI $\lambda$5890        & 10 Oct 2019 & 2.3$\pm$0.3 & $-$ & $-$ & $-$ \\
NaI $\lambda$5890        & 10 Nov 2019 & 1.8$\pm$0.3 & $-$ & $-$680$\pm$100/800$\pm$180 & $-$ \\ %left edge is 5877.0+/-2.0 Ang, right edge is 5912.0+/-3.5 Ang
%
%lines at 5167.321, 5172,684, 5183.604 rest in air -> 5167.65, 5183.93 in WD systemic frame; include 7 km/s systematic error component
MgI $\lambda$5167    & 10 Oct 2019 & 4.5$\pm$0.6 & $-$ & $-$ & $-$ \\  
MgI $\lambda$5167    & 10 Nov 2019 & 5.6$\pm$0.8 & $-$ & $-$735$\pm$75/755$\pm$115 & $-$ \\  %left edge is 5155.0+/-1.3 Ang, right edge is 5197.0+/-2.0 Ang
%line at 8806.757 rest in air -> 8807.315 in WD systemic frame; include 7 km/s systematic error component
%
MgI $\lambda$8806     & 10 Nov 2019 & 4.8$\pm$1.4 & $-$ & $-$630$\pm$60/670$\pm$50 & 1300$\pm$78 \\
\enddata
\tablecomments{See notes to Table \ref{tabline0006}.}
\end{deluxetable}

\clearpage

\begin{deluxetable}{lccccc}
\tabletypesize{\small}
\tablecolumns{6}
\tablewidth{0pt}
\tablecaption{\large{GaiaJ0644$-$0352 Emission Line Measurements \label{tabline0644}} }
\tablehead{ 
  \colhead{Transition} & 
  \colhead{UT Date} &
  \colhead{Eq.\ Width} &
  \colhead{Peak Separation} &
  \colhead{v$_{\rm max}$sin$i$} &
  \colhead{Full Width} \\
  \colhead{} & 
  \colhead{} &
  \colhead{(\AA )} &
  \colhead{(km\,s$^{-1}$)} &
  \colhead{(km\,s$^{-1}$)} &
  \colhead{(km\,s$^{-1}$)}
}
\startdata
%line at 8500.35 rest in vacuum -> 8501.80 in WD systemic frame; include 5 km/s systematic error component
CaII $\lambda$8498     & 09 Dec 2019 & 1.5$\pm$0.4 & $-$ & $-$370$\pm$45/530$\pm$105 & 900$\pm$115 \\  %left edge is 8491.3+/-1.3 Ang, right edge is 8516.9+/-3.0 Ang
%
%line at 8544.44 rest in vacuum -> 8545.90 in WD systemic frame; include 5 km/s systematic error component on any velocity measurement
CaII $\lambda$8542     & 09 Dec 2019 & 1.0$\pm$0.4 & $-$ & $-$370$\pm$60/460$\pm$40 & 835$\pm$72 \\  %left edge is 8535.3+/-1.7 Ang, right edge is 8559.1+/-1.2 Ang
%
%line at 8664.52 rest in vacuum -> 8666.00 in WD systemic frame; include 5 km/s systematic error component on any velocity measurement
CaII $\lambda$8662     & 09 Dec 2019 & 1.1$\pm$0.5 & $-$ & $-$330$\pm$25/475$\pm$30 & 805$\pm$39 \\  %left edge is 8656.4+/-0.7 Ang, right edge is 8679.7+/-0.8 Ang
%
%some lines between 5312 and 5367 Ang?
\enddata
\tablecomments{See notes to Table \ref{tabline0006}.}
\end{deluxetable}

\clearpage

\begin{deluxetable}{lccccc}
\tabletypesize{\small}
\tablecolumns{6}
\tablewidth{0pt}
\tablecaption{\large{WD0842+572 Emission Line Measurements \label{tabline0842}} }
\tablehead{ 
  \colhead{Transition} & 
  \colhead{UT Date} &
  \colhead{Eq.\ Width} &
  \colhead{Peak Separation} &
  \colhead{v$_{\rm max}$sin$i$} &
  \colhead{Full Width} \\
  \colhead{} & 
  \colhead{} &
  \colhead{(\AA )} &
  \colhead{(km\,s$^{-1}$)} &
  \colhead{(km\,s$^{-1}$)} &
  \colhead{(km\,s$^{-1}$)}
}
\startdata
%line at 4572.3767 rest in vacuum -> 4572.30 in WD systemic frame; include 5 km/s systematic error component
MgI] $\lambda$4571      & 18 Mar 2019 & 0.6$\pm$0.2 & $-$ & $-$ & $-$ \\
MgI] $\lambda$4571      & 24 Apr 2019 & 0.4$\pm$0.1 & $-$ & $-$ & $-$ \\
MgI] $\lambda$4571      & 09 Dec 2019 & 0.4$\pm$0.1 & 240$\pm$5 & $-$150$\pm$20/140$\pm$14 & 290$\pm$24 \\ %left edge is 4570.0+/-0.3 Ang, right edge is 4574.4+/-0.2 Ang -- left peak is 4570.45+/-0.015 Ang, right peak is 4574.09+/-0.01 Ang
%
%line at 5185.0478 rest in vacuum -> 5184.96 in WD systemic frame; include 5 km/s systematic error component
MgI $\lambda$5184      & 09 Dec 2019 & 0.6$\pm$0.1 & 230$\pm$5 & $-$135$\pm$11/135$\pm$10 & 270$\pm$14 \\ %left edge is 5182.6+/-0.2 Ang, right edge is 5187.3+/-0.1 Ang -- left peak is 5182.93+/-0.01 Ang, right peak is 5186.93+/-0.01 Ang
%
%line at 3906.629 rest in vacuum -> 3906.56 in WD systemic frame; include 5 km/s systematic error component
SiI $\lambda$3905     & 15 Jun 2020  & 0.5$\pm$0.3 & $-$ & $-$180$\pm$45/150$\pm$20 & 330$\pm$50 \\ %left edge is 3904.2+/-0.6 Ang, right edge is 3908.5+/-0.25 Ang
%
%6574.7 Ang - unambiguous
%left edge is 6571.4+/-0.25 Ang, right edge is 6577.5+/-0.2 Ang
%Fe II 6575.3085? 6575.20 in WD systemic frame
%  => -175/105 km/s, too far red
%Ca I 6574.60? 6574.488 in WD systemic frame (CaI 6572.78 air)
%  => -140/135 km/s, match and ground state!
%line at 6574.60 rest in vacuum - > 6574.49 in WD systemic frame; include 5 km/s systematic error component
CaI] $\lambda$6573     & 09 Dec 2019 & 0.2$\pm$0.1 & 230$\pm$5 & $-$140$\pm$12/135$\pm$10 & 275$\pm$15 \\  %left edge is 6571.4+/-0.25 Ang, right edge is 6577.5+/-0.2 Ang
%
%line at 3934.77 rest in vacuum -> 3934.70 in WD systemic frame; include 5 km/s systematic error component
CaII $\lambda$3933     & 15 Jun 2020  & 0.1$\pm$0.1 & $-$ & $-$170$\pm$40/120$\pm$25 & 290$\pm$47 \\
\enddata
\tablecomments{See notes to Table \ref{tabline0006}.}
\end{deluxetable}

\clearpage

\begin{deluxetable}{lccccc}
\tabletypesize{\small}
\tablecolumns{6}
\tablewidth{0pt}
\tablecaption{\large{WD1622+587 Emission Line Measurements \label{tabline1622}} }
\tablehead{ 
  \colhead{Transition} & 
  \colhead{UT Date} &
  \colhead{Eq.\ Width} &
  \colhead{Peak Separation} &
  \colhead{v$_{\rm max}$sin$i$} &
  \colhead{Full Width} \\
  \colhead{} & 
  \colhead{} &
  \colhead{(\AA )} &
  \colhead{(km\,s$^{-1}$)} &
  \colhead{(km\,s$^{-1}$)} &
  \colhead{(km\,s$^{-1}$)}
}
\startdata
%lines at 7774.08, 7776.31, and 7777.53 Ang in vacuum -> 7773.097, 7776.55 Ang in WD systemic frame; include 5 km/s systematic error component
OI $\lambda$7772        & 12 Jul 2019 & 1.3$\pm$0.1 & $-$ & $-$ & $-$ \\
OI $\lambda$7772        & 16 Jul 2019 & 1.9$\pm$0.7 & $-$ & $-$460$\pm$95/445$\pm$95 & $-$ \\ %left edge is 7761.2+/-2.5 Ang, right edge is 7788.1+/-2.5 Ang
%
%line at 8500.35 rest in vacuum -> 8499.27 in WD systemic frame; include 5 km/s systematic error component
CaII $\lambda$8498     & 16 Jul 2019 & 2.4$\pm$0.6 & $-$ & $-$420$\pm$105/430$\pm$70 & 850$\pm$127 \\  %left edge is 8487.3+/-3.0 Ang, right edge is 8511.5+/-2.0 Ang
%
%line at 8544.44 rest in vacuum -> 8543.36 in WD systemic frame; include 5 km/s systematic error component on any velocity measurement
CaII $\lambda$8542     & 16 Jul 2019 & 2.7$\pm$1.3 & $-$ & $-$550$\pm$140/480$\pm$105 & 1030$\pm$175 \\  %left edge is 8527.6+/-4.0 Ang, right edge is 8557.0+/-3.0 Ang
%
%line at 8664.52 rest in vacuum -> 8663.42 in WD systemic frame; include 5 km/s systematic error component on any velocity measurement
CaII $\lambda$8662     & 16 Jul 2019 & 1.1$\pm$0.9 & $-$ & $-$300$\pm$120/470$\pm$70 & 770$\pm$139 \\  %left edge is 8654.75+/-3.5 Ang, right edge is 8677.0+/-2.0 Ang
%
%line at 5277.4650 rest in vacuum -> 5276.80 in WD systemic frame; include 5 km/s systematic error component
FeII $\lambda$5276     & 16 Jul 2019  & 0.3$\pm$0.2 & $-$ & $-$400$\pm$115/570$\pm$115 & 970$\pm$161 \\ %left edge is 5269.75+/-2.0 Ang, right edge is 5286.8+/-2.0 Ang
%
%line at 5318.0889 rest in vacuum -> 5317.42 in WD systemic frame; include 5 km/s systematic error component
FeII $\lambda$5317    & 16 Jul 2019  & 0.4$\pm$0.2 & $-$ & $-$445$\pm$115/540$\pm$140 & 985$\pm$181 \\
\enddata
\tablecomments{See notes to Table \ref{tabline0006}.}
\end{deluxetable}

\clearpage

\begin{deluxetable}{lccccc}
\tabletypesize{\small}
\tablecolumns{6}
\tablewidth{0pt}
\tablecaption{\large{GaiaJ2100+2122 Emission Line Measurements \label{tabline2100}} }
\tablehead{ 
  \colhead{Transition} & 
  \colhead{UT Date} &
  \colhead{Eq.\ Width} &
  \colhead{Peak Separation} &
  \colhead{v$_{\rm max}$sin$i$} &
  \colhead{Full Width} \\
  \colhead{} & 
  \colhead{} &
  \colhead{(\AA )} &
  \colhead{(km\,s$^{-1}$)} &
  \colhead{(km\,s$^{-1}$)} &
  \colhead{(km\,s$^{-1}$)}
}
\startdata
%lines at 7774.08, 7776.31, and 7777.53 Ang in vacuum -> 7773.26, 7776.71 Ang in WD systemic frame; include 5 km/s systematic error component
OI $\lambda$7772        & 24 May 2019 & 0.4$\pm$0.2 & $-$ & $-$ & $-$ \\
OI $\lambda$7772        & 16 Jul 2019   & 0.6$\pm$0.2 & $-$ & $-$390$\pm$60/350$\pm$80 & $-$ \\ %left edge is 7763.1+/-1.5 Ang, right edge is 7785.7+/-2.0 Ang
%
%line at 8500.35 rest in vacuum -> 8499.45 in WD systemic frame; include 5 km/s systematic error component
CaII $\lambda$8498     & 24 May 2019 & 1.6$\pm$0.2 & $-$ & $-$ & 740$\pm$190 \\ %left edge is 8483.7+/-3.5 Ang, right edge is 8504.7+/-4.0 Ang
CaII $\lambda$8498     & 16 Jul 2019 & 0.8$\pm$0.2 & $-$ & $-$385$\pm$80/460$\pm$70 & 845$\pm$105 \\  %left edge is 8488.55+/-2.2 Ang, right edge is 8512.55+/-2.0 Ang
%
%line at 8544.44 rest in vacuum -> 8543.54 in WD systemic frame; include 5 km/s systematic error component on any velocity measurement
CaII $\lambda$8542     & 24 May 2019 & 1.7$\pm$0.3 & $-$ & $-$ & 795$\pm$115 \\ %left edge is 8529.6+/-2.0 Ang, right edge is 8552.2+/-2.5 Ang
CaII $\lambda$8542     & 16 Jul 2019 & 0.5$\pm$0.1 & $-$ & $-$455$\pm$140/480$\pm$140 & 935$\pm$199 \\  %left edge is 8530.55+/-4.0 Ang, right edge is 8557.2+/-4.0 Ang
%
%line at 8664.52 rest in vacuum -> 8663.60 in WD systemic frame; include 5 km/s systematic error component on any velocity measurement
CaII $\lambda$8662     & 24 May 2019 & 1.3$\pm$0.1 & $-$ & $-$ & 780$\pm$145 \\ %left edge is 8647.8+/-3.5 Ang, right edge is 8670.3+/-2.3 Ang
CaII $\lambda$8662     & 16 Jul 2019 & 0.3$\pm$0.2 & $-$ & $-$405$\pm$105/465$\pm$70 & 870$\pm$125 \\
\enddata
\tablecomments{See notes to Table \ref{tabline0006}.}
\end{deluxetable}


\begin{thebibliography}{87}
\expandafter\ifx\csname natexlab\endcsname\relax\def\natexlab#1{#1}\fi

\bibitem[\protect\astroncite{{Balayan}}{1997{\natexlab{a}}}]{balayan97a}
{Balayan}, S.~K. 1997{\natexlab{a}}, {\em Astrophysics\/}, {\bf 40}, 101

\bibitem[\protect\astroncite{{Balayan}}{1997{\natexlab{b}}}]{balayan97b}
--- 1997{\natexlab{b}}, {\em Astrophysics\/}, {\bf 40}, 211

\bibitem[\protect\astroncite{{Bear} \& {Soker}}{2013}]{bear13}
{Bear}, E. \& {Soker}, N. 2013, {\em \na\/}, {\bf 19}, 56

\bibitem[\protect\astroncite{{B{\'e}dard} {\em et~al.\/}}{2020}]{bedard20}
{B{\'e}dard}, A., {Bergeron}, P., {Brassard}, P., \& {Fontaine}, G. 2020, {\em \apj\/} accepted, arXiv:2008.07469

\bibitem[\protect\astroncite{{Bergeron} {\em et~al.\/}}{1992}]{bergeron92}
{Bergeron}, P., {Saffer}, R.~A., \& {Liebert}, J. 1992, {\em \apj\/}, {\bf
  394}, 228

\bibitem[\protect\astroncite{{Bergeron} {\em et~al.\/}}{2019}]{bergeron19}
{Bergeron}, P., {\em et~al.\/} 2019, {\em \apj\/}, {\bf 876}, 67

\bibitem[\protect\astroncite{{Bochkarev} \& {Rafikov}}{2011}]{bochkarev11}
{Bochkarev}, K.~V. \& {Rafikov}, R.~R. 2011, {\em \apj\/}, {\bf 741}, 36

\bibitem[\protect\astroncite{{Bogn{\'a}r} {\em et~al.\/}}{2018}]{bognar18}
{Bogn{\'a}r}, Z., {Kalup}, C., {S{\'o}dor}, {\'A}., {Charpinet}, S., \&
  {Hermes}, J.~J. 2018, {\em \mnras\/}, {\bf 478}, 2676

\bibitem[\protect\astroncite{{Brinkworth} {\em et~al.\/}}{2012}]{brinkworth12}
{Brinkworth}, C., {Gaensicke}, B., {Girven}, J., {Hoard}, D., {Marsh}, T.,
  {Parsons}, S., \& {Koester}, D. 2012, {\em \apj\/}, {\bf 750}, 86

\bibitem[\protect\astroncite{{Cauley} {\em et~al.\/}}{2018}]{cauley18}
{Cauley}, P.~W., {Farihi}, J., {Redfield}, S., {Bachman}, S., {Parsons}, S.~G.,
  \& {G{\"a}nsicke}, B.~T. 2018, {\em \apjl\/}, {\bf 852}, L22

\bibitem[\protect\astroncite{{Debes} \& {Sigurdsson}}{2002}]{debes02}
{Debes}, J.~H. \& {Sigurdsson}, S. 2002, {\em \apj\/}, {\bf 572}, 556

\bibitem[\protect\astroncite{{Dennihy} {\em et~al.\/}}{2018}]{dennihy18}
{Dennihy}, E., {Clemens}, J.~C., {Dunlap}, B.~H., {Fanale}, S.~M., {Fuchs},
  J.~T., \& {Hermes}, J.~J. 2018, {\em \apj\/}, {\bf 854}, 40

\bibitem[\protect\astroncite{{Dufour} {\em et~al.\/}}{2005}]{dufour05}
{Dufour}, P., {Bergeron}, P., \& {Fontaine}, G. 2005, {\em \apj\/}, {\bf 627},
  404

\bibitem[\protect\astroncite{{Dufour} {\em et~al.\/}}{2010}]{dufour10}
{Dufour}, P., {Kilic}, M., {Fontaine}, G., {Bergeron}, P., {Lachapelle}, F.,
  {Kleinman}, S.~J., \& {Leggett}, S.~K. 2010, {\em \apj\/}, {\bf 719}, 803

\bibitem[\protect\astroncite{{Dufour} {\em et~al.\/}}{2012}]{dufour12}
{Dufour}, P., {Kilic}, M., {Fontaine}, G., {Bergeron}, P., {Melis}, C., \&
  {Bochanski}, J. 2012, {\em \apj\/}, {\bf 749}, 6

\bibitem[\protect\astroncite{{Dufour} {\em et~al.\/}}{2007}]{dufour07}
{Dufour}, P., {\em et~al.\/} 2007, {\em \apj\/}, {\bf 663}, 1291

\bibitem[\protect\astroncite{{Farihi}}{2016}]{farihi16}
{Farihi}, J. 2016, {\em \nar\/}, {\bf 71}, 9

\bibitem[\protect\astroncite{{Farihi} {\em et~al.\/}}{2013}]{farihi13}
{Farihi}, J., {G{\"a}nsicke}, B.~T., \& {Koester}, D. 2013, {\em Science\/},
  {\bf 342}, 218

\bibitem[\protect\astroncite{{Farihi} {\em et~al.\/}}{2012}]{farihi12a}
{Farihi}, J., {G{\"a}nsicke}, B.~T., {Steele}, P.~R., {Girven}, J., {Burleigh},
  M.~R., {Breedt}, E., \& {Koester}, D. 2012, {\em \mnras\/}, {\bf 421}, 1635

\bibitem[\protect\astroncite{{Farihi} {\em et~al.\/}}{2009}]{farihi09}
{Farihi}, J., {Jura}, M., \& {Zuckerman}, B. 2009, {\em \apj\/}, {\bf 694}, 805

\bibitem[\protect\astroncite{{Farihi} {\em et~al.\/}}{2018}]{farihi18}
{Farihi}, J., {\em et~al.\/} 2018, {\em \mnras\/}, {\bf 481}, 2601

\bibitem[\protect\astroncite{{Fortin-Archambault} {\em
  et~al.\/}}{2020}]{fortinarchambault20}
{Fortin-Archambault}, M., {Dufour}, P., \& {Xu}, S. 2020, {\em \apj\/}, {\bf
  888}, 47

\bibitem[\protect\astroncite{{Frewen} \& {Hansen}}{2014}]{frewen14}
{Frewen}, S.~F.~N. \& {Hansen}, B.~M.~S. 2014, {\em \mnras\/}, {\bf 439}, 2442

\bibitem[\protect\astroncite{{G{\"a}nsicke} {\em et~al.\/}}{2012}]{gaensicke12}
{G{\"a}nsicke}, B.~T., {Koester}, D., {Farihi}, J., {Girven}, J., {Parsons},
  S.~G., \& {Breedt}, E. 2012, {\em \mnras\/}, {\bf 424}, 333

\bibitem[\protect\astroncite{{G{\"a}nsicke} {\em et~al.\/}}{2008}]{gaensicke08}
{G{\"a}nsicke}, B.~T., {Koester}, D., {Marsh}, T.~R., {Rebassa-Mansergas}, A.,
  \& {Southworth}, J. 2008, {\em \mnras\/}, {\bf 391}, L103

\bibitem[\protect\astroncite{{G{\"a}nsicke} {\em et~al.\/}}{2007}]{gaensicke07}
{G{\"a}nsicke}, B.~T., {Marsh}, T.~R., \& {Southworth}, J. 2007, {\em
  \mnras\/}, {\bf 380}, L35

\bibitem[\protect\astroncite{{G{\"a}nsicke} {\em et~al.\/}}{2006}]{gaensicke06}
{G{\"a}nsicke}, B.~T., {Marsh}, T.~R., {Southworth}, J., \&
  {Rebassa-Mansergas}, A. 2006, {\em Science\/}, {\bf 314}, 1908

\bibitem[\protect\astroncite{{G{\"a}nsicke} {\em et~al.\/}}{2019}]{gaensicke19}
{G{\"a}nsicke}, B.~T., {Schreiber}, M.~R., {Toloza}, O., {Fusillo}, N. P.~G.,
  {Koester}, D., \& {Manser}, C.~J. 2019, {\em \nat\/}, {\bf 576}, 61

\bibitem[\protect\astroncite{{Genest-Beaulieu} \&
  {Bergeron}}{2019{\natexlab{a}}}]{genestbeaulieu19a}
{Genest-Beaulieu}, C. \& {Bergeron}, P. 2019{\natexlab{a}}, {\em \apj\/}, {\bf
  871}, 169

\bibitem[\protect\astroncite{{Genest-Beaulieu} \&
  {Bergeron}}{2019{\natexlab{b}}}]{genestbeaulieu19b}
--- 2019{\natexlab{b}}, {\em \apj\/}, {\bf 882}, 106

\bibitem[\protect\astroncite{{Gentile Fusillo} {\em
  et~al.\/}}{2017}]{gentilefusillo17}
{Gentile Fusillo}, N.~P., {G{\"a}nsicke}, B.~T., {Farihi}, J., {Koester}, D.,
  {Schreiber}, M.~R., \& {Pala}, A.~F. 2017, {\em \mnras\/}, {\bf 468}, 971

\bibitem[\protect\astroncite{{Gentile Fusillo} {\em
  et~al.\/}}{2019}]{gentilefusillo19}
{Gentile Fusillo}, N.~P., {\em et~al.\/} 2019, {\em \mnras\/}, {\bf 482}, 4570

\bibitem[\protect\astroncite{{Gianninas} {\em et~al.\/}}{2006}]{gianninas06}
{Gianninas}, A., {Bergeron}, P., \& {Fontaine}, G. 2006, {\em \aj\/}, {\bf
  132}, 831

\bibitem[\protect\astroncite{{Gianninas} {\em et~al.\/}}{2011}]{gianninas11}
{Gianninas}, A., {Bergeron}, P., \& {Ruiz}, M.~T. 2011, {\em \apj\/}, {\bf
  743}, 138

\bibitem[\protect\astroncite{{Grishin} \& {Veras}}{2019}]{grishin19}
{Grishin}, E. \& {Veras}, D. 2019, {\em \mnras\/}, {\bf 489}, 168

\bibitem[\protect\astroncite{{Guo} {\em et~al.\/}}{2015}]{guo15}
{Guo}, J., {Tziamtzis}, A., {Wang}, Z., {Liu}, J., {Zhao}, J., \& {Wang}, S.
  2015, {\em \apjl\/}, {\bf 810}, L17

\bibitem[\protect\astroncite{{Hartmann} {\em et~al.\/}}{2011}]{hartmann11}
{Hartmann}, S., {Nagel}, T., {Rauch}, T., \& {Werner}, K. 2011, {\em \aap\/},
  {\bf 530}, A7

\bibitem[\protect\astroncite{{Hartmann} {\em et~al.\/}}{2016}]{hartmann16}
--- 2016, {\em \aap\/}, {\bf 593}, A67

\bibitem[\protect\astroncite{{Hook} {\em et~al.\/}}{2004}]{hook04}
{Hook}, I.~M., {J{\o}rgensen}, I., {Allington-Smith}, J.~R., {Davies}, R.~L.,
  {Metcalfe}, N., {Murowinski}, R.~G., \& {Crampton}, D. 2004, {\em \pasp\/},
  {\bf 116}, 425

\bibitem[\protect\astroncite{{Horne} \& {Marsh}}{1986}]{horne86}
{Horne}, K. \& {Marsh}, T.~R. 1986, {\em \mnras\/}, {\bf 218}, 761

\bibitem[\protect\astroncite{{Johnson} \& {Soderblom}}{1987}]{johnson87}
{Johnson}, D.~R.~H. \& {Soderblom}, D.~R. 1987, {\em \aj\/}, {\bf 93}, 864

\bibitem[\protect\astroncite{{Jura}}{2003}]{jura03b}
{Jura}, M. 2003, {\em \apjl\/}, {\bf 584}, L91

\bibitem[\protect\astroncite{{Jura}}{2008}]{jura08}
--- 2008, {\em \aj\/}, {\bf 135}, 1785

\bibitem[\protect\astroncite{{Kelson}}{2003}]{kelson03}
{Kelson}, D.~D. 2003, {\em \pasp\/}, {\bf 115}, 688

\bibitem[\protect\astroncite{{Kelson} {\em et~al.\/}}{2000}]{kelson00}
{Kelson}, D.~D., {Illingworth}, G.~D., {van Dokkum}, P.~G., \& {Franx}, M.
  2000, {\em \apj\/}, {\bf 531}, 159

\bibitem[\protect\astroncite{{Kenyon} \& {Bromley}}{2017}]{kenyon17}
{Kenyon}, S.~J. \& {Bromley}, B.~C. 2017, {\em \apj\/}, {\bf 850}, 50

\bibitem[\protect\astroncite{{Kilic} {\em et~al.\/}}{2020}]{kilic20}
{Kilic}, M., {Bergeron}, P., {Kosakowski}, A., {Brown}, W.~R., {Agueros},
  M.~A., \& {Blouin}, S. 2020, {\em \apj\/}, {\bf 898}, 84

\bibitem[\protect\astroncite{{Klein} {\em et~al.\/}}{2010}]{klein10}
{Klein}, B., {Jura}, M., {Koester}, D., {Zuckerman}, B., \& {Melis}, C. 2010,
  {\em \apj\/}, {\bf 709}, 950

\bibitem[\protect\astroncite{{Klein} {\em et~al.\/}}{2020}]{klein20}
{Klein}, B., {\em et~al.\/} 2020, {\em \apj\/}, {\bf 900}, 2

\bibitem[\protect\astroncite{{Li} {\em et~al.\/}}{2017}]{li17}
{Li}, L., {Zhang}, F., {Kong}, X., {Han}, Q., \& {Li}, J. 2017, {\em \apj\/},
  {\bf 836}, 71

\bibitem[\protect\astroncite{{Liebert} {\em et~al.\/}}{2005}]{liebert05}
{Liebert}, J., {Bergeron}, P., \& {Holberg}, J.~B. 2005, {\em \apjs\/}, {\bf
  156}, 47

\bibitem[\protect\astroncite{{Limoges} {\em et~al.\/}}{2015}]{limoges15}
{Limoges}, M.~M., {Bergeron}, P., \& {L{\'e}pine}, S. 2015, {\em \apjs\/}, {\bf
  219}, 19

\bibitem[\protect\astroncite{{Malamud} \& {Perets}}{2020}]{malamud20}
{Malamud}, U. \& {Perets}, H.~B. 2020, {\em \mnras\/}, {\bf 492}, 5561

\bibitem[\protect\astroncite{{Maldonado} {\em et~al.\/}}{2020a}]{maldonado20a}
{Maldonado}, R.~F., {Villaver}, E., {Mustill}, A.~J., {Chavez}, M., \&
  {Bertone}, E. 2020a,  {\em \mnras\/}, {\bf 497}, 4091

\bibitem[\protect\astroncite{{Maldonado} {\em et~al.\/}}{2020b}]{maldonado20b}
{Maldonado}, R.~F., {Villaver}, E., {Mustill}, A.~J., {Chavez}, M., \&
  {Bertone}, E. 2020b, {\em \mnras\/}, arXiv:2009.10844

\bibitem[\protect\astroncite{{Manser} {\em
  et~al.\/}}{2016{\natexlab{a}}}]{manser16b}
{Manser}, C.~J., {G{\"a}nsicke}, B.~T., {Koester}, D., {Marsh}, T.~R., \&
  {Southworth}, J. 2016{\natexlab{a}}, {\em \mnras\/}, {\bf 462}, 1461

\bibitem[\protect\astroncite{{Manser} {\em
  et~al.\/}}{2016{\natexlab{b}}}]{manser16}
{Manser}, C.~J., {\em et~al.\/} 2016{\natexlab{b}}, {\em \mnras\/}, {\bf 455},
  4467

\bibitem[\protect\astroncite{{Manser} {\em et~al.\/}}{2019}]{manser19}
--- 2019, {\em Science\/}, {\bf 364}, 66

\bibitem[\protect\astroncite{{Manser} {\em et~al.\/}}{2020}]{manser20}
--- 2020, {\em \mnras\/}, {\bf 493}, 2127

\bibitem[\protect\astroncite{{Marshall} {\em et~al.\/}}{2008}]{marshall08}
{Marshall}, J.~L., {\em et~al.\/} 2008, in {\em Society of Photo-Optical
  Instrumentation Engineers (SPIE) Conference Series\/}, vol. 7014 of {\em
  Society of Photo-Optical Instrumentation Engineers (SPIE) Conference
  Series\/}

\bibitem[\protect\astroncite{{Matlovi{\v{c}}} {\em
  et~al.\/}}{2020}]{matlovic20}
{Matlovi{\v{c}}}, P., {T{\'o}th}, J., {Korno{\v{s}}}, L., \& {Loehle}, S. 2020,
  {\em \icarus\/}, {\bf 347}, 113817

\bibitem[\protect\astroncite{{McCook} \& {Sion}}{1987}]{mccook87}
{McCook}, G.~P. \& {Sion}, E.~M. 1987, {\em \apjs\/}, {\bf 65}, 603

\bibitem[\protect\astroncite{{Melis} \& {Dufour}}{2017}]{melis17}
{Melis}, C. \& {Dufour}, P. 2017, {\em \apj\/}, {\bf 834}, 1

\bibitem[\protect\astroncite{{Melis} {\em et~al.\/}}{2010}]{melis10}
{Melis}, C., {Jura}, M., {Albert}, L., {Klein}, B., \& {Zuckerman}, B. 2010,
  {\em \apj\/}, {\bf 722}, 1078

\bibitem[\protect\astroncite{{Melis} {\em et~al.\/}}{2011}]{melis11}
{Melis}, C., {\em et~al.\/} 2011, {\em \apj\/}, {\bf 732}, 90

\bibitem[\protect\astroncite{{Melis} {\em et~al.\/}}{2012}]{melis12}
--- 2012, {\em \apjl\/}, {\bf 751}, L4

\bibitem[\protect\astroncite{{Metzger} {\em et~al.\/}}{2012}]{metzger12}
{Metzger}, B.~D., {Rafikov}, R.~R., \& {Bochkarev}, K.~V. 2012, {\em \mnras\/},
  {\bf 423}, 505

\bibitem[\protect\astroncite{{Mustill} {\em et~al.\/}}{2018}]{mustill18}
{Mustill}, A.~J., {Villaver}, E., {Veras}, D., {G{\"a}nsicke}, B.~T., \&
  {Bonsor}, A. 2018, {\em \mnras\/}, {\bf 476}, 3939

\bibitem[\protect\astroncite{{Nixon} {\em et~al.\/}}{2020}]{nixon20}
{Nixon}, C.~J., {Pringle}, J.~E., {Coughlin}, E.~R., {Swan}, A., \& {Farihi},
  J. 2020, {\em arXiv e-prints\/}, arXiv:2006.07639

\bibitem[\protect\astroncite{{Rafikov}}{2011{\natexlab{a}}}]{rafikov11a}
{Rafikov}, R.~R. 2011{\natexlab{a}}, {\em \apjl\/}, {\bf 732}, L3

\bibitem[\protect\astroncite{{Rafikov}}{2011{\natexlab{b}}}]{rafikov11b}
--- 2011{\natexlab{b}}, {\em \mnras\/}, {\bf 416}, L55

\bibitem[\protect\astroncite{{Rebassa-Mansergas} {\em
  et~al.\/}}{2019}]{rebassa19}
{Rebassa-Mansergas}, A., {Solano}, E., {Xu}, S., {Rodrigo}, C.,
  {Jim{\'e}nez-Esteban}, F.~M., \& {Torres}, S. 2019, {\em \mnras\/}, {\bf
  489}, 3990

\bibitem[\protect\astroncite{{Swan} {\em et~al.\/}}{2019}]{swan19}
{Swan}, A., {Farihi}, J., \& {Wilson}, T.~G. 2019, {\em \mnras\/}, {\bf 484},
  L109

\bibitem[\protect\astroncite{{Swan} {\em et~al.\/}}{2020}]{swan20}
{Swan}, A., {Farihi}, J., {Wilson}, T.~G., \& {Parsons}, S.~G. 2020, {\em \mnras\/}, {\bf 496}, 5233

\bibitem[\protect\astroncite{{van Hoof}}{2018}]{vanhoof18}
{van Hoof}, P. A.~M. 2018, {\em Galaxies\/}, {\bf 6}, 63

\bibitem[\protect\astroncite{{Vanderburg} {\em et~al.\/}}{2015}]{vanderburg15}
{Vanderburg}, A., {\em et~al.\/} 2015, {\em \nat\/}, {\bf 526}, 546

\bibitem[\protect\astroncite{{Veras} {\em et~al.\/}}{2014}]{veras14}
{Veras}, D., {Leinhardt}, Z.~M., {Bonsor}, A., \& {G{\"a}nsicke}, B.~T. 2014,
  {\em \mnras\/}, {\bf 445}, 2244

\bibitem[\protect\astroncite{{Veras} {\em et~al.\/}}{2016}]{veras16}
{Veras}, D., {Mustill}, A.~J., {G{\"a}nsicke}, B.~T., {Redfield}, S.,
  {Georgakarakos}, N., {Bowler}, A.~B., \& {Lloyd}, M.~J.~S. 2016, {\em
  \mnras\/}, {\bf 458}, 3942

\bibitem[\protect\astroncite{{Vogt} {\em et~al.\/}}{1994}]{vogt94}
{Vogt}, S.~S., {\em et~al.\/} 1994, in {\em Proc. SPIE Instrumentation in
  Astronomy VIII, David L. Crawford; Eric R. Craine; Eds., Volume 2198, p.
  362\/}, edited by D.~L. {Crawford} \& E.~R. {Craine}, vol. 2198 of {\em
  Presented at the Society of Photo-Optical Instrumentation Engineers (SPIE)
  Conference\/},  362

\bibitem[\protect\astroncite{{Wang} {\em et~al.\/}}{2019}]{wang19}
{Wang}, T.-g., {\em et~al.\/} 2019, {\em \apjl\/}, {\bf 886}, L5

\bibitem[\protect\astroncite{{Wilson} {\em et~al.\/}}{2014}]{wilson14}
{Wilson}, D.~J., {G{\"a}nsicke}, B.~T., {Koester}, D., {Raddi}, R., {Breedt},
  E., {Southworth}, J., \& {Parsons}, S.~G. 2014, {\em \mnras\/}, {\bf 445},
  1878

\bibitem[\protect\astroncite{{Wilson} {\em et~al.\/}}{2015}]{wilson15}
{Wilson}, D.~J., {G{\"a}nsicke}, B.~T., {Koester}, D., {Toloza}, O., {Pala},
  A.~F., {Breedt}, E., \& {Parsons}, S.~G. 2015, {\em \mnras\/}, {\bf 451},
  3237

\bibitem[\protect\astroncite{{Xu} \& {Jura}}{2014}]{xujura14}
{Xu}, S. \& {Jura}, M. 2014, {\em \apjl\/}, {\bf 792}, L39

\bibitem[\protect\astroncite{{Xu} {\em et~al.\/}}{2016}]{xu16}
{Xu}, S., {Jura}, M., {Dufour}, P., \& {Zuckerman}, B. 2016, {\em \apjl\/},
  {\bf 816}, L22

\bibitem[\protect\astroncite{{Xu} {\em et~al.\/}}{2013}]{xu13}
{Xu}, S., {Jura}, M., {Klein}, B., {Koester}, D., \& {Zuckerman}, B. 2013, {\em
  \apj\/}, {\bf 766}, 132

\bibitem[\protect\astroncite{{Xu} {\em et~al.\/}}{2020}]{xu20}
{Xu}, S., {Lai}, S., \& {Dennihy}, E. 2020, {\em ArXiv e-prints\/}, arXiv:2009.00193

\bibitem[\protect\astroncite{{Xu} {\em et~al.\/}}{2019}]{xu19}
{Xu}, S., {\em et~al.\/} 2019, {\em \aj\/}, {\bf 158}, 242

\bibitem[\protect\astroncite{{Zuckerman} {\em et~al.\/}}{2007}]{zuckerman07}
{Zuckerman}, B., {Koester}, D., {Melis}, C., {Hansen}, B.~M., \& {Jura}, M.
  2007, {\em \apj\/}, {\bf 671}, 872

\bibitem[\protect\astroncite{{Zuckerman} {\em et~al.\/}}{2003}]{zuckerman03}
{Zuckerman}, B., {Koester}, D., {Reid}, I.~N., \& {H{\"u}nsch}, M. 2003, {\em
  \apj\/}, {\bf 596}, 477

\bibitem[\protect\astroncite{{Zuckerman} {\em et~al.\/}}{2010}]{zuckerman10}
{Zuckerman}, B., {Melis}, C., {Klein}, B., {Koester}, D., \& {Jura}, M. 2010,
  {\em \apj\/}, {\bf 722}, 725

\end{thebibliography}
\end{document}